\documentclass[twocolumn]{aastex62}

\usepackage[utf8]{inputenc}
\DeclareUnicodeCharacter{2212}{-}% support older LaTeX versions
\usepackage{graphicx}
\graphicspath{{./}}
\usepackage{amsmath}
\usepackage{afterpage}
\usepackage{enumitem}
\usepackage{xcolor}
\usepackage{float}
\setenumerate{itemsep=0mm}

\usepackage{amsmath}
\usepackage{amssymb}
\usepackage{xspace}
\usepackage{xifthen}
\usepackage{eso-pic}

\newcommand{\Gaia}{{\it Gaia}\xspace}

\definecolor{forestgreen}{HTML}{228B22}
\definecolor{urlblue}{HTML}{000000}

\mathchardef\mhyphen="2D

\newcommand{\roughly}{\ensuremath{ {\sim}\,} }

\newlength{\dhatheight}

\newcommand{\code}[1]{\texttt{#1}\xspace}

\newcommand{\unit}[1]{\ensuremath{\mathrm{\,#1}}\xspace}

\newcommand{\Gyr}{\unit{Gyr}}
\newcommand{\Myr}{\unit{Myr}}

\newcommand{\degree}{\ensuremath{{}^{\circ}}\xspace}
\newcommand{\degrees}{\degree}

\newcommand{\pc}{\unit{pc}}
\newcommand{\kpc}{\unit{kpc}}
\newcommand{\Mpc}{\unit{Mpc}}

\newcommand{\Msun}{\unit{M_\odot}}

\newcommand{\Lsun}{\unit{L_\odot}}

\newcommand{\secref}[1]{Section~\ref{sec:#1}}

\newcommand{\tabref}[1]{Table~\ref{tab:#1}}

\newcommand{\figref}[1]{Figure~\ref{fig:#1}}

\newcommand{\bandvar}[2][]{%
  \ifthenelse{\isempty{#1}}{\var{#2}}{\var{#2\_#1}}%
}

\newcommand{\ra}{{\ensuremath{\alpha_{2000}}}\xspace}
\newcommand{\dec}{{\ensuremath{\delta_{2000}}}\xspace}
\newcommand{\age}{{\ensuremath{\tau}}\xspace}

\newcommand{\ellip}{\ensuremath{\epsilon}\xspace}
\newcommand{\PA}{\ensuremath{\mathrm{P.A.}}\xspace}

\newcommand{\HEALPix}{\code{HEALPix}}
\newcommand{\healpix}{\HEALPix}

\newcommand{\emcee}{\code{emcee}}
\newcommand{\ugali}{\code{ugali}}
\newcommand{\simple}{\code{simple}}
\newcommand{\var}[1]{\ensuremath{\texttt{\MakeUppercase{#1}}}\xspace}

\providecommand\physrep{\ref@jnl{Phys.~Rep.}}%
\providecommand\apjs{\ref@jnl{ApJS}}%
\providecommand{\jcap}{\ref@jnl{JCAP}}%
\shorttitle{Discovery and Confirmation of an Ultra-Faint Dwarf Galaxy in Pegasus}

\shortauthors{DELVE Collaboration}

\begin{document}

\reportnum{\footnotesize 
FERMILAB-PUB-22-211-PPD}

\title{Pegasus IV: Discovery and Spectroscopic Confirmation of an Ultra-Faint Dwarf Galaxy in the Constellation Pegasus}

\author[0000-0003-1697-7062]{W.~Cerny}
\affiliation{Kavli Institute for Cosmological Physics, University of Chicago, Chicago, IL 60637, USA}
\affiliation{Department of Astronomy and Astrophysics, University of Chicago, Chicago IL 60637, USA}

\author[0000-0002-4733-4994]{J.~D.~Simon}
\affiliation{Observatories of the Carnegie Institution for Science, 813 Santa Barbara Street, Pasadena, CA 91101, USA}

\author[0000-0002-9110-6163]{T.~S.~Li}
\affiliation{Department of Astronomy and Astrophysics, University of Toronto, 50 St. George Street, Toronto ON, M5S 3H4, Canada}

\author[0000-0001-8251-933X]{A.~Drlica-Wagner}
\affiliation{Fermi National Accelerator Laboratory, P.O.\ Box 500, Batavia, IL 60510, USA}
\affiliation{Kavli Institute for Cosmological Physics, University of Chicago, Chicago, IL 60637, USA}
\affiliation{Department of Astronomy and Astrophysics, University of Chicago, Chicago IL 60637, USA}

\author[0000-0002-6021-8760]{A.~B.~Pace}
\affiliation{McWilliams Center for Cosmology, Carnegie Mellon University, 5000 Forbes Avenue, Pittsburgh, PA 15213, USA}

\author[0000-0002-9144-7726]{C.~E.~Mart\'inez-V\'azquez}
\affiliation{Gemini Observatory, NSF's NOIRLab, 670 N. A'ohoku Place, Hilo, HI 96720, USA}
\affiliation{Cerro Tololo Inter-American Observatory, NSF's National Optical-Infrared Astronomy Research Laboratory, Casilla 603, La Serena, Chile}

\author[0000-0002-7134-8296]{A.~H.~Riley}
\affiliation{George P. and Cynthia Woods Mitchell Institute for Fundamental Physics and Astronomy, Texas A\&M University, College Station, TX 77843, USA}
\affiliation{Department of Physics and Astronomy, Texas A\&M University, College Station, TX 77843, USA}

\author[0000-0001-9649-4815]{B.~Mutlu-Pakdil}
\affiliation{Kavli Institute for Cosmological Physics, University of Chicago, Chicago, IL 60637, USA}
\affiliation{Department of Astronomy and Astrophysics, University of Chicago, Chicago IL 60637, USA}

\author[0000-0003-3519-4004]{S.~Mau}
\affiliation{Department of Physics, Stanford University, 382 Via Pueblo Mall, Stanford, CA 94305, USA}
\affiliation{Kavli Institute for Particle Astrophysics \& Cosmology, P.O.\ Box 2450, Stanford University, Stanford, CA 94305, USA}

\author[0000-0001-6957-1627]{P.~S.~Ferguson}
\affiliation{Department of Physics, University of Wisconsin-Madison, Madison, WI 53706, USA}

\author[0000-0002-8448-5505]{D.~Erkal}
\affiliation{Department of Physics, University of Surrey, Guildford GU2 7XH, UK}

\author[0000-0002-0810-5558]{R.~R.~Munoz}
\affiliation{Departamento de Astronom\'ia, Universidad de Chile, Camino El Observatorio 1515, Las Condes, Santiago, Chile}

\author[0000-0003-4383-2969]{C.~R.~Bom}
\affiliation{Centro Brasileiro de Pesquisas F\'isicas, Rua Dr. Xavier Sigaud 150, 22290-180 Rio de Janeiro, RJ, Brazil}

\author[0000-0002-3936-9628]{J.~L.~Carlin}
\affiliation{Rubin Observatory/AURA, 950 North Cherry Avenue, Tucson, AZ, 85719, USA}

\author{D.~Carollo}
\affiliation{INAF-Osservatorio Astronomico di Trieste, I-34143 Trieste, Italy}

\author[0000-0003-1680-1884]{Y.~Choi}
\affiliation{Space Telescope Science Institute, 3700 San Martin Drive, Baltimore, MD 21218, USA}

\author[0000-0002-4863-8842]{A.~P.~Ji}
\affiliation{Department of Astronomy and Astrophysics, University of Chicago, Chicago IL 60637, USA}
\affiliation{Kavli Institute for Cosmological Physics, University of Chicago, Chicago, IL 60637, USA}

\author{D.~Mart\'{i}nez-Delgado}
\affiliation{Instituto de Astrof\'{i}sica de Andaluc\'{i}a, CSIC, E-18080 Granada, Spain}

\author{V.~Manwadkar}
\affiliation{Department of Astronomy and Astrophysics, University of Chicago, Chicago IL 60637, USA}

\author[0000-0002-7483-7327]{A.~E.~Miller}
\affiliation{Leibniz-Institut für Astrophysik Potsdam (AIP), An der Sternwarte 16, D-14482 Potsdam, Germany}
\affiliation{Institut f\"{u}r Physik und Astronomie, Universit\"{a}t Potsdam, Haus 28, Karl-Liebknecht-Str. 24/25, D-14476 Golm (Potsdam), Germany}

\author{N.~E.~D.~No\"el} 
\affiliation{Department of Physics, University of Surrey, Guildford GU2 7XH, UK}

\author[0000-0002-1594-1466]{J.~D.~Sakowska}
\affiliation{Department of Physics, University of Surrey, Guildford GU2 7XH, UK}

\author[0000-0003-4102-380X]{D.~J.~Sand}
\affiliation{Department of Astronomy/Steward Observatory, 933 North Cherry Avenue, Room N204, Tucson, AZ 85721-0065, USA}

\author[0000-0003-1479-3059]{G.~S.~Stringfellow}
\affiliation{Center for Astrophysics and Space Astronomy, University of Colorado, 389 UCB, Boulder, CO 80309-0389, USA}

\author[0000-0002-9599-310X]{E.~J.~Tollerud}
\affiliation{Space Telescope Science Institute, 3700 San Martin Drive, Baltimore, MD 21218, USA}

\author[0000-0003-4341-6172]{A.~K.~Vivas}
\affiliation{Cerro Tololo Inter-American Observatory, NSF's National Optical-Infrared Astronomy Research Laboratory,\\ Casilla 603, La Serena, Chile}

\author{J.~A.~Carballo-Bello}
\affiliation{Instituto de Alta Investigaci\'on, Sede Esmeralda, Universidad de Tarapac\'a, Av. Luis Emilio Recabarren 2477, Iquique, Chile}

\author{D.~Hernandez-Lang}
\affiliation{Ludwig-Maximilians-Universität München, Scheinerstraße 1, München, Germany}

\author[0000-0001-5160-4486]{D.~J.~James}
\affiliation{ASTRAVEO, LLC, PO Box 1668, Gloucester, MA 01931}

\author{J.~L.~Nilo Castellon}
\affiliation{Departamento de Astronom\'ia, Universidad de La Serena, Avenida Juan Cisternas 1200, La Serena, Chile}
\affiliation{Direcci\'on de Investigaci\'on y Desarrollo, Universidad de La Serena, Av. Ra\'ul Bitr\'an Nachary N. 1305, LaSerena, Chile}

\author[0000-0002-7134-8296]{K.~A.~G.~Olsen}
\affiliation{NSF's National Optical-Infrared Astronomy Research Laboratory, 950 N. Cherry Ave., Tucson, AZ 85719, USA}

\author[0000-0001-6455-9135]{A.~Zenteno}
\affiliation{Cerro Tololo Inter-American Observatory, NSF's National Optical-Infrared Astronomy Research Laboratory,\\ Casilla 603, La Serena, Chile}

\correspondingauthor{William Cerny}
\email{williamcerny@uchicago.edu}

% \color{red}
\collaboration{(DELVE Collaboration)}
% \color{black}

\begin{abstract}
We report the discovery of Pegasus~IV, an ultra-faint dwarf galaxy found in archival data from the Dark Energy Camera processed by the DECam Local Volume Exploration Survey. Pegasus~IV is a  compact, ultra-faint stellar system ($r_{1/2} = 41^{+8}_{-6}$ pc; $M_V = -4.25 \pm 0.2$ mag) located at a heliocentric distance of $90^{+4}_{-6} \kpc$. Based on spectra of seven non-variable member stars observed with Magellan/IMACS, we confidently resolve Pegasus~IV's velocity dispersion, measuring $\sigma_{v} = 3.3^{+1.7}_{-1.1}$~$\text{ km s}^{-1}$ (after excluding three velocity outliers); this implies a mass-to-light ratio of $M_{1/2}/L_{V,1/2} = 167^{+224}_{-99} M_{\odot}/L_{\odot}$ for the system. From the five stars with the highest signal-to-noise spectra, we also measure a systemic metallicity of $\rm [Fe/H] = -2.67^{+0.25}_{-0.29}$ dex, making Pegasus~IV one of the most metal-poor ultra-faint dwarfs.  We tentatively resolve a non-zero metallicity dispersion for the system.
These measurements provide strong evidence that Pegasus~IV is a dark-matter-dominated dwarf galaxy, rather than a star cluster. We measure Pegasus~IV's proper motion using data from \textit{Gaia} Early Data Release 3, finding ($\mu_{\alpha*}, \mu_{\delta}) = (0.33\pm 0.07, -0.21 \pm 0.08) \text{  mas yr}^{-1}$. When combined with our measured systemic velocity, this proper motion suggests that Pegasus~IV is on an elliptical, retrograde orbit, and is currently near its orbital apocenter. Lastly, we identify three potential RR Lyrae variable stars within Pegasus~IV, including one candidate member located more than ten half-light radii away from the system's centroid. The discovery of yet another ultra-faint dwarf galaxy strongly suggests that the census of Milky Way satellites is still incomplete, even within 100 kpc. 

\end{abstract}


\section{Introduction}
\label{sec:intro}
Ultra-faint dwarf galaxies represent some of the most extreme galaxies in the known universe: they are the smallest, least luminous, least metal-enriched, and most dark-matter dominated galaxies yet discovered \citep[e.g.,][]{2006ApJ...650L..51M, 2012AJ....144....4M, 2019ARA&A..57..375S}. These systems were formed at high redshift, likely before the epoch of reionization, and thus serve as well-preserved ``fossils" that trace the assembly and chemical enrichment histories of their host galaxies \citep[e.g.,][]{2005ApJ...635..931B, 2009ApJ...693.1859B, 2010AN....331..474F,2014ApJ...786...74F,2014ApJ...796...91B}. By virtue of their high dark matter content and comparatively minimal baryonic components, these systems are pristine laboratories for studying the nature of dark matter itself. For example, nearby ultra-faint dwarf galaxies are promising sites for the indirect detection of dark matter annihilation or decay through gamma-ray signals \citep[e.g.,][]{PhysRevD.89.042001,2017ApJ...834..110A,2018RPPh...81e6901S}, and the kinematics of stars in these galaxies offer the ability to test the cold dark matter paradigm's prediction for the inner density profile of dark matter halos \citep[e.g.,][]{1995ApJ...447L..25B,2021A&A...651A..80Z,2021arXiv211209374Z}. Additionally, the number and distribution of these systems around the Milky Way can also be leveraged to gain further insight into dark matter microphysics \citep[e.g.,][]{2012MNRAS.420.2318L,2017ARA&A..55..343B, 2021PhRvL.126i1101N}.

The considerable wealth of information about galaxy formation and dark matter encoded in ultra-faint dwarf galaxies has motivated extensive efforts toward their discovery and characterization. Although these galaxies are expected to be the most common class of galaxy by number, their extremely low luminosity has limited their study to the very local universe, where these systems have been discovered exclusively as resolved satellites of the Milky Way, the Magellanic Clouds, and the closest galaxies in the Local Volume (within $\sim 5 \Mpc$). Dedicated searches using deep, wide-area photometric catalogs from digital sky surveys have proven to be extremely successful, resulting in the discovery of more than 60 of these systems to date \citep[e.g.,][]{2005AJ....129.2692W,2006ApJ...650L..41Z,2007ApJ...662L..83W,2007ApJ...654..897B,2014MNRAS.441.2124B,2015ApJ...804L..44K,Bechtol:2015,Koposov_2015,Drlica-Wagner:2015,2015ApJ...802L..18L,2016MNRAS.463..712T,2018MNRAS.479.5343K,Torrealba:2019}. In turn, the characterization of these systems has benefited from follow-up spectroscopy, which can provide robust measurements of the metallicity and mass-to-light ratios of these systems \citep[e.g.,][]{2005ApJ...630L.141K,2007ApJ...670..313S,2008ApJ...685L..43K,2013ApJ...768..172C,2020ApJ...892..137S,2021ApJ...920...92J}.

\par  Despite the explosion of discoveries in the last two decades, cold dark matter simulations predict that numerous ultra-faint Milky Way satellites remain to be discovered, even in regions of sky covered by previous sky surveys \citep[e.g.,][]{2014ApJ...795L..13H,2018MNRAS.479.2853N,2020ApJ...893...48N,2021arXiv211204511M}. This prediction has recently been affirmed by the discovery of three new Milky Way satellite galaxies by the Hyper Suprime-Cam Subaru Strategic Program \citep{Homma:2016,Homma:2018,Homma:2019} and four additional satellites (including both dwarf galaxy candidates and globular clusters) by the DECam Local Volume Exploration (DELVE; \citealt{2021ApJS..256....2D, 2020ApJ...890..136M, 2021ApJ...910...18C,2021ApJ...920L..44C}). 
\par In this work, we present the discovery and characterization of yet another ultra-faint Milky Way satellite by DELVE. This new system, Pegasus~IV, lies at the very northern edge of sky accessible to the Dark Energy Camera (DECam; \citealt{2015AJ....150..150F})
in a region previously covered at a shallower depth by the Sloan Digital Sky Survey (SDSS; \citealt{2000AJ....120.1579Y} and the Panoramic Survey Telescope and Rapid Response System 1 survey (PS1; \citealt{2016arXiv161205560C}). We use medium-resolution Magellan/IMACS spectroscopy to measure the metallicities and line-of-sight velocities of candidate member stars. We resolve a stellar velocity dispersion and confirm that this system is a dark-matter-dominated ultra-faint dwarf galaxy. 
\par This paper is organized as follows. In \secref{search}, we describe the DELVE survey, its photometric catalogs, and our ongoing search for undiscovered ultra-faint stellar systems. We also introduce the newly discovered system Pegasus~IV. In \secref{ugali}, we characterize the morphology and stellar population of Pegasus~IV through maximum-likelihood fits to DELVE photometric data. In \secref{spec}, we measure the velocities of stars in the field of Pegasus~IV and use the resolved velocity dispersion to infer its mass and dark matter content. We also measure [Fe/H] metallicities and find tentative evidence for a metallicity spread. In \secref{discussion}, we discuss the implications of these results for the Pegasus~IV's classification, leverage \Gaia proper motions and our velocity measurements to constrain its orbit, and highlight the presence of three RR Lyrae variable stars.  In \secref{summary}, we summarize these results and describe avenues for future study.

\section{DELVE Data and Satellite Search}
\label{sec:search}
\subsection{DELVE Data}
\label{sec:data}
The DELVE survey is an ongoing multi-component observational campaign seeking to achieve deep, contiguous coverage of the high-Galactic-latitude southern sky in the $g,r,i,z$ bands by combining 126 nights of new observations  with existing public archival DECam data. DELVE is split into three main survey components dedicated to studying the resolved stellar substructures and satellite populations of the Milky Way (DELVE-WIDE), the Magellanic Clouds (DELVE-MC), and four nearby galaxies with stellar mass similar to the Magellanic Clouds (DELVE-DEEP). To date, DELVE has taken ${\sim}20,000$ new exposures toward this goal, and is expected to finish collecting observations in the 2022B semester. A more detailed description of the DELVE science goals, observing strategy, and progress can be found in \citet{2021ApJS..256....2D}. 
\par For this work, we used a new internal photometric catalog for DELVE-WIDE covering nearly the entire sky accessible to DECam with $\delta_{\rm J2000} < +30\degree$ and $|b| > 10\degree$, excluding the Dark Energy Survey footprint. This new catalog will be described in detail in a forthcoming paper (A. Drlica-Wagner et al.\, in prep.); we describe the critical components here. We began by selecting all available DELVE and publicly-available exposures with exposure times between $30$ and $350$ seconds and effective exposure time scale factors $t_{\rm eff} >0.3$ \citep[see][]{Neilsen:2015}. After this selection, we were left with a total of ${\sim}40,000$ exposures, the largest contributors to which were the Dark Energy Camera Legacy Survey (DECaLS;  \citealt{2019AJ....157..168D}), the DECam eROSITA Survey (DeROSITAS)\footnote{http://astro.userena.cl/derositas/}, and DELVE itself. DELVE-WIDE primarily collects $g,i$ band observations, and the $r,z$ data come primarily from the former two survey programs.
\par We processed all exposures consistently using the DES Data Management Pipeline (DESDM; \citealt{2018PASP..130g4501M}), which reduces and detrends DECam images using custom seasonally-averaged bias and flat images, and performs background subtraction. Automated source detection and point-spread-function photometry was performed on individual reduced CCD images using SourceExtractor \citep{Bertin:1996} and PSFEx \citep{Bertin:2011}. Stellar positions were then calibrated against \Gaia Data Release 2 \citep{2018A&A...616A...1G} using \code{SCAMP} \citep{Bertin:2006}, and the photometry was calibrated on a CCD-by-CCD basis using zeropoints derived from the ATLAS Refcat2 catalog \citep{Tonry:2018} that were transformed into the DECam photometric system (see Appendix B of \citealt{2021ApJS..256....2D}). Lastly, the resulting calibrated \code{SourceExtractor} catalogs for each individual CCD image were merged into a unified multi-band object catalog following the procedure introduced in \citet{Drlica-Wagner:2015}.
\par Reddening due to interstellar dust was calculated for each object in the resultant catalog from a bilinear interpolation of the maps of \citet{Schlegel:1998} with the rescaling from \citet{Schlafly:2011}. Bandpass-specific extinctions were then derived using the coefficients used for DES DR1 \citep{2018ApJS..239...18A}. Hereafter, we utilize the subscript ``0'' to denote extinction-corrected magnitudes.

\subsection{Satellite Search}
We performed a matched-filter search for old, metal-poor stellar systems in the DELVE-WIDE catalog described above using the \code{simple} algorithm\footnote{\url{https://github.com/DarkEnergySurvey/simple}}  \citep{Bechtol:2015}, which has been succesfully leveraged to discover more than twenty Milky Way satellites to date.
We began by dividing the DELVE-WIDE catalog described in \secref{data} into HEALPix \citep{Gorski:2005} pixels at $\var{nside} = 32$ ($\roughly3.4~\text{ deg}^2$ per pixel). For each pixel, we selected stars consistent with an old $(\tau = 12.5 \Gyr$), metal poor ($Z = 0.0001$) PARSEC isochrone \citep{Bressan:2012}, which we scanned in distance modulus from 16.0 mag to 23.0 mag in intervals of 0.5 mag. Specifically, at each step in the distance modulus grid, we selected all stars with colors consistent with the isochrone locus in color--magnitude space following $\Delta(g-r)_0 < \sqrt{0.1^{2} + \sigma_{g}^2+ \sigma_{r}^2}$. Stars were defined as sources satisfying the criterion
{\par\small\begin{align*}
|\var{SPREAD\_MODEL\_G}| < 0.003 + \var{SPREADERR\_MODEL\_G},
\end{align*}\normalsize}
\noindent where the variable \var{SPREAD\_MODEL} and its associated error, \var{SPREADERR\_{\allowbreak}MODEL}, are calculated from a likelihood ratio between the best-fitting local PSF model and a more extended model derived from the same PSF model that is additionally convolved with a circular exponential disk model  \citep{2012ApJ...757...83D}.
After these selections, the resulting filtered stellar density field was smoothed by a $2'$ Gaussian kernel, and local density peaks were identified by iteratively raising a density threshold until fewer than ten distinct peaks remained. Lastly, we computed the Poisson significance of each peak relative to the local background field. Informed by previous searches using \code{simple}, we inspected diagnostic plots for all candidates above a significance threshold of $5.5\sigma$.

\begin{figure*}
    \centering
    \includegraphics[width = \textwidth]{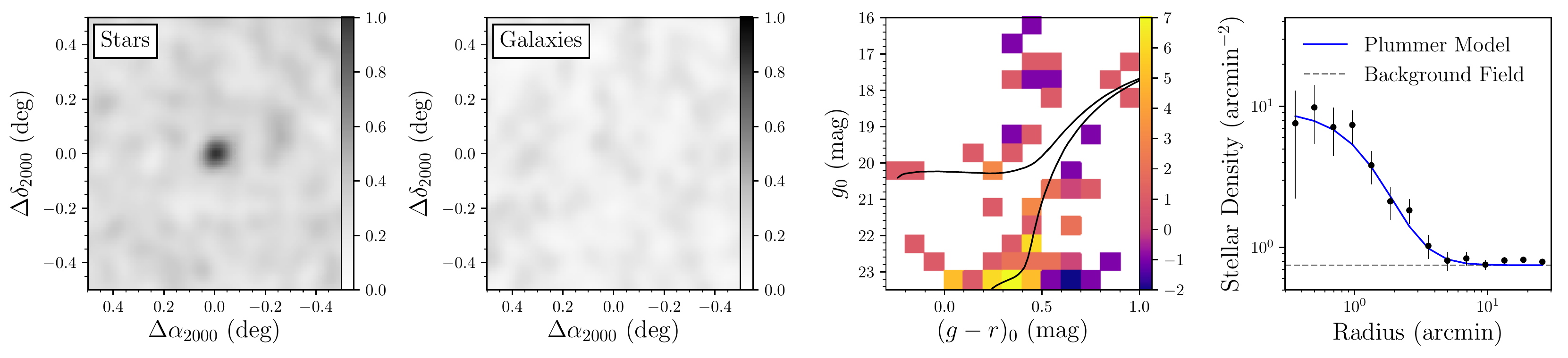}
    \caption{Diagnostic plots for Pegasus~IV similar to those visually inspected in the search results from \simple (except using the DECam follow-up exposures described in \secref{disco}). (Left) Smoothed spatial distribution of isochrone-filtered stars within a $0.25$ deg$^2$ region centered on Pegasus~IV. (Center Left) A similar plot to the leftmost panel, except showing the smoothed spatial distribution of galaxies. (Center Right) Hess diagram for a $r_h = 1.6'$ region centered on Pegasus~IV after subtracting the background signal from a concentric equal-area annulus at $11r_h$. The best-fit \citet{Bressan:2012} isochrone from the \ugali parameter fit (\secref{ugali}) is shown in black. (Right) Radial density profile of stars passing the isochrone filter. The errors are derived from the standard deviation of stellar counts in a given annulus divided by the area of that annulus. The best-fit Plummer model (see \secref{ugali}) is shown in blue. The dashed gray line corresponds to the background field stellar density.}
    \label{fig:diagnostic}
\end{figure*}

\subsection{Discovery of Pegasus~IV}
\label{sec:disco}
During visual inspection of the search results produced by \code{simple}, we identified a candidate stellar system near ($\alpha_{\rm J2000},\delta_{\rm J2000}) = (328.54\degree,26.62\degree$) at a significance of $6.2\sigma$.\footnote{This significance was an underestimate, as a result of a relatively poor initial distance modulus fit from \simple Our \ugali likelihood analysis (\secref{ugali}) later suggested a test statistic (TS) of $\rm TS = 198$, corresponding to a Gaussian significance of $\roughly~14.1\sigma$.} Within the candidate pool, this system was exceptional because it appeared to display seven stars at $g_0 \sim 20.5$ spanning a range of photometric color -- a feature indicative of a blue horizontal branch. Querying this candidate's centroid in the SIMBAD database \citep{2000A&AS..143....9W} revealed the existence of two RR Lyrae variable stars within a radius of $2\arcmin$, both of which were independently identified by the PS1 RR Lyrae catalog \citep{2017AJ....153..204S} and the \Gaia DR2 variability catalogs \citep{2018A&A...618A..30H,2019A&A...622A..60C}. 
\par These identifications strongly merited further investigation of the candidate system. However, the relatively shallow depth of the discovery data was found to be insufficient to draw firm conclusions about the nature and properties of this system. Therefore, we obtained additional $g,r,i$ imaging of the candidate system during regular DELVE observing and in DECam engineering time in August 2021. These newer observations consisted of $333$ second exposures centered on the candidate, improving the depth by $\sim 0.4$ mag in each band compared to the discovery data. These deeper exposures were then incorporated into a newer iteration of the DELVE catalog (prepared identically to the catalog described in \secref{data}), and this newer catalog was used for all analyses and figures in the following sections.
\par In \figref{diagnostic}, we present diagnostic plots for the candidate stellar system similar to those generated for each overdensity identified by \simple. These include the smoothed distribution of isochrone-filtered stars and galaxies (leftmost and center-left panels, respectively), a background-subtracted Hess diagram (center-right panel), and a radial profile for the system (rightmost panel), including the best-fit \citet{1911MNRAS..71..460P} model derived in \secref{ugali}. 
\par Our analyses described in the following sections strongly suggest that this system is an ultra-faint dwarf galaxy, rather than a star cluster. Therefore, following the historical naming convention for confirmed dwarf galaxy satellites of the Milky Way, we refer to the system as Pegasus~IV throughout this work.

\begin{figure*}
    \centering
    \includegraphics[width = .9\textwidth]{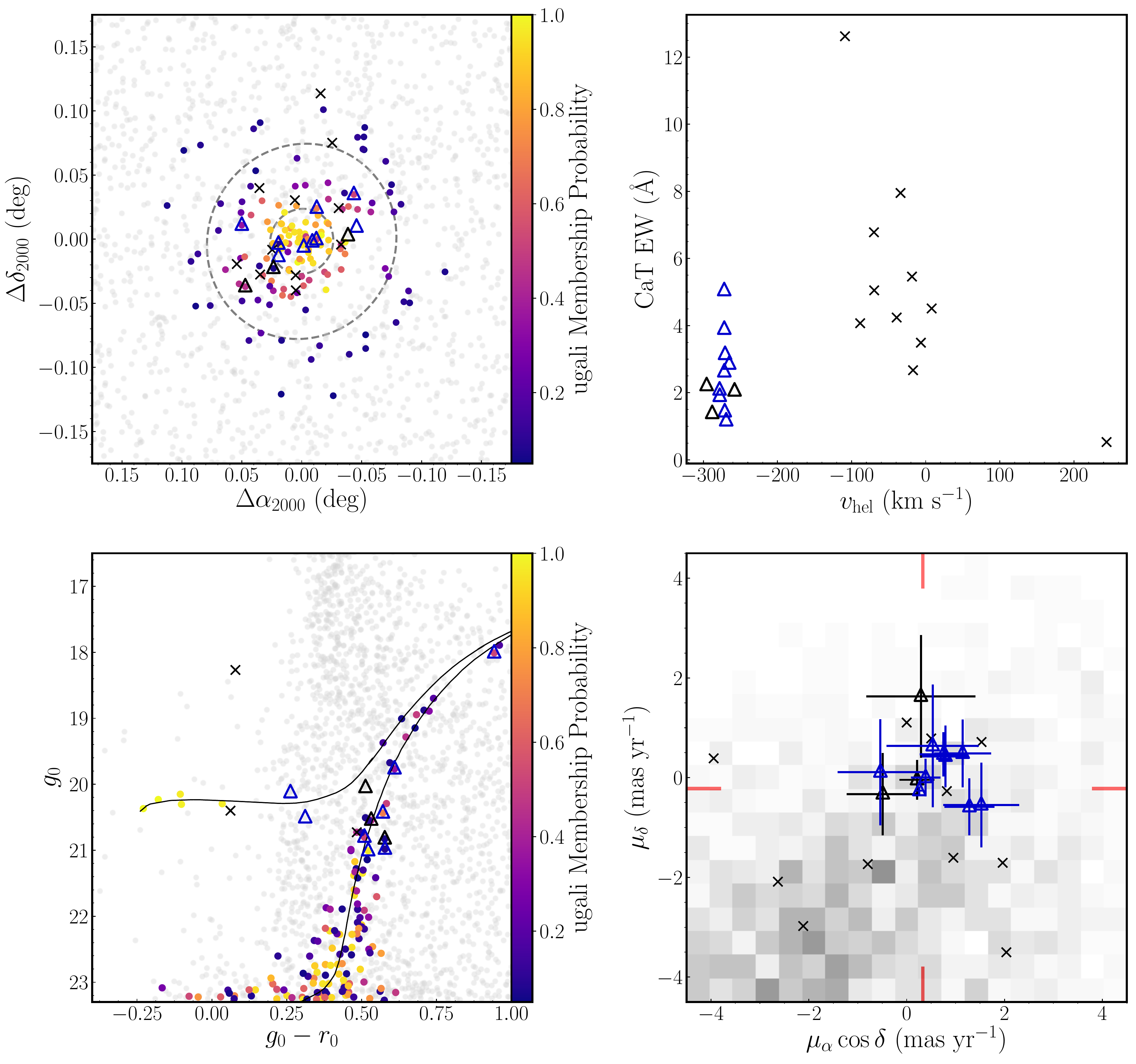}
    \caption{(Top Left) Spatial distribution of stars within a small region (radius of $\sim 0.25\degree$) centered on Pegasus~IV. Stars with \ugali probability $p_{\ugali} > 0.05$ are colored by their membership probability, while stars with $p_{\ugali} < 0.05$ are colored in gray.  Blue triangles denote the 9 clear candidate spectroscopic members, while black triangles denote candidate spectroscopic members with uncertain status. Black crosses denote stars observed spectroscopically but deemed non-members. Contours representing $1r_h$ and $3r_h$ are overplotted with a gray dashed line. (Top Right) Calcium Triplet Equivalent Width (CaT EW, in angstroms) vs.\ heliocentric radial velocity ($v_{\rm hel}$, in $\rm km \text{ } s^{-1}$) for all 23 stars observed with IMACS at S/N $> 3$.  (Bottom Left)  Color--magnitude diagram for the same region shown in the top left panel, with the same color/symbol scheme as the preceding panels. The two candidate horizontal branch variable stars are seen as blue triangles at $g_0 - r_0 \sim 0.25$. One member lacking DECam photometry is excluded here. The majority of nonmembers (especially those selected based on \Gaia alone) have redder colors than the axis range shown here.  (Bottom Right) \Gaia proper motions of the stars observed spectroscopically with IMACS, overlaid over a 2D proper motion histogram of all \Gaia sources within a radius of $\sim 0.25\degree$. The candidate spectroscopic members cluster closely near the systemic mean proper motion of $(\mu_{\alpha*}, \mu{\delta}) = (0.33, -0.22)~\rm mas~ yr^{-1}$, denoted by the red hatches (see \secref{proper_motion}).}
    \label{fig:mainfig}.
\end{figure*}

\section{Morphological Properties of Pegasus~IV}
\label{sec:ugali}
To determine Pegasus~IV's morphological properties and the nature of its stellar population, we used the maximum likelihood approach implemented in the Ultra-faint GAlaxy
LIkelihood toolkit (\code{ugali}\footnote{https://github.com/DarkEnergySurvey/ugali/}; \citealt{Bechtol:2015,Drlica-Wagner:2020}). Pegasus~IV's structure was modelled with a \citet{1911MNRAS..71..460P} stellar density profile, and a \citet{Bressan:2012} isochrone was fit to its observed color--magnitude diagram. We simultaneously constrained the centroid coordinates ($\alpha_{2000},\delta_{2000}$), angular semi-major axis length ($a_h$), ellipticity ($\ellip$), position angle East of North (P.A.) of the Plummer profile and the distance modulus $(m − M)_0$, age ($\tau$), and metallicity ($Z$) of the isochrone, in addition to the stellar richness ($\lambda$), which measures the total number of observable stars in the system.  To do so, we explored this multi-dimensional parameter space using the affine-invariant Markov Chain Monte Carlo sampler \code{emcee} \citep{Foreman-Mackey:2013}, and derived parameter estimates and uncertainties from the median and 16th/84th percentiles of the resulting posterior distributions.  Although we later measured a spectroscopic metallicity for the system, we did not fix the metallicity for this parameter fit, both to maintain consistency with the literature and to avoid potential systematic offsets in metallicity between isochrone models and the spectroscopic metallicities.

\par We report the values associated with each of these parameters, in addition to several properties derived from these results, above the first divider in \tabref{properties}. These extra derived properties include the system's azimuthally-averaged angular half-light radius ($r_h$), defined as $r_h = a_h \sqrt{1 − \ellip}$ and the system's absolute magnitude ($M_V$), integrated stellar luminosity ($L_V$), and stellar mass ($M_*$). The absolute V-band magnitude was derived following \citet{Martin:2008}, and both the stellar mass and stellar luminosity were computed by integrating along the best-fit isochrone assuming a \citet{Chabrier:2001} initial mass function.

\par The results from this parameter fit suggested that Pegasus~IV is a relatively small ($r_{1/2} = $ 41 pc), round (ellipticity consistent with zero) stellar system at a heliocentric distance of $D_{\odot} \sim 90$ kpc.  In the top left panel of \figref{mainfig},
we plot the spatial distribution of stars in a small region centered on Pegasus~IV. Stars with \ugali membership probabilities $p_{\ugali} > 5\%$ are colored by their membership probability; stars below this threshold are plotted in gray. Ellipses denoting $r_h$ and $3r_h$ are plotted with a gray dashed line. The bottom left panel displays a color--magnitude diagram covering the same area, with the same color scheme. The best-fit \citet{Bressan:2012} isochrone ($\rm [Fe/H] = -1.96$ dex) from the \ugali fit is shown as a solid black line. While this isochrone is more metal-rich than the spectroscopic metallicity we derive in the following sections, we note that the posterior distribution for the metallicity was bounded below at $Z = 0.0001$ ($\rm [Fe/H] = -2.2 $ dex), corresponding to the lowest metallicity in the \citet{Bressan:2012} library. The upper limit on  Pegasus~IV's metallicity from the \ugali fit was $\rm [Fe/H] = -1.92$ dex (at 95\% confidence), and thus our later identification of a lower metallicity for the system is not surprising.

\begin{deluxetable*}{l c c c c}
\tablecolumns{5}
\tablewidth{0pt}
\tabletypesize{\small}
\tablecaption{\label{tab:properties} Measured and Derived Properties of Pegasus IV}
\tablehead{\colhead{Parameter} & \colhead{Description} &  \colhead{Value} & \colhead{Units} & \colhead{Section}}
\startdata
\ra & Centroid Right Ascension &  $328.539^{+0.003}_{-0.004}$ & deg & \ref{sec:ugali}\\
\dec & Centroid Declination &  $26.620^{+0.003}_{-0.003}$ & deg & \ref{sec:ugali} \\
$a_h$ & Angular Semi-Major Axis Length & $1.60^{+0.29}_{-0.25}$  & arcmin & \ref{sec:ugali}  \\
$a_{1/2}$ & Physical Semi-Major Axis Length & ${42^{+8}_{-6}} $  &  pc & \ref{sec:ugali}  \\
$r_h$ & Azimuthally-Averaged Angular Half-Light Radius & $1.55^{+0.29}_{-0.24}$ & arcmin & \ref{sec:ugali} \\
$r_{1/2}$ & Azimuthally-Averaged Physical Half-Light Radius & ${41^{+8}_{-6}} $  & pc & \ref{sec:ugali}  \\
$\epsilon$ & Ellipticity & $< 0.41$\tablenotemark{a} & ... & \ref{sec:ugali}\\
\PA  & Position Angle of Major Axis (East of North) & $115^{+27}_{-41}$ & deg & \ref{sec:ugali} \\
$(m-M)_0$  & Distance Modulus & $19.77^{+ 0.03}_{-0.03} \pm 0.1$\tablenotemark{b} & mag & \ref{sec:ugali}, \ref{sec:rrl}\\
$D_{\odot}$  & Heliocentric Distance & $90^{+4}_{-6}$  & kpc & \ref{sec:ugali} \\
$\age$ & Age & $>12.5$\tablenotemark{c} & $\Gyr$ & \ref{sec:ugali} \\
$M_V$ & Absolute (Integrated) $V$-band Magnitude & $-4.25 \pm 0.2$\tablenotemark{d} & mag  & \ref{sec:ugali}  \\
$L_V$ & Luminosity & $4800^{+800}_{-700}$  & $\Lsun$ & \ref{sec:ugali} \\
$M_{*}$ & Stellar Mass & $4400^{+800}_{-600}$ & $\Msun$ & \ref{sec:ugali} \\
$E(B-V)$ & Mean Reddening Within the Half-Light Radius & 0.06 & mag & \ref{sec:ugali} \\
\hline
$N_{\rm spec}$ &  Number of Spectroscopic Members &   $9$ & ... & \ref{sec:specmem} \\ 
$v_{\rm hel}$ &  Systemic Radial Velocity in Heliocentric Frame &  $-273.6^{+1.6}_{-1.5}$ & km s$^{-1}$ & \ref{sec:dynamics}\\
$v_{\rm GSR}$ &  Systemic Radial Velocity in the Galactic Standard of Rest &  $-53.8 \pm 1.5$ & km s$^{-1}$ & \ref{sec:dynamics}\\
$\sigma_v$ & Velocity Dispersion &  $3.3^{+1.7}_{-1.1}$ & km s$^{-1}$  & \ref{sec:dynamics} \\
$M_{1/2}$ & Dynamical Mass within $r_{1/2}$ & $4.0^{+5.1}_{-2.3} \times 10^{5} $ & $\Msun$ & \ref{sec:dynamics}\\ 
$M_{1/2}/L_{V,1/2}$ & Mass-to-Light Ratio within $r_{1/2}$ & $167^{+224}_{-99}$ & $\Msun/\Lsun$ & \ref{sec:dynamics}\\ 
\hline
$[\rm Fe/H]_{\rm spec}$ &  Mean Spectroscopic Metallicity & $-2.67^{+0.25}_{-0.29}$ & dex & \ref{sec:mdisp}\\
$\sigma_{[\rm Fe/H]}$ &  Metallicity Dispersion among Spectroscopic Members & $0.46^{+0.29}_{-0.17}$ & dex & \ref{sec:mdisp}\\
\hline 
$\mu_{\alpha *}$ & Proper Motion in Right Ascension & $0.33\pm0.07$ & mas yr$^{-1}$ & \ref{sec:proper_motion}\\ % cos(\delta)
$\mu_{\delta}$ & Proper Motion in Declination &  $-0.21\pm0.08$ & mas yr$^{-1}$& \ref{sec:proper_motion}\\
$d_{\rm GC}$ & Galactocentric Distance & 89 & \kpc &\ref{sec:orbit}\\
$r_{\rm apo}$ & Orbital Apocenter & $94^{+8}_{-7}$ & \kpc & \ref{sec:orbit}\\
$r_{\rm peri}$ & Orbital Pericenter & $32_{-14}^{+18}$ & \kpc & \ref{sec:orbit}\\
$ e $ & Orbital Eccentricity & $0.49^{+0.17}_{-0.16}$ & ... & \ref{sec:orbit}\\
\hline 
$\log_{10} J(0.2\degree)$ & Integrated $J$-factor within a solid angle of 0.2\degree & $17.8\pm0.8$ & GeV$^2$cm$^{−5}$ & \ref{sec:jfactor} \\
$\log_{10} J(0.5\degree)$ & Integrated $J$-factor within a solid angle of 0.5\degree & $17.9\pm0.8$ & GeV$^2$cm$^{−5}$ & \ref{sec:jfactor} \\
\enddata
% \tablecomments{The quoted uncertainties were derived from the highest density interval containing the peak and 68\% of the marginalized posterior distribution.}
\tablenotetext{a}{The posterior distribution peaked near $\ellip = 0$. We therefore quote an upper limit at the 95\% confidence level.}
\tablenotetext{b}{Following \citet{Drlica-Wagner:2015}, we assume a systematic uncertainty of $\pm0.1$ mag on the distance modulus to account for uncertainties in isochrone modeling.}
\tablenotetext{c}{The posterior distribution peaked near $\tau= 13.5 \Gyr$, corresponding to the oldest age in our PARSEC isochrone grid. We therefore quote a lower limit at the 95\% confidence level.}
\tablenotetext{d}{The uncertainty in the absolute visual magnitude was calculated following \citet{Martin:2008} and does not include the uncertainty on the distance.}

\end{deluxetable*}

\section{Stellar Velocities and Metallicities from Magellan/IMACS Spectroscopy}
\label{sec:spec}
\subsection{Observations and Data Reduction}

\par To confirm that Pegasus~IV is a bound stellar system, and to determine its kinematic and dynamical properties, we observed the system with the 6.5m Magellan-Baade Telescope and the Inamori-Magellan Areal Camera and Spectrograph (IMACS; \citealt{2011PASP..123..288D}) on a two-night observing run spanning September 12--13, 2021.
Following previous studies of ultra-faint dwarf galaxies using IMACS, we used the instrument's f/4 camera and the 1200 $\ell/$mm grating blazed at $9000$ $\rm \AA$ \hspace{.25em}\citep[e.g.,][]{JDS17}.
The resulting spectra spanned a wavelength range of $\sim 7500-9000$ \AA at $R \sim 11,000$, sufficient for precise velocity and metallicity measurements from the Calcium Triplet (CaT) absorption feature centered at roughly $8500$ \AA.  
\par We observed a single multislit mask centered on the system, which featured 32 $0.7" \times 5"$ slits. Targets were chosen in the following order. Firstly, we selected red giant branch (RGB) and horizontal branch (HB) stars consistent with a \citet{2016ApJS..222....8D} isochrone with age $\tau = 12.5$ Gyr and metallicity $\rm [Fe/H] = -2.3$ in our DECam photometry, informed by past studies of ultra-faint dwarf galaxies. We then added bright stars that we identified as possible members on the basis of a preliminary mixture model analysis of their proper motions in $\Gaia$ EDR3 (see \secref{proper_motion}). Lastly, to fill remaining available space on the slitmask, we added several stars from \Gaia that lacked DECam photometry.
% identified as likely members by a preliminary fit using \code{ugali} 
\par Due to the northern declination of Pegasus~IV ($\delta_{2000} \sim +27\degree$) and the southern latitude of Las Campanas Observatory, we were only able to observe Pegasus~IV at airmass $\lesssim 1.8$ with Magellan/IMACS for a little over an hour on each night. 
On each night, we collected two science exposures (1800s + 2400s), followed by (Kr, Ar, Ne, He) arc lamp calibration frames and flat frames.  
The typical seeing for these observations was $1.1"$ on  September 12 and $0.75"$ on September 13.

\par We reduced the IMACS spectroscopic observations following the procedure described by \citet{JDS17}. In brief, this process first involved using the Cosmos reduction pipeline \citep{2011PASP..123..288D,2017ascl.soft05001O}  to map slits on the IMACS detector plane and achieve a preliminary wavelength solution based on the arc lamp data. Then, a modified version of the DEEP2 data reduction pipeline \citep{2012ascl.soft03003C,2013ApJS..208....5N} was used to extract and calibrate the one-dimensional spectrum for each star. We then combined the spectra from the four exposures using inverse-variance weighting.

\subsection{Velocity Measurements}
\label{sec:measureV}
We measured stellar radial velocities from the IMACS spectra following the method introduced in \citet{2017ApJ...838....8L}.
This method involves fitting the reduced spectrum of each star with velocity templates by shifting the template through a range of velocities to find the velocity $v_{\rm obs}$ that maximizes the likelihood
\begin{equation}
    \mathcal{L} = -\frac{1}{2} \sum_{\lambda = \lambda_{1}}^{\lambda_{2}} \frac{\left[f_{\rm spec}(\lambda) - f_{\rm temp}\left(\lambda\left(1 + \frac{v_{\rm obs}}{c}\right)\right)\right]^{2}}{\sigma_{\rm spec}^{2}}.
\end{equation}

\noindent Here, $f_{\rm spec}(\lambda)$ and $\sigma_{\rm spec}^{2}(\lambda)$ represent a normalized spectrum and its corresponding variance, and $f_{\rm temp}$ represents a normalized velocity template spectrum. 
Because we measured velocities specifically from the CaT absorption feature, we set the wavelength bounds of the spectral fit to be $\lambda_{1} = 8450\,{\rm \AA}$ and $\lambda_{2} = 8685\,{\rm \AA}$. 
All of our IMACS spectra were fit with three velocity templates:  HD122563, a very metal-poor RGB star; HD26297, a more metal-rich RGB star; and HD161817, a blue horizontal branch star. We report the velocity measurement from the template that produced the largest likelihood at the best-fit velocity.
\par For each spectrum-template combination, we ran the MCMC sampler implemented by \code{emcee} to sample the likelihood function above. To ensure robust sampling, we used 25 walkers each taking 2000 steps, with the first 500 steps for each walker discarded as burn-in. 
Then, for each star, we took the median and the standard deviation (after
$5\sigma$ clipping) of the velocity posterior distribution for the best-fit template as the measured
velocity $v_{\rm obs}$ and velocity error $\sigma_{v_{\rm obs}}$ respectively.
\par We next applied a telluric correction to this measured velocity $v_{\rm obs}$ to account for the miscentering of stars within slits, which can lead to small ($< 10 \text{ km s}^{-1}$) offsets in the measured velocities of stars (see e.g., \citealt{2007ApJ...663..960S}). To derive the correction for each spectrum, we re-ran the identical template-fitting MCMC procedure described above except with a telluric template, setting $\lambda_{1} = 7550 \rm \text{ } \AA$ and $\lambda_{2} = 7700 \rm \text{ } \AA$. 
The median and standard deviation of the resulting posterior distribution then provided the magnitude of the telluric correction $v_{\rm tell}$ and its associated variance $\sigma^{2}_{v_{\rm tell}}$.
\par The corrected velocity of each star, $v$, was calculated as $v = v_{\rm obs} - v_{\rm tell}$ with an associated uncertainty of $\sigma_{v_{\rm stat}} = \sqrt{\sigma^{2}_{v_{\rm obs}} + \sigma^{2}_{v_{\rm tell}}}$. The error $\sigma_{v_{\rm stat}}$ is purely statistical in nature, and is directly correlated with the $S/N$ of each individual spectrum. Informed by previous studies that considered the repeatability of IMACS velocities between successive nights \citep[e.g.,][]{JDS17,2018ApJ...857..145L}, we also added a $1.0 \text{ km s}^{-1}$ systematic error term in quadrature to each velocity measurement error.  
\par In summary, the above steps resulted in velocities $v$ for each star, each with a single associated uncertainty. These velocities were then transformed into the heliocentric frame. For the rest of this work, we denote the resulting heliocentric velocities as $v_{\rm hel}$. In total, we were able to measure reliable velocities for 23 unique stars at $S/N > 3$.
\subsection{Metallicity Measurements}
We measured the metallicity of red giant branch member stars in Pegasus~IV through the equivalent widths (EWs) of the CaT lines. We modelled each of the three CaT lines for each star with a Gaussian-plus-Lorentzian profile  \citep[e.g.,][]{2014A&A...572A..82H,2015ApJ...808...95S}, and converted their summed EWs to $\rm [Fe/H]$ metallicities using the calibration relation from \citet{2013MNRAS.434.1681C}. This relation requires an absolute V-band magnitude for each star, and thus we first converted from the DELVE $g,r$-band photometry to this system using the relation provided in  \citet{Bechtol:2015}, and then subtracted the distance modulus derived from the \code{ugali} fit (\secref{ugali}).
% For one star
The resulting error on the metallicity for each star was fully propagated from a combination of four sources: (1) uncertainty in the EW measurements, including a $0.2 {\rm \AA}$ systematic uncertainty floor \citep{2018ApJ...857..145L}; (2) uncertainties in the coefficients from the \citet{2013MNRAS.434.1681C} relation; (3) uncertainties in the DELVE photometry, and (4) uncertainty associated with the  distance modulus from \ugali. The first of these sources of error is dominant for all but the brightest star.
\par In general, accurate CaT EW measurements require higher signal-to-noise than accurate velocity measurements. Visual inspection of the spectra for stars in our sample revealed that the CaT fits for stars with low signal-to-noise were of poor quality, and thus we opted to impose a $S/N > 5$ cut for metallicity measurements. In total, we measured metallicities for 11 stars above this threshold.
\subsection{Spectroscopic Membership Determination}
\label{sec:specmem}
\par From the 23 spectra with $S/N > 3$ for which we measured velocities, we identified a clear clustering of twelve stars with radial velocities $-300\lesssim v_{\rm hel} \lesssim -250 \text{ km s}^{-1}$, including nine within the narrower range of $-282 \text{ km s}^{-1} \lesssim v_{\rm hel} \lesssim -262 \text{ km s}^{-1}$ (see top right panel of \figref{mainfig}). These twelve stars were separated in velocity from all other measured stars with $S/N >3$ by a gap of $>100 \text{ km s}^{-1}$, and were all located within $4' \text{ } (\sim 2.5r_h)$ of our derived centroid for Pegasus~IV. We summarize the key properties of these 12 stars in \tabref{specmem}.
\par To assess which stars among this sample of 12 were plausible Pegasus~IV members as opposed to Milky Way contaminants, we subjectively inspected these stars' proper motions from \Gaia EDR3, locations in color--magnitude space from the DELVE photometry, and heliocentric velocities and metallicities from the IMACS spectroscopy (where possible). We found that all 12 stars displayed self-consistent proper motions (within $1-2 \sigma$) and were photometrically consistent with an old, metal-poor isochrone (see bottom panels of \figref{mainfig}). Thus, we found no reason to reject any stars as members on the basis of color or proper motion information.

\begin{figure}
    \centering
    \includegraphics[width = \columnwidth]{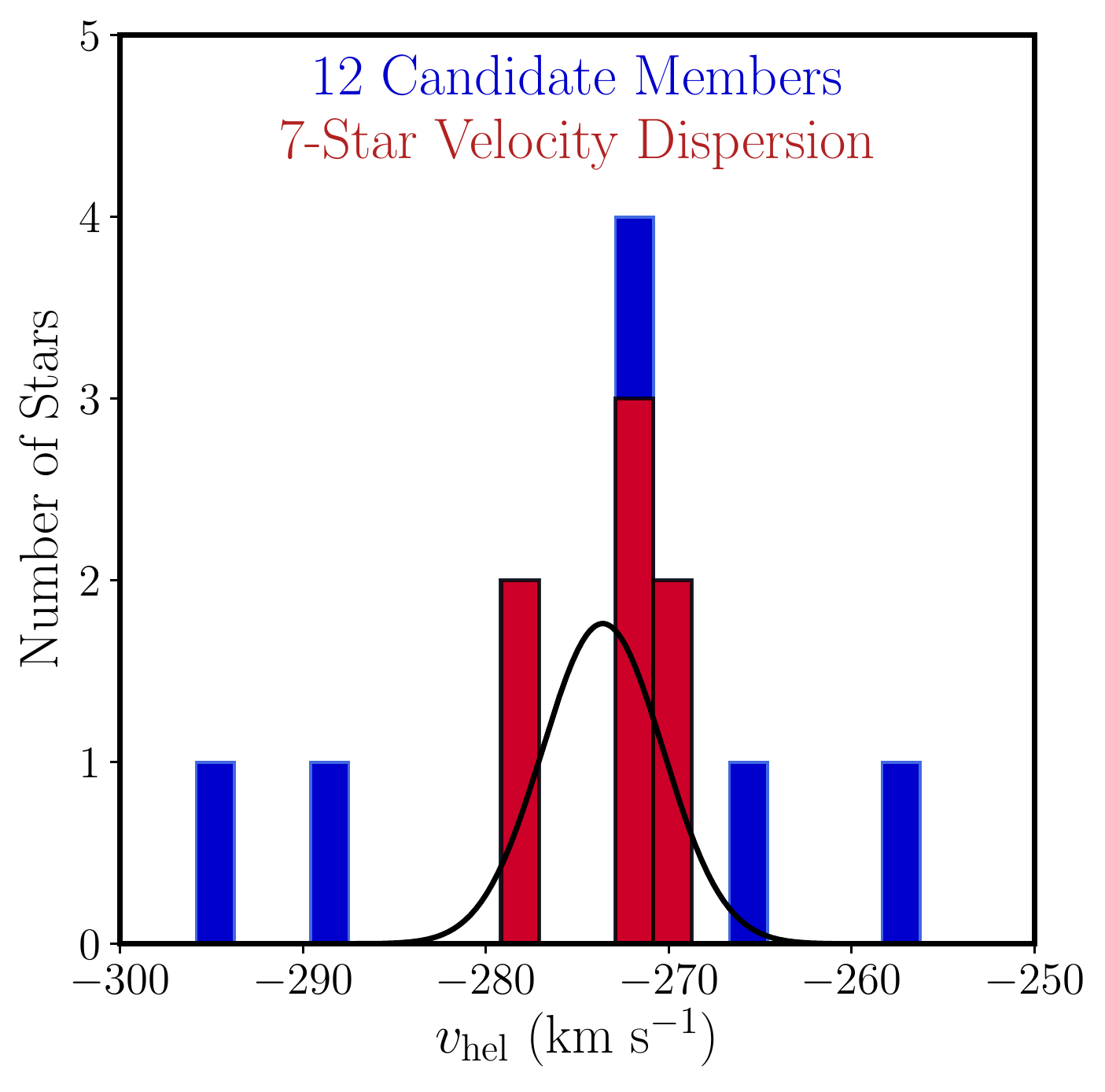}
    \caption{Histogram of radial velocities for the 12 stars identified as candidate members of Pegasus~IV (blue), including the 7-star subsample used for our dynamical analysis (red). The best-fit velocity dispersion model, which was derived from those  7 stars, is shown as a black Gaussian curve. The two stars that appear consistent with this model, but that are excluded from the red histogram, correspond to the two spectroscopically-observed candidate HB variable stars.}
    \label{fig:histo}
\end{figure}

\par The velocities of these 12 stars appeared to show a considerable spread, ranging from $-258 \text{ km s}^{-1} \lesssim v_{\rm hel}  \lesssim -296 \text{ km s}^{-1}$. As can be seen in \figref{histo}, nine of these stars lay within $10 \text{ km s}^{-1}$ of the apparent mode near $v_{\rm hel} \sim -272 \text{ km s}^{-1}$. The remaining three stars fell significantly outside of this range, lying at $v_{\rm hel} \sim [-258, -288, -296] \text{ km s}^{-1}$. Even if Pegasus~IV truly exhibits a large velocity dispersion, these stars' separation from the peak of the observed velocity distribution suggested that they are either non-members or are binary star members of Pegasus~IV that were observed at an orbital phase that places them far from their center-of-mass velocity.\footnote{No detectable variation in velocity for binary stars is expected in our data between the two successive nights of our observations.} 
\par The first of the aforementioned velocity outliers (at $v_{\rm hel} \sim -257 \text{ km s}^{-1}$) was found to have $\rm [Fe/H] = -2.80 \pm 0.33$ dex -- consistent with the mean metallicity of this system (see \secref{mdisp}).  In general, isolated halo stars at this metallicity are relatively rare \citep[e.g,][]{2009A&A...507..817S, 2020MNRAS.492.4986Y}. Thus, we presume that this star is a candidate binary member star of Pegasus~IV, rather than a non-member, but emphasize that this assumption has a significant impact on the measured velocity dispersion (see \secref{dynamics}). We further caution that the metallicity estimate for this star assumes that the star is located at the distance of Pegasus~IV, and will be underestimated should the star prove to be an foreground contaminant. In contrast to the case of the velocity dispersion, however, this star has only a small effect on our estimation of Pegasus~IV's metallicity dispersion (see \secref{mdisp}).
\par For the remaining two velocity outliers, both of which appeared to lie along the RGB of our best-fit isochrone, we were unable to confidently distinguish whether these stars are binary members or foreground non-members in the absence of multi-epoch velocity measurements. To assess the likelihood that these two stars could be Milky Way foreground contaminants, we ran a simulation using the web interface to the Besan\c{c}on Galactic model \citep{2003A&A...409..523R}. We first queried the model to produce a catalog of stellar magnitudes and kinematic measurements for simulated stars within a $1$ deg$^2$ region centered on Pegasus~IV. We then transformed the resultant magnitudes from the SDSS photometric system to the DECam photometric system using the equations provided by \citet{2018ApJS..235...33D}. Then, we computed the expected surface density of Milky Way stars within a radius $r < 30'$ that were consistent with the RGB of our target selection isochrone, had heliocentric radial velocities $-300 \text{ km s}^{-1} \leq v_{\rm hel} \leq -250 \text{ km s}^{-1}$, and had small proper motions ($|\mu| < 4 \text{ mas yr}^{-1}$  in each direction). After multiplying this surface density by the area of the region that the IMACS slitmask covered ($\sim 100$ arcmin$^2$), we found that $\sim1$ foreground star is expected in our spectroscopic sample within this velocity range. Our observation of two stars with outlying radial velocities is slightly inconsistent with this prediction, potentially suggesting that one or both of these stars is a binary member of Pegasus IV. We reiterate that the membership status of these two stars remains highly uncertain.

\par Lastly, to assess whether the brightest star was indeed a member star despite its relatively high metallicity ($\rm [Fe/H] = -2.03 \pm 0.11$ dex; first row of \tabref{specmem}), we measured the equivalent width of its Mg I $\lambda8807\,{\rm \AA}$ absorption line. As described by \citet{2012A&A...539A.123B}, this line can be used in conjunction with the CaT to discriminate between foreground Milky Way contaminants (primarily main-sequence stars) and dwarf galaxy members (red giants). Fitting the Mg I line with a Gaussian profile, we calculated the equivalent width to be $0.16 \pm 0.02\,{\rm \AA}$ (statistical error only). Given the star’s CaT equivalent width of $5.1 \pm 0.1 \pm 0.2 {\rm \AA}$ , this confidently places the star in the red giant regime defined by Equation 1 of \citet{2012A&A...539A.123B}, and thus we concluded that it is very likely that this star is a true RGB member of Pegasus~IV.\footnote{We opted not to conduct a similar analysis on the remaining candidate member stars because of their significantly lower S/N. Furthermore, \citet{2012A&A...539A.123B} suggest that the  contaminant/giant populations become less distinguishable in the Mg EW /  CaT EW plane for stars with metallicities approaching $\rm [Fe/H] = - 3.0$ dex, which would render this approach ineffective for the remaining stars for which we measured $S/N > 5$ metallicities.}

\par In summary, we identified nine clear spectroscopic member stars, in addition to one candidate binary member and two potential members with considerably uncertain status. Of the nine clear members, seven are RGB stars, and two appear to lie on the HB. The clear members are shown as blue triangles in \figref{mainfig}, while the potential members with uncertain status are shown as black triangles. One of the two spectroscopically-observed HB stars is classified as a RR-Lyrae-type variable in the PS1 and \Gaia RR Lyrae catalogs (see \secref{rrl}), and the other appears to show some signs of photometric variability in our data. We include all 12 of these stars in \tabref{specmem} as candidate members, but include a comment in the final column to highlight each of the uncertain cases.

% \movetabledown=2.7in
% \begin{rotatetable*}
\begin{deluxetable*}{l c c c c c c c c }
\tablecolumns{9}
\tablewidth{0pt}
\tabletypesize{\normalsize}
\tablecaption{\label{tab:specmem} Properties of Pegasus IV spectroscopic member candidates, ordered by decreasing IMACS spectrum S/N}
\tablehead{
\colhead{\Gaia EDR3 SourceID} & \colhead{R.A.}   &  \colhead{Decl.}  &  \colhead{$g_{0}$} &  \colhead{$r_{0}$}  & \colhead{S/N} &  \colhead{$v_{\rm hel}$}   &  \colhead{[Fe/H]} &  \colhead{Comment} \\
 \colhead{} & \colhead{(deg)}   &  \colhead{(deg)}  &  \colhead{(mag)} &  \colhead{(mag)} &  \colhead{}   &  \colhead{(km/s)}  &  \colhead{(dex)} &  \colhead{}
}
\startdata
1796890071833434112 & 328.52536 & 26.62094 & 18.01 & 17.07  & 52.4  & -271.9 $ \pm $ 1.0 & -2.03 $ \pm $  0.11 & RGB\\
1796887219975171328 & 328.59498 & 26.63219 & 19.77 & 19.16  & 10.6 & -278.1 $ \pm $ 1.3  & -2.85 $ \pm $ 0.36  & RGB\\
1796888907896667520 & 328.48786 & 26.63070 & \ldots\tablenotemark{a} & \ldots\tablenotemark{a}  & 7.9 & -277.9 $ \pm $ 1.7 & -2.89 $ \pm $ 0.35 & RGB\\
1796888834882857216 &  328.49587 & 26.62398 & 20.05 & 19.54  & 7.8 & -257.9 $ \pm $ 2.0 & -2.80\tablenotemark{b} $ \pm $ 0.33  & Binary/Non-member?\\
1796890381071133568 &  328.52503 & 26.64546 & 20.44 & 19.87  & 5.5 & -269.2 $ \pm $ 2.2 & -3.32 $ \pm $ 0.38 & RGB\\
1796890071833414784 & 328.53716 & 26.61518 & 20.13 & 19.87  & 4.7 & -265.1 $ \pm $ 1.8 & \ldots & HB; Variable?\\
1796886807658193536 & 328.59171 & 26.58421 & 20.54 & 20.01  & 4.6 & -295.6 $ \pm $ 2.5 & \ldots & Non-member? \\
1796887082536156928 & 328.56077 & 26.60775 & 20.51 & 20.20  & 3.6 & -271.8 $ \pm $ 3.7  & \ldots & HB; RR Lyrae\\
1796891171345139456 & 328.49033 & 26.65611 & 20.80 & 20.29  & 3.5 & -271.2 $ \pm $ 4.15 & \ldots & RGB \\
1796887048176397952 & 328.56543 & 26.59852 & 20.83 & 20.25  & 3.5 & -288.1 $ \pm $ 2.82  & \ldots & Non-member? \\
1796887151255658752 & 328.56076 & 26.61745 & 20.98 & 20.40  & 3.3 & -270.7 $ \pm $ 3.23 & \ldots & RGB \\
1796890071833423872 & 328.52913 & 26.61916 & 21.00 & 20.49  & 3.2 & -271.75 $ \pm $ 3.69 & \ldots & RGB \\
\enddata
\tablecomments{R.A. and Decl. coordinates are taken from $\Gaia$ EDR3. The $g,r$-band photometric measurements are taken from DELVE (with one exception; see below), and correspond to AB magnitudes in the DECam photometric system. The reported signal-to-noise ratios ($S/N$) refer to the IMACS spectroscopic data. We intend to release an extended version of this table that includes spectroscopically-observed non-members upon publication.}
\tablenotetext{a}{This star was not in the DELVE photometric catalog, as it was obscured by a charge-bleed artifact caused by a nearby bright star. For the purpose of deriving this star's [Fe/H] metallicity, we instead calculated a $V$-band magnitude for this star using the \Gaia EDR3 photometry and assumed a conservative error of $ \pm  0.1$ mag. This magnitude was then used in the \citet{2013MNRAS.434.1681C} relation.} 
\tablenotetext{b}{This metallicity assumes that the star is a member of Pegasus IV, and therefore that its distance is $\sim 90$ kpc. }

\end{deluxetable*}
% \end{rotatetable*} 

\subsection{Velocity Dispersion and Mass}
\label{sec:dynamics}
To constrain the systemic velocity ($v_{\rm hel}$) and velocity dispersion ($\sigma_v$) of Pegasus~IV, we sampled the two-parameter Gaussian likelihood function defined by Equation 8 of \citet{2006AJ....131.2114W} using \code{emcee}. We applied a uniform prior on $v_{\rm hel}$ with range $[-250,\, -300]\,{\rm \text{ } km\,s^{-1}}$, and a uniform prior on $\log(\sigma_v)$ with range $[-2,\,2]$.
For our primary kinematic measurements, we included only the seven clear (non-outlier) RGB member stars described in \secref{specmem}.  We excluded the two candidate variable stars on the HB from our kinematic sample, since the pulsation of variable stars causes their apparent velocities to vary over time.
\par
\begin{figure*}
    \centering
    \includegraphics[width = .45\textwidth]{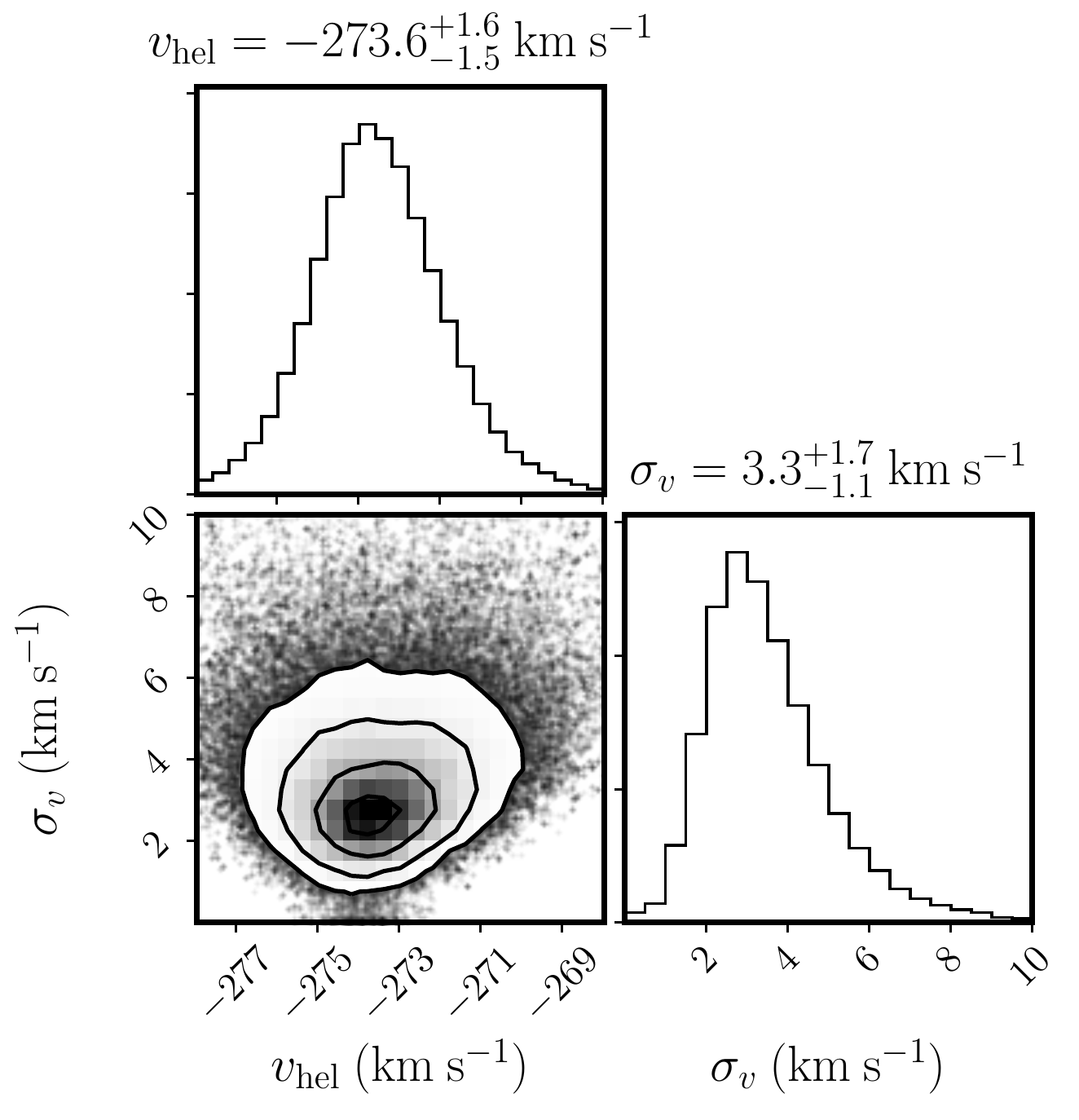}
    \includegraphics[width = .45\textwidth]{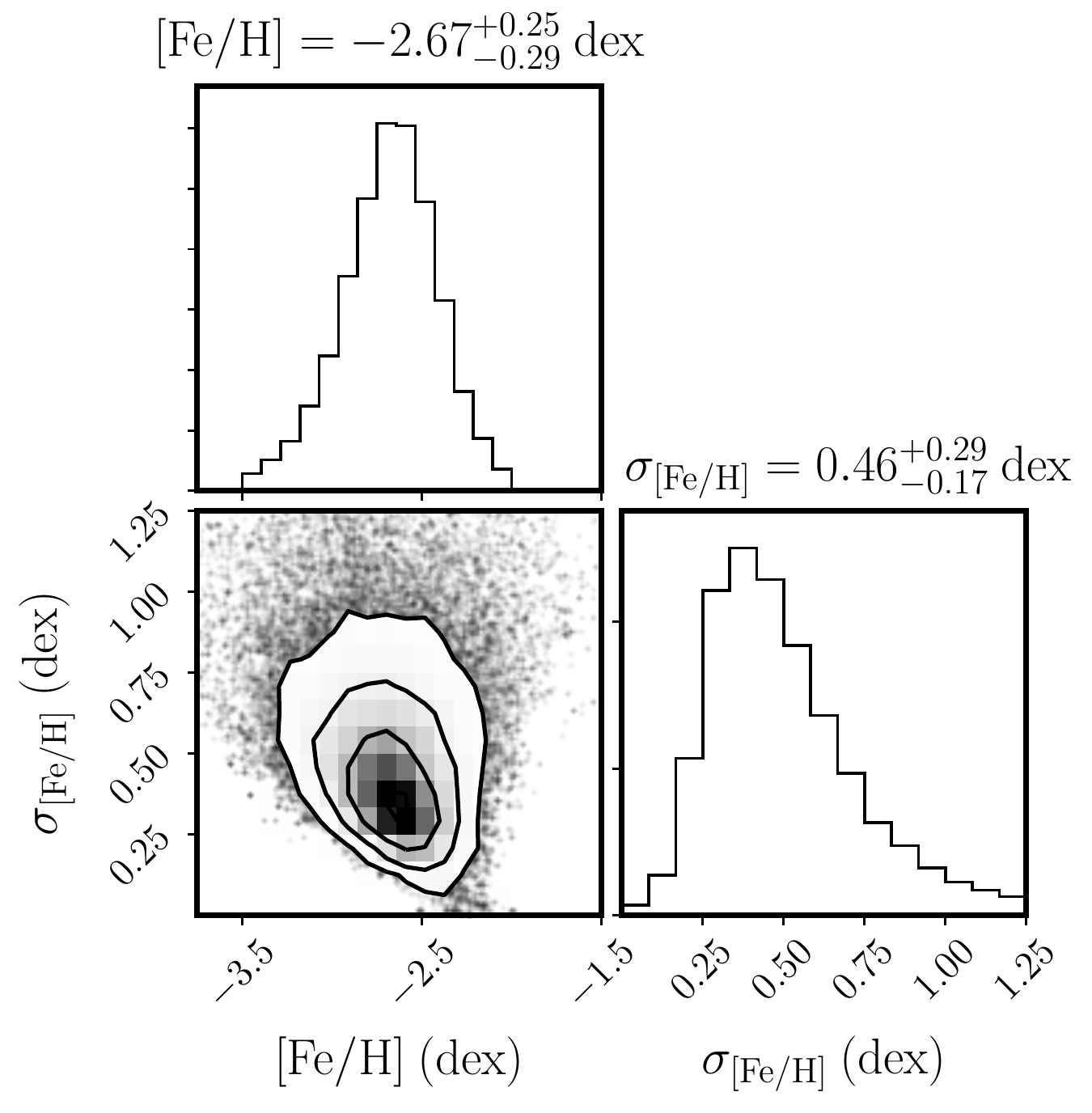}
    \caption{(Left) Two-dimensional posterior probability distributions for the systemic velocity and velocity dispersion, derived through the MCMC sampling procedure described in \secref{dynamics}. (Right) Similar posterior distributions for the metallicity and metallicity dispersion of Pegasus~IV (see \secref{mdisp}). The left panel used our nominal sample of seven non-variable, non-velocity-outlier stars, while the right panel used the five brightest candidate members. Four stars overlap between these samples.}
    \label{fig:dispersions}
\end{figure*}

 Using these seven stars, and applying the priors described above, we measured Pegasus~IV's systemic velocity to be $v_{\rm hel} = -273.6^{+1.6}_{-1.5} \text{ km s}^{-1}$ with a velocity dispersion of $\sigma_v = 3.3^{+1.7}_{-1.1} \text{ km s}^{-1}$. The resulting posterior probability distributions from the MCMC sampling are shown in the left panel of \figref{dispersions}, and the best-fit model is depicted in black over the velocity histogram in \figref{histo}.
To assess the impact of our prior on this measurement, we also explored adopting a flat prior on $\sigma_v$, rather than $\log(\sigma_v)$.  Holding all else constant, this change of prior resulted in changes to the systemic velocity and velocity dispersion that were significantly smaller than the quoted errors of our primary measurements.
\par Our measured velocity dispersion of $\sigma_{v} = 3.3^{+1.7}_{-1.1} \text{ km s}^{-1}$ is clearly non-zero, implying that we confidently resolved the internal dynamics of Pegasus~IV. However, the value of this dispersion was found to be sensitive to the exact member used in our velocity dispersion fit. In particular, we observed that including the metal-poor outlier at $v_{\rm hel} \sim -257 \text{ km s}^{-1}$ (while retaining our default priors) raised the velocity dispersion to $\sigma_{v_{\rm hel}} = 6.0^{+2.0}_{-1.3} \text{ km s}^{-1}$, consistent within $\lesssim 1.5\sigma$ of the measured dispersion from our nominal seven-star sample. Similarly, including all 12 candidate member stars would raise the velocity dispersion to $\sigma_v = 10.0_{-1.9}^{+2.8} \text{ km s}^{-1}$ (after relaxing the $\log(\sigma_v)$ prior to [-3,3]). Given that adopting these alternate member samples only increased the resulting velocity dispersion, our primary results derived from the seven-star sample can be considered as the most conservative estimate of the dark matter content of Pegasus~IV. This ensures that our ultimate conclusion that Pegasus~IV is a dark-matter dominated dwarf galaxy (see \secref{class}) is insensitive to assumptions about the nature of these apparent velocity outliers. 

\par Therefore, under the assumption that Pegasus~IV is a dispersion-supported system in dynamical equilbrium, we proceeded to estimate the system's dynamical mass using the mass estimator introduced in Equation 2 of \citet{2010MNRAS.406.1220W}:
\begin{equation}
M_{1/2} \approx 930\left(\frac{\sigma_{v}^{2}}{{\rm km}^2 \rm s^{-2}}\right) \left(\frac{r_{1/2}}{\rm pc}\right)  M_{\odot}.
\end{equation}
Using our measured dispersion from the nominal seven-star sample and the half-light radius from \secref{ugali}, we found Pegasus~IV's enclosed mass within $r_{1/2}$ to be $4.0^{+5.1}_{-2.3} \times 10^{5} M_{\odot}$. The mass-to-light ratio within one half-light radius is therefore $M_{1/2}/L_{V, 1/2} = 166^{+224}_{-99} M_{\odot}/L_{\odot}$.

% \subsection{J-Factor}

\subsection{Metallicity and Metallicity Spread}
\label{sec:mdisp}
To measure Pegasus~IV's mean metallicity ($\rm [Fe/H]_{\rm spec}$) and metallicity dispersion $(\sigma_{\rm [Fe/H]})$, we applied a simple Gaussian likelihood model that was nearly identical to the model used for the velocity and velocity dispersion. We adopted a uniform prior on $\log(\sigma_{\rm [Fe/H]})$ with range $[-2,\,2]$, and again performed MCMC sampling using \code{emcee}. By default, we opted only to use the five stars with $S/N >5$ (including the binary candidate at $v_{\rm hel} \sim -257 \text{ km s}^{-1}$).\footnote{The spectra and corresponding CaT fits for each of these stars are shown in Appendix \secref{spectra}.} For these stars, we found $\rm [Fe/H]_{\rm spec} = -2.67^{+0.25}_{-0.29}$ dex and metallicity dispersion $\sigma_{\rm [Fe/H]} = 0.46^{+0.29}_{-0.17}$ dex. The resulting posterior probability distributions are shown in the righthand panel of \figref{dispersions}. 
\par Given the small sample of stars with $S/N > 5$ (five in total), we performed a jacknife test \citep{1982jbor.book.....E} to assess the robustness of our measured metallicity and metallicity dispersion. We removed individual stars , one at a time, and re-ran the MCMC sampling. After doing so, we found that the brightest, most metal-rich star (first row of \tabref{specmem}) had a particularly strong influence on the metallicity and metallicity dispersion. Removing this star resulted in a more metal-poor systemic mean metallicity of $\rm [Fe/H] = -2.95 \pm 0.19$ dex and a much smaller dispersion of $0.06^{+0.16}_{-0.04}$ dex, in $\sim 2\sigma$ tension with the five-star measurement. By contrast, removing each of the three stars at $\rm [Fe/H] \sim -2.85$ dex (second, third, and fourth row of  \tabref{specmem}) minimally affected the mean metallicity and minorly increased the measured dispersion (well-within the uncertainties on the five-star sample dispersion). Lastly, removing the most-metal poor star (fifth row of  \tabref{specmem}, $\rm [Fe/H] \sim -3.3 $ dex) increased the mean metallicity to $\rm [Fe/H] = -2.49^{+0.27}_{-0.32}$ dex and resulted in a slightly smaller dispersion of $\sigma_{\rm [Fe/H]} = 0.39^{+0.35}_{-0.20}$ dex. Both of these are consistent within $1\sigma$ of the five-star result.
\par These results, in aggregate, suggest that Pegasus~IV is a metal-poor stellar system with a tentative detection of a non-zero metallicity dispersion. The magnitude of the dispersion is highly contingent on the membership of the brightest star. Since our measurement of the \ion{Mg}{1}\,$\lambda8807~{\rm \AA}$ line for this star gives no reason to doubt its membership, we opt to report the metallicity and corresponding dispersion from the five-star sample, namely, $\rm [Fe/H]_{\rm spec} = -2.67^{+0.25}_{-0.29}$ dex and $\sigma_{\rm [Fe/H]} = 0.46^{+0.29}_{-0.17}$, but we emphasize that the value of our measured dispersion is tentative and should be interpreted cautiously due to the small sample size. We also note that this includes the star with velocity of $v \sim 257 \rm \text{ km s}^{-1}$, which we assumed to be a true member, but excluded in our kinematic analysis given the likelihood that this star is an unresolved binary.

\par Regardless of the input stellar sample, the mean metallicity of Pegasus~IV places it among the most metal-poor ultra-faint dwarfs known, which include Reticulum~III, Bootes~II, Tucana~II, Horologium~I, Draco~II, and Reticulum~II, the metallicities of which range from $-2.81 < {\rm [Fe/H]} < -2.65$ dex. Our measurement suggests that Pegasus~IV is slightly more metal-poor than other dwarf galaxies of similar absolute magnitude (see right panel of \figref{pop_comparison}), but this difference does not appear to be statistically significant.
\par Our measured metallicity dispersion ($\sigma_{\rm [Fe/H]} = 0.46^{+0.29}_{-0.17}$ dex), while relatively uncertain, is comparable to the dispersions observed in other ultra-faint dwarf galaxies at similar absolute magnitude, i.e., Columba I, Coma Berenices I, Leo V, Pisces II, and Ursa Major II, which have $\sigma_{\rm [Fe/H]} = 0.71, 0.43, 0.30, 0.48, 0.67$ dex, respectively \citep{2019A&A...623A.129F, 2019ARA&A..57..375S,2021ApJ...920...92J, 2015ApJ...810...56K}. Pegasus~IV's metallicity dispersion can also be compared to the intrinsic iron abundance spreads observed in the Milky Way's (bright) globular cluster population, which \citet{2019ApJS..245....5B} found to have a median metallicity dispersion of $\sigma_{\rm [Fe/H]} = 0.045$ dex across a sample of 55 clusters with high-resolution spectra. 
\subsection{Proper Motion}
\label{sec:proper_motion}
We computed the systemic proper motion of Pegasus~IV using the precise astrometry provided by \Gaia EDR3 \citep{Gaia_Brown_2021A&A...649A...1G}. 
We analyzed three different proper motion models to measure the systemic proper motion.
The first was a mixture model composed of dwarf and Milky Way components and utilizes spatial position and proper motion \citep{Pace2019ApJ...875...77P}.
This model was run prior to the acquisition of both deeper photometry and spectroscopy, and its results informed our spectroscopic target selection. 
It was run with preliminary spatial parameters and only used stars with DECam photometry.  
With this model, we found $\mu_{\alpha *}=0.39\pm0.17\,{\rm mas\,yr^{-1}}$ and $\mu_{\delta}=0.01\pm0.18\,{\rm mas\,yr^{-1}}$. The model also reports the number of probable members with proper motion measurements, which was found to be $N=12.3\pm1.4$ stars. 
Due to the color--magnitude selection window, this model missed the brightest member, which explains the worse precision compared to the following models. 

The second proper motion model was similar to the first, but used only a fixed sample of spectroscopic members as input. 
We used a multi-variate Gaussian distribution to model the dwarf, 
and we sampled the posterior probability using \texttt{emcee}.
With this model, we found $\mu_{\alpha *}=0.33\pm0.07\,{\rm mas\,yr^{-1}}$, $\mu_{\delta}=-0.21\pm0.08\,{\rm mas\,yr^{-1}}$, assuming a fixed sample of $N=9$ stars (consisting of the seven stars used for the dynamical analysis, in addition to the two spectroscopically-observed variable star candidates). 
The third model used a similar mixture model, but built on \citet{Pace2019ApJ...875...77P} by incorporating spectroscopic information.  We pre-assigned the membership of stars with spectroscopy, which assists in determining both the dwarf and Milky Way proper motion distributions. We did not exclude stars missing DECam photometry and instead applied a loose {\it Gaia} color--magnitude selection for these stars. 
With this model, we found $\mu_{\alpha *}=0.33\pm0.07\,{\rm mas\,yr^{-1}}$ and $\mu_{\delta}=-0.22\pm0.08\,{\rm mas\,yr^{-1}}$, and $N=13.1\pm0.6$.
The proper motion from this model is almost identical to the spectroscopic-member-only results from the second model, likely because the additional members are generally faint (mostly HB stars) and do not significantly influence the measurement.  
We note that the majority of the systemic proper motion precision comes from the brightest member. The proper motion error of this star is $\sigma_{\mu_{\alpha *}}=0.08 \text{ }{\rm mas\,yr^{-1}}$ (similar to the systemic proper motion error) and its inclusion decreases the systemic proper motion error by $\sim40\%$.
We opted to use the systemic proper motion derived from the spectroscopic members as our preferred measurement for further analysis of Pegasus~IV's kinematics, since this measurement is least likely to be biased by contaminant stars. We do note, however, that differences have been observed between dwarf galaxy proper motions derived from spectroscopic samples and those derived without (e.g., \citealt{2018A&A...620A.155M}).

\section{Discussion}
\label{sec:discussion}
\subsection{Classification of Pegasus~IV}
\label{sec:class}
\begin{figure*}
    \centering
    \includegraphics[width = \textwidth]{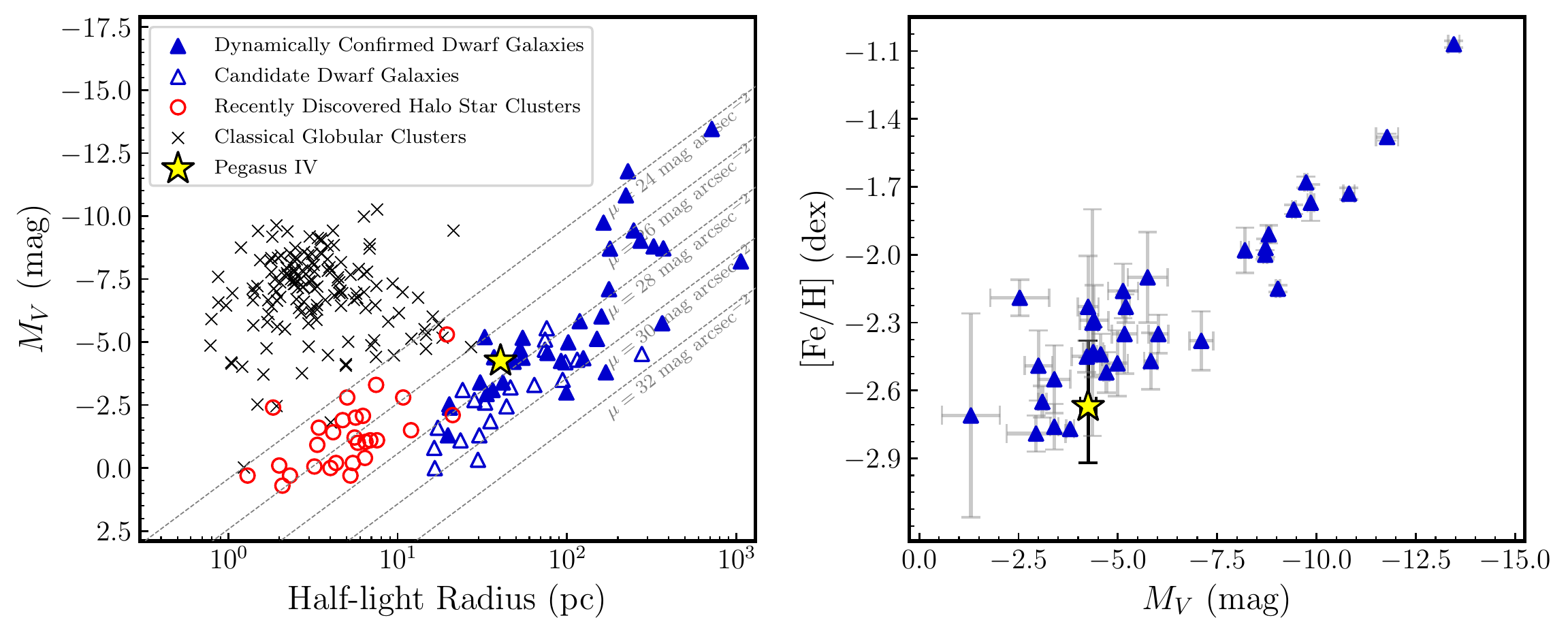}
    \caption{(Left) Absolute V-band magnitude ($M_V$) vs. azimuthally-averaged physical half-light radius ($r_{1/2}$) for the population of known Milky Way globular clusters, faint halo star clusters, and candidate and confirmed dwarf galaxies. Pegasus~IV's morphological properties are consistent with the population of candidate and confirmed ultra-faint dwarf galaxies. (Right) Absolute V-band magnitude vs. mean iron abundance $(\rm [Fe/H])$ for the population of dynamically-confirmed ultra-faint dwarf galaxies. Pegasus~IV appears to be more metal-poor compared to the population of known dwarfs at the same absolute magnitude, although its mean [Fe/H] metallicity is relatively uncertain. A full reference list for both panels is included in Appendix \secref{refs}.}
    \label{fig:pop_comparison}
\end{figure*}

Recent discoveries of ultra-faint Milky Way satellites have broadly consisted of two classes of objects: dark-matter-dominated dwarf galaxies and likely baryon-dominated halo star clusters. We find that Pegasus~IV is significantly more consistent with the former class of objects on the basis of its size, mass-to-light ratio, and metallicity dispersion.
Specifically, Pegasus~IV's half-light radius is larger than the population of known globular clusters (see left panel of \figref{pop_comparison}). More conclusively, Pegasus~IV's large mass-to-light ratio ($M_{1/2}/L_{V,1/2} =  167^{+224}_{-99}\ M_{\odot}/L_{\odot}$) is inconsistent with the known population of halo star clusters, which typically exhibit mass-to-light ratios of $\sim 1-3 \text{ } M_{\odot}/L_{\odot}$ \citep[e.g.,][]{2020MNRAS.492.3859D}. Lastly, the system's tentatively-resolved metallicity dispersion suggests that it has undergone multiple generations of star formation and/or that its gravitational potential well is deep enough to have retained supernova ejecta, both of which are indicative of a dark-matter-dominated dwarf galaxy \citep[e.g.,][]{Willman:2012}.

\par We note that the conclusion that Pegasus~IV is an ultra-faint dwarf galaxy could be further tested in the future through higher-resolution spectroscopic observations of its bright member stars. Such spectra would allow for measurements of the galaxy's $\alpha-$element and neutron-capture element abundances, both of which can independently offer further insight into the classification of this system \citep[e.g,][]{2019ApJ...870...83J}. Alternately, deeper medium resolution spectra could provide iron abundances for a large sample of stars, allowing for a more robust measurement of the system's metallicity dispersion.

\par 
\subsection{Orbit}
\label{sec:orbit}

To determine Pegasus~IV's orbital properties, we integrated 500 realizations of its orbit using the \code{gala} Python package \citep{2017JOSS....2..388P}. For each realization, we determined Pegasus~IV's initial conditions $\{\alpha_{\rm J2000},\delta_{\rm J2000}, D_{\odot}, \mu_{\alpha *},\mu_{\delta}, v_{\odot}\}$ by sampling from the error distributions of its observed position and kinematics (\tabref{properties}), which we approximated as Gaussian for all parameters.  We then rewound Pegasus~IV's orbit back in time for 5 \Gyr in the presence of \code{gala}'s default Milky Way model, which includes a spherical nucleus and bulge, a Miyamoto-Nagai disk \citep{1975PASJ...27..533M}, and a spherical Navarro-Frenk-White dark matter halo (NFW; \citealt{Navarro1996ApJ...462..563N}). 
\par At the conclusion of each integration, we recorded \code{gala}'s estimate for Pegasus~IV's apocenter ($r_{\rm apo}$), pericenter ($r_{\rm peri}$), eccentricity ($e$), orbital angular momentum perpendicular to the Galactic disk ($L_z$), and total energy ($E$).  From the median, 16th, and 84th percentile of the distributions for these quantities across the 500 realizations, we find:
\begin{itemize}
    \item $r_{\rm apo} = 94^{+8}_{-7} \kpc$ \hspace{2.0em} $r_{\rm peri} = 32_{-14}^{+18} \kpc$
    \item $e = 0.49^{+0.17}_{-0.16}$
    \item $L_z = 6.3^{+2.9}_{-2.6} \kpc^{2} \Myr^{-1} $
    \item $E = -0.049^{+0.005}_{-0.004} \kpc^{2} \Myr^{-2}$.
\end{itemize}

\par
\begin{figure*}[!ht]
    \centering
    \includegraphics[width = \textwidth]{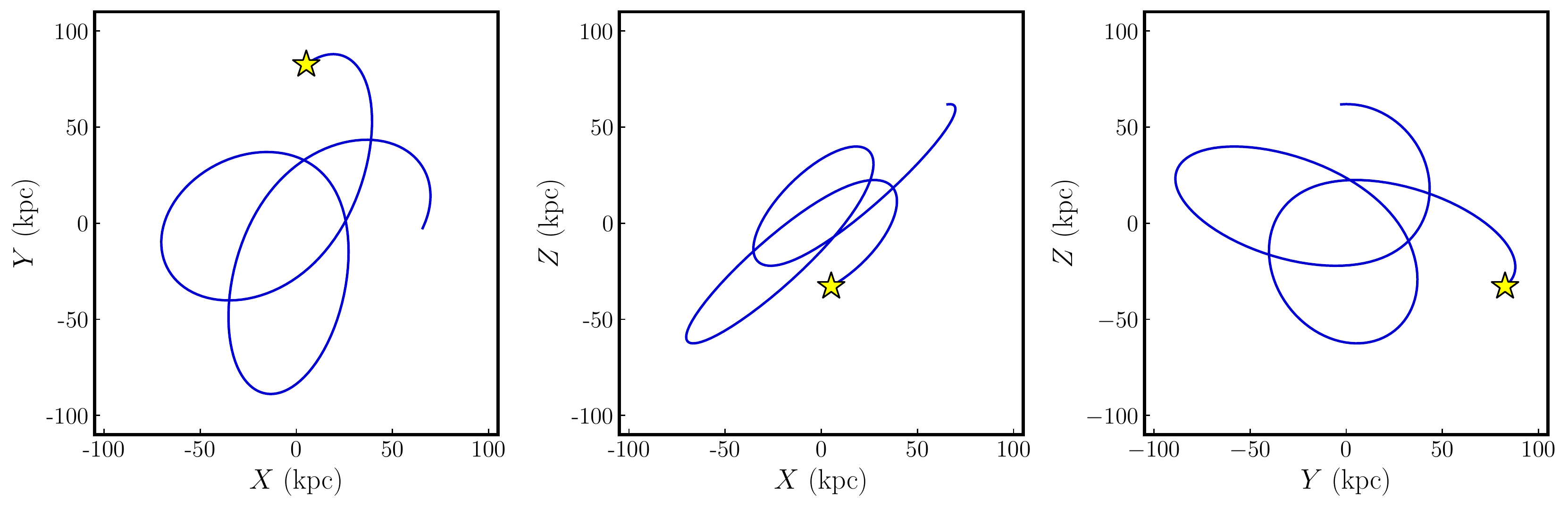}
    \caption{Projections of Pegasus~IV's fiducial orbit for the last 5 \Gyr in the Galactocentric X-Y, X-Z, and Y-Z planes (left, center, and right panels, respectively). Pegasus~IV's current position is depicted as a gold star. }
    \label{fig:orbitmodel}
\end{figure*}

 In \figref{orbitmodel}, we depict the last $5 \Gyr$ of Pegasus~IV's orbit (in various projections) assuming the velocity, distance, and proper motions reported in \tabref{properties} as initial conditions. Notably, the model predicted that Pegasus~IV  passed its apocenter within the last $\sim 200 \Myr$ and experienced its last pericentric passage $\sim 1 \Gyr$ ago. 
\par To contextualize Pegasus~IV's proximity to its orbital apocenter, we computed the ratio: 
$f = (d_{\rm GC} - r_{\rm peri})/(r_{\rm apo} - r_{\rm peri})$
following \citet{2018A&A...619A.103F}. This ratio quantifies a satellite's proximity to its pericenter ($f=0$) or apocenter ($f=1$). Assuming Pegasus~IV's distance to the Galactic center is $d_{\rm GC} = 89 \kpc$ and adopting the apocenter/pericenter distances given above, we found $f= 0.92$. This value for the ratio $f$ places Pegasus~IV in a regime that is underpopulated compared to the predictions from simple orbital dynamics (for example, see Figure 5 of \citealt{2021arXiv211006950L}). Our discovery of Pegasus~IV in a previously-surveyed region of the sky may support the hypothesis that the dearth of known Milky Way satellite galaxies observed near their apocenters ($f \sim 1$) is an observational selection effect \citep[e.g.,][]{Simon_2018,2018A&A...619A.103F,2021arXiv211006950L}.

\subsection{Association with Local Group Structures}
\par A number of recently discovered ultra-faint dwarf galaxies have been proposed to be associated with the Large Magellanic Cloud  \citep[LMC;  e.g.,][]{Koposov_2015,Drlica-Wagner:2015, Patel_etal_2020,Erkal_etal_2020,2022MNRAS.511.2610C}.
To assess whether Pegasus IV is a satellite of the LMC, we rewound the system in the combined presence of the LMC and Milky Way potential using the technique described in \cite{Erkal_etal_2020}. For the Milky Way potential, we used the potential fits of \cite{McMillan_2017}. We note that we did not select the highest likelihood potential but instead sampled the Milky Way from the posterior chains of \cite{McMillan_2017} to account for uncertainties in the potential. We modeled the LMC as a Hernquist profile \citep{Hernquist_1990} with a mass of $1.38\times10^{11} M_\odot$ and a scale radius of $16.08$ kpc, motivated by the results of \cite{Erkal_etal_2019}. In these simulations, we treated the LMC and Milky Way as particles sourcing their respective potentials and thus account for the reflex motion of the Milky Way in response to the LMC \citep[e.g.][]{Gomez_etal_2015}. We modeled the dynamical friction of the Milky Way on the LMC using the approximations in \cite{Jethwa_etal_2016}. For the LMC's present-day proper motions, distance, and radial velocity we used values provided by \cite{Kallivayalil_etal_2013}, \cite{Pietrzyski_etal_2019}, and \cite{van_der_Marel_etal_2002}, respectively.
 
\par In order to account for uncertainties, we Monte Carlo sampled the present-day observables of Pegasus IV, the Milky Way potential, and the LMC's present-day observables 10,000 times and rewound the satellite for 5 Gyr.\footnote{This model produced estimates for Pegasus~IV's apocenter and pericenter that agreed with the results from the Milky-Way-potential-only integration (\secref{orbit}) to well within the quoted uncertainties reported in \tabref{properties}.} We computed the energy of Pegasus~IV relative to the LMC 5 Gyr ago \citep[as in][]{Erkal_etal_2020}, and found that Pegasus IV has a 0.07\% chance of having originally been energetically bound to the LMC, suggesting that it is not an LMC satellite. We also considered the approach of \citet{Patel_etal_2020} and determined the closest passage of Pegasus~IV to the LMC and compared their relative speed to the escape speed of the LMC. With this approach, we found that Pegasus~IV passes the LMC at $61\pm12$ kpc, with a relative speed of $363\pm19$ km/s. This is $\sim 3$ times the escape speed of the LMC, which also suggests that Pegasus IV is not an LMC satellite.

\par A substantial fraction of the known Milky Way satellite galaxies lie on a thin, co-rotating plane nearly perpendicular to the Milky Way's stellar disk dubbed the Vast Polar Structure \citep[VPOS;][]{2012MNRAS.423.1109P,2015MNRAS.453.1047P, 2018A&A...619A.103F, HLi:2021}.
Adopting the same VPOS parameters as \citet{HLi:2021}, namely the assumed normal ($l_\text{MW}, b_\text{MW}) = (169.3\degree, -2.8\degree) $ and angular tolerance $\theta_\text{inVPOS} = 36.87 \degree$, we found it unlikely that Pegasus~IV is a VPOS member.
The observed angle between the VPOS and the satellite's orbital pole is $\theta_\text{VPOS} = 52.3^{+19.8}_{-19.5} \degree$ and the probability that the orbital pole lies within $\theta_\text{inVPOS}$ of the VPOS normal is $\sim 20$\%.
While this does not rule out the possibility that Pegasus~IV is a VPOS member, the currently available phase space measurements do not favor this scenario.
% Pegasus~IV is co-orbiting relative to the VPOS normal.

\par Lastly, we considered whether Pegasus~IV might be associated with debris from the Sagittarius dwarf spheroidal galaxy (Sgr) and its extended stellar stream. Considering the Sgr model and associated coordinate system from \citet{2010ApJ...714..229L}, we found that Pegasus~IV is located at an angle of $\beta = -52.9 \degree$ from the Sgr debris plane. We found a comparably large separation when considering the newer Sgr model from \citet{2021MNRAS.501.2279V}, who additionally incorporated the impact of the LMC when modelling Sgr's debris stream. We therefore conclude that Pegasus~IV is unlikely to be associated with Sgr.

\subsection{Astrophysical J-factor/D-factor}
\label{sec:jfactor}

The Milky Way dwarf spheroidal satellite galaxies are excellent targets for searches for dark matter annihilation or decay products due to their close proximity, astrophysical backgrounds, and large mass-to-light ratios \citep[e.g.,][]{2015PhRvL.115w1301A}. 
The astrophysical component of the dark matter flux from annihilation (decay) is known as the J-factor (D-factor) and depends on the squared (linear) dark matter density along the line of sight. 
\par Our framework to calculate J-factors and D-factors follows \citet{Pace2019MNRAS.482.3480P} and is similar to other previous analyses of dwarf spheroidal galaxies \citep[e.g., ][]{Bonnivard2015MNRAS.446.3002B, GeringerSameth2015ApJ...801...74G}. Briefly, we solved for the velocity dispersion in the spherical Jeans equations and compared it to the velocity dispersion from the spectroscopic members to determine the dark matter density profile. We assumed the dark-matter-dominated mass follows an NFW profile, while the stellar distribution follows a Plummer profile. We assumed that stellar anisotropy is constant with radius.
We used the results derived in Section~\ref{sec:ugali} for the distance, structural parameters ($a_h$, $\epsilon$), and associated uncertainties, which were transformed into Gaussian priors. For more details, see \citet{Pace2019MNRAS.482.3480P}.

We applied this methodology to the same seven star (non-variable) member sample used for our dynamical analysis in \secref{dynamics}. We calculated integrated J-factors of  $\log_{10}{J}=17.7\pm0.8, 17.8\pm0.8, 17.9\pm0.8$ for solid angles of $\theta=0.1\degree, 0.2\degree, 0.5\degree$ in logarithmic units of ${\rm GeV^2~cm^{-5}}$. 
The integrated D-factors are  $\log_{10}{D}=16.9\pm0.4, 17.3\pm0.5, 17.8\pm0.6$ for solid angles of $\theta=0.1\degree, 0.2\degree, 0.5\degree$ in logarithmic units of ${\rm GeV~cm^{-2}}$. 
The predicted J-factor is $\log_{10} J(0.5\degrees) \sim 17.6$ based on velocity dispersion, heliocentric distance, and half-light radius scaling relations and agrees with the full dynamical analysis  \citep{Pace2019MNRAS.482.3480P}. 
This J-factor is not large compared to other ultra-faint dwarfs due primarily to the relatively large distance of Pegasus~IV.
We note that if Pegasus~IV were located at its pericenter ($d=32~\kpc$), its J-factor would be comparable to the largest J-factors measured for other dwarf galaxies: $\log_{10}{J(0.5\degree)}\sim18.8$.

\subsection{Distance from Two RR Lyrae Variable Stars}
\label{sec:rrl}
RR-Lyrae-type variable stars (RRLs) are excellent tracers of old, metal-poor stellar populations in the Milky Way halo, and have been identified in nearly every ultra-faint dwarf galaxy \citep[e.g.,][]{2008ApJ...675L..73G,2013AJ....146...94B,2017ApJ...845L..10M,2018ApJ...861...23J,2019MNRAS.490.2183M,2019ApJ...875..120J,2020ApJS..247...35V,2021AJ....162..253M}. According to the empirical relation derived by \citet{2019MNRAS.490.2183M}, ultra-faint dwarf galaxies with the same absolute magnitude as Pegasus~IV ($M_V = -4.25$) are expected to have between 2 and 4 RRLs.
\par As introduced in \secref{disco}, we identified two RRLs in the \Gaia and PS1 RRL catalogs within a two arcminute radius of Pegasus~IV's centroid at the time of discovery. The first of these stars (\Gaia DR2/EDR3 $\var{source\_id}$: 1796887082536156928; \Gaia $G = 20.08$ mag) was labelled as an RRab star in both catalogs with a period of 0.7088 days (averaging between the individual catalogs, which agreed at the level of 0.0001 days). This star was identified as a spectroscopic member in \secref{specmem} on the basis of its radial velocity. The second of these stars (\Gaia DR2/EDR3 $\var{source\_id}$: 1796890209272433792; $G = 20.24$ mag) was labelled as an RRc-type variable with period 0.31373 days (again averaging between \Gaia and PS1, which agreed within 0.00001 days for this star); we do not have a spectrum for this star.
\par Under the assumption that these stars were \textit{bona fide} RRL member stars of Pegasus~IV, we estimated their absolute magnitudes using the empirical calibration given in \citet{2018MNRAS.481.1195M}:
\begin{equation}
     M_{G} = \left(0.32 \pm 0.04\right) \rm [Fe/H] + (1.11 \pm 0.06).
\end{equation}
Assuming that the (unknown) RRL metallicities are sampled from the Pegasus~IV metallicity distribution function (MDF), which we approximate as a Gaussian centered on $\rm [Fe/H] = -2.67$ dex with variance $\sigma = 0.46$ dex, we found that the expected absolute magnitude of the two stars is $M_{G} = 0.26 \pm 0.19$ mag, where the uncertainties include contributions from both the sampled RRL metallicity and the errors associated with the coefficients in the \citet{2018MNRAS.481.1195M} relation. From this absolute magnitude, the resulting distance modulus for each of the RRLs was then derived from
\begin{equation}
    (m-M)_{0} = G - (R_{G} \times (E(B-V)) - M_{G},
\end{equation}
where $R_G$ is the ratio of total-to-selective absorption for the \Gaia $G$ filter, which we assumed to be $R_G = 2.45$ \citep{Wang_2019}. Taking $E(B-V) = 0.06$ mag for both stars (\tabref{properties}), we found: 
$(m-M)_{0}  = 19.67 \pm 0.19$ from the first RRL and $ (m-M)_{0} = 19.83 \pm 0.19$ for the second, neglecting the errors on $G$ and $E(B-V)$ as they were subdominant to the error on $M_G$. The average of these distance moduli is $(m-M)_{0} = 19.75 \pm 0.13$, in excellent agreement with the distance modulus derived from isochrone-fitting, $(m-M)_{0} =19.77 \pm 0.03 \text{ (stat)} \pm 0.1 \text{ (sys)}$ (\secref{ugali}).

\subsection{A Distant RRL Member?}
\par The \Gaia and PS1 RR Lyrae catalogs include an additional RRL located at ($\alpha_{\rm J2000},\delta_{\rm J2000}) = (328.834\degree,\allowbreak 26.602\degree$), corresponding to a 15.8' separation from Pegasus~IV's centroid, or roughly ten half-light radii ($\sim 0.42$ kpc). This star (\Gaia DR2/EDR3 $\var{source\_id}$: 1796879729552126080; \Gaia $G = 20.12$ mag) was flagged in the PS1 catalog as an RRc with a period $P = 0.400555$~days. Its \Gaia EDR3 proper motion ($\mu_\alpha *, \mu_\delta) = (+0.114 \pm 0.460, -0.328 \pm 0.541)\,{\rm mas\, yr}^{-1}$ is consistent with the systemic mean proper motion derived in \secref{proper_motion}: $(+0.33 \pm 0.07, -0.21 \pm 0.08)\,{\rm mas\, yr}^{-1}$. The distance modulus of this star according to the  \citet{2018MNRAS.481.1195M} relation is $(m-M)_0 = 19.71 \pm 0.19$, lying between the distance moduli derived for the other two RRLs discussed in the previous section, and in equally good agreement with the distance modulus derived through isochrone fitting. 
\par These properties suggest that this RRL may be related to Pegasus IV, despite its extreme angular separation. To quantify the possibility that this star is a field RR Lyrae, as opposed to a true Pegasus~IV member, we integrated the RR Lyrae number density radial profile given in \citet{2018ApJ...855...43M} between galactocentric distances of 80 to 100 kpc. We found that only 0.0075 RR Lyrae stars are expected in a 0.25~$\rm deg^2$ region around Pegasus~IV. Thus, it is very unlikely that this star is a field star, as opposed to a true Pegasus~IV member. 
\par RRLs with large angular separations have been observed in the vicinity of several ultra-faint dwarf galaxies \citep[e.g,][]{2020ApJS..247...35V,2021ApJ...911..109S}, and have been proposed to be tidally-stripped members of these galaxies.\footnote{We also note that \citet{2021NatAs...5..392C} discoverd and confirmed multiple member stars at extremely large separations from the Tucana~II dwarf galaxy, highlighting that yet more member stars may be discoverable in the outskirts of Pegasus~IV.} To assess whether tidal stripping is needed to explain the position of this RRL relative to Pegasus IV, we calculated the system's Jacobi radius following Equation 8.91 of \citet{2008gady.book.....B}. As explained by \citet{2008gady.book.....B}, the Jacobi radius approximately corresponds to the expected maximum observed extent of a satellite system in a circular orbit. Adopting the dynamical and structural properties from \tabref{properties}, and assuming the simple power-law Milky Way potential from \citet{2016ApJ...829..108E}, we found that the Jacobi radius for Pegasus~IV is $\sim 0.6$ kpc -- larger than the projected separation of this RRL from the main body of Pegasus IV ($\sim 0.42$ kpc). However, if we instead perform this calculation assuming that Pegasus~IV is at its pericenter distance ($r_{\rm peri} = 32$~kpc), the Jacobi radius is found to be $\sim 0.26$ kpc, smaller than the observed projected separation. We note, though, that these Jacobi radii are significant underestimates, as they are calculated using the dynamical mass within $r_{1/2}$ in absence of a total mass estimate for Pegasus~IV.
\par This latter Jacobi radius estimate admits the possibility that the distant RRL was tidally stripped from the main body of Pegasus~IV at a previous pericentric passage, although the close clustering of the confirmed spectroscopic members somewhat disfavors this interpretation. Ultimately, it is difficult to confirm or dispute this star's connection to Pegasus~IV without a radial velocity measurement. Wider-area spectroscopic member samples may allow for searches for features suggestive of tidal disruption (e.g., velocity gradients), which would add credence to the tidal origin of this distant star if present.  Improved distance estimation for each of the RRLs may also offer further insight into the consistency of this star with the majority of Pegasus~IV's members.

\par Lastly, we note also that there may be yet more RRL members of Pegasus~IV, as the \Gaia and PS1 RRL catalogs are incomplete at faint magnitudes \citep[e.g.,][]{2020MNRAS.496.3291M}. Our team has recently obtained deeper Gemini North / GMOS imaging of Pegasus~IV (GN-2021B-FT-111; PI: C. Martinez-Vazquez). 
We therefore defer a more extensive search for RRLs in the central region of Pegasus IV to a future study leveraging these data.
These new data will also help disambiguate the nature of the second spectroscopically-observed horizontal branch star, which appeared to show some signs of variability in the sparsely sampled DELVE data.

\section{Summary}
We have presented the discovery of Pegasus~IV, an ultra-faint dwarf galaxy found in a wide-area search of DELVE data. Through a maximum-likelihood fit to the system's morphology and observed color--magnitude diagram, we found that Pegasus~IV is an old, metal-poor stellar system with a half-light radius of $r_{1/2} = 41 \pc$ and an absolute magnitude $M_V = -4.25$. With Magellan/IMACS medium-resolution spectra for a small sample of member stars, we resolved the internal kinematics of the system, finding a velocity dispersion of $\sigma_v = 3.3^{+1.7}_{-1.1} \text{ km s}^{-1}$, implying a mass-to-light ratio for the system of $M_{1/2}/L_{V,1/2} = 167^{+224}_{-99} M_{\odot}/L_{\odot}$. We used the CaT absorption lines in the same spectra to derive iron abundances for five stars, which suggested that Pegasus~IV is very metal-poor ($\rm [Fe/H] = -2.67$) and exhibits a metallicity spread that further suggests its nature as a dwarf galaxy.  We also measured Pegasus~IV's proper motion using data from \Gaia EDR3, which, in conjunction with the system's measured velocity of $v_{\rm hel} = -273.6 \text{ km s}^{-1}$, suggested that Pegasus~IV is on a retrograde orbit, and just passed its orbital apocenter. Lastly, we constrained the distance to Pegasus~IV using a metallicity--absolute magnitude relation for two RR Lyrae stars found in the system, confirming that the system is located at a heliocentric distance of $\sim 90 \kpc$ as determined through isochrone fitting.
\par Our discovery of Pegasus~IV in data from DECam is consistent with the prediction that many ultra-faint Milky Way satellites remain to be discovered, not only in previously unsearched regions, but also in regions of sky previously covered by current-generation surveys. Survey efforts including DELVE-WIDE will likely continue to play an important role in this ongoing satellite census. Illustratively, \citet{2021arXiv211204511M} recently forecasted that DELVE-WIDE may discover $34^{+17}_{-13}$ ultra-faint dwarf galaxies with $M_V < 0$ and $r_{1/2} > 10\text{ pc}$ across its nominal footprint ($\delta_{2000} < 0\degree$; $|b| > 10\degree$), assuming that DELVE will achieve comparable sensitivity to searches over third-year Dark Energy Survey data (DES Y3; \citealt{Drlica-Wagner:2020}). Furthermore, the upcoming Vera C. Rubin Observatory Legacy Survey of Space and Time \citep{2019ApJ...873..111I} is expected to discover hundreds of ultra-faint dwarf galaxies both around the Milky Way and beyond \citep[e.g.,][]{2014ApJ...795L..13H,2021ApJ...918...88M,2021A&A...654A..40T,2021arXiv211204511M}. This growing sample of ultra-faint dwarf galaxies will undoubtedly provide new constraints on the properties of dark matter and will offer key insight into the process of galaxy formation on the smallest scales.
\label{sec:summary}

\section{Acknowledgments}
The DELVE project is partially supported by Fermilab LDRD project L2019-011 and the NASA Fermi Guest Investigator Program Cycle 9 No. 91201. WC gratefully acknowledges support from the University of Chicago Quad Undergraduate Research Scholars program and from the Carnegie Astrophysics Summer Student Internship (CASSI) program, during which training in spectroscopic data analysis was acquired.
ABP acknowledges support from NSF grant AST-1813881. BMP is supported by an NSF Astronomy and Astrophysics Postdoctoral Fellowship under award AST-2001663.
This research received support from the National Science Foundation (NSF) under grant no.\ NSF DGE-1656518 through the NSF Graduate Research Fellowship received by SM. R. R. M. gratefully acknowledges support by the ANID BASAL project FB210003. JAC-B acknowledges support from ANID FONDECYT Regular 1220083. 

This project used data obtained with the Dark Energy Camera (DECam), which was constructed by the Dark Energy Survey (DES) collaboration.
Funding for the DES Projects has been provided by 
the DOE and NSF (USA),   
MISE (Spain),   
STFC (UK), 
HEFCE (UK), 
NCSA (UIUC), 
KICP (U. Chicago), 
CCAPP (Ohio State), 
MIFPA (Texas A\&M University),  
CNPQ, 
FAPERJ, 
FINEP (Brazil), 
MINECO (Spain), 
DFG (Germany), 
and the collaborating institutions in the Dark Energy Survey, which are
Argonne Lab, 
UC Santa Cruz, 
University of Cambridge, 
CIEMAT-Madrid, 
University of Chicago, 
University College London, 
DES-Brazil Consortium, 
University of Edinburgh, 
ETH Z{\"u}rich, 
Fermilab, 
University of Illinois, 
ICE (IEEC-CSIC), 
IFAE Barcelona, 
Lawrence Berkeley Lab, 
LMU M{\"u}nchen, and the associated Excellence Cluster Universe, 
University of Michigan, 
NSF's National Optical-Infrared Astronomy Research Laboratory, 
University of Nottingham, 
Ohio State University, 
OzDES Membership Consortium
University of Pennsylvania, 
University of Portsmouth, 
SLAC National Lab, 
Stanford University, 
University of Sussex, 
and Texas A\&M University.

This work has made use of data from the European Space Agency (ESA) mission {\it Gaia} (\url{https://www.cosmos.esa.int/gaia}), processed by the {\it Gaia} Data Processing and Analysis Consortium (DPAC, \url{https://www.cosmos.esa.int/web/gaia/dpac/consortium}).
Funding for the DPAC has been provided by national institutions, in particular the institutions participating in the {\it Gaia} Multilateral Agreement.

\par Based on observations at Cerro Tololo Inter-American Observatory, NSF's National Optical-Infrared Astronomy Research Laboratory (2019A-0305; PI: Drlica-Wagner), which is operated by the Association of Universities for Research in Astronomy (AURA) under a cooperative agreement with the National Science Foundation.
\par This research has made use of NASA’s Astrophysics Data System Bibliographic Services.
\par This manuscript has been authored by Fermi Research Alliance, LLC, under contract No.\ DE-AC02-07CH11359 with the US Department of Energy, Office of Science, Office of High Energy Physics. The United States Government retains and the publisher, by accepting the article for publication, acknowledges that the United States Government retains a non-exclusive, paid-up, irrevocable, worldwide license to publish or reproduce the published form of this manuscript, or allow others to do so, for United States Government purposes.

\facility{Blanco, \Gaia, Magellan-IMACS}
\software{\code{SourceExtractor} \citep{Bertin:1996} \code{PSFEx} \citep{Bertin:2011}, \emcee \citep{Foreman-Mackey:2013}, \code{gala} \citep{2017JOSS....2..388P}, \healpix \citep{Gorski:2005},\footnote{\url{http://healpix.sourceforge.net}} \code{healpy},\footnote{\url{https://github.com/healpy/healpy}} \ugali \citep{Bechtol:2015}\footnote{\url{https://github.com/DarkEnergySurvey/ugali}}, \code{simple} \citep{Bechtol:2015}.\footnote{\url{https://github.com/DarkEnergySurvey/simple}}}

\bibliography{main}

\begin{thebibliography}{}
\expandafter\ifx\csname natexlab\endcsname\relax\def\natexlab#1{#1}\fi
\providecommand{\url}[1]{\href{#1}{#1}}

\bibitem[{{Abbott} {et~al.}(2018){Abbott}, {Abdalla}, {Allam}, {Amara},
  {Annis}, {Asorey}, {Avila}, {Ballester}, {Banerji}, {Barkhouse}, {Baruah},
  {Baumer}, {Bechtol}, {Becker}, {Benoit-L{\'e}vy}, {Bernstein}, {Bertin},
  {Blazek}, {Bocquet}, {Brooks}, {Brout}, {Buckley-Geer}, {Burke}, {Busti},
  {Campisano}, {Cardiel-Sas}, {Carnero Rosell}, {Carrasco Kind}, {Carretero},
  {Castander}, {Cawthon}, {Chang}, {Chen}, {Conselice}, {Costa}, {Crocce},
  {Cunha}, {D'Andrea}, {da Costa}, {Das}, {Daues}, {Davis}, {Davis}, {De
  Vicente}, {DePoy}, {DeRose}, {Desai}, {Diehl}, {Dietrich}, {Dodelson},
  {Doel}, {Drlica-Wagner}, {Eifler}, {Elliott}, {Evrard}, {Farahi}, {Fausti
  Neto}, {Fernandez}, {Finley}, {Flaugher}, {Foley}, {Fosalba}, {Friedel},
  {Frieman}, {Garc{\'\i}a-Bellido}, {Gaztanaga}, {Gerdes}, {Giannantonio},
  {Gill}, {Glazebrook}, {Goldstein}, {Gower}, {Gruen}, {Gruendl}, {Gschwend},
  {Gupta}, {Gutierrez}, {Hamilton}, {Hartley}, {Hinton}, {Hislop}, {Hollowood},
  {Honscheid}, {Hoyle}, {Huterer}, {Jain}, {James}, {Jeltema}, {Johnson},
  {Johnson}, {Kacprzak}, {Kent}, {Khullar}, {Klein}, {Kovacs}, {Koziol},
  {Krause}, {Kremin}, {Kron}, {Kuehn}, {Kuhlmann}, {Kuropatkin}, {Lahav},
  {Lasker}, {Li}, {Li}, {Liddle}, {Lima}, {Lin}, {L{\'o}pez-Reyes}, {MacCrann},
  {Maia}, {Maloney}, {Manera}, {March}, {Marriner}, {Marshall}, {Martini},
  {McClintock}, {McKay}, {McMahon}, {Melchior}, {Menanteau}, {Miller},
  {Miquel}, {Mohr}, {Morganson}, {Mould}, {Neilsen}, {Nichol}, {Nogueira},
  {Nord}, {Nugent}, {Nunes}, {Ogando}, {Old}, {Pace}, {Palmese},
  {Paz-Chinch{\'o}n}, {Peiris}, {Percival}, {Petravick}, {Plazas}, {Poh},
  {Pond}, {Porredon}, {Pujol}, {Refregier}, {Reil}, {Ricker}, {Rollins},
  {Romer}, {Roodman}, {Rooney}, {Ross}, {Rykoff}, {Sako}, {Sanchez}, {Sanchez},
  {Santiago}, {Saro}, {Scarpine}, {Scolnic}, {Serrano}, {Sevilla-Noarbe},
  {Sheldon}, {Shipp}, {Silveira}, {Smith}, {Smith}, {Smith}, {Soares-Santos},
  {Sobreira}, {Song}, {Stebbins}, {Suchyta}, {Sullivan}, {Swanson}, {Tarle},
  {Thaler}, {Thomas}, {Thomas}, {Troxel}, {Tucker}, {Vikram}, {Vivas},
  {Walker}, {Wechsler}, {Weller}, {Wester}, {Wolf}, {Wu}, {Yanny}, {Zenteno},
  {Zhang}, {Zuntz}, {DES Collaboration}, {Juneau}, {Fitzpatrick}, {Nikutta},
  {Nidever}, {Olsen}, {Scott}, \& {NOAO Data Lab}}]{2018ApJS..239...18A}
{Abbott}, T.~M.~C., {Abdalla}, F.~B., {Allam}, S., {et~al.} 2018, \apjs, 239,
  18

\bibitem[{Ackermann {et~al.}(2014)Ackermann, Albert, Anderson, Baldini, Ballet,
  Barbiellini, Bastieri, Bechtol, Bellazzini, Bissaldi, Bloom, Bonamente,
  Bouvier, Brandt, Bregeon, Brigida, Bruel, Buehler, Buson, Caliandro, Cameron,
  Caragiulo, Caraveo, Cecchi, Charles, Chekhtman, Chiang, Ciprini, Claus,
  Cohen-Tanugi, Conrad, D'Ammando, de~Angelis, Dermer, Digel, do~Couto~e Silva,
  Drell, Drlica-Wagner, Essig, Favuzzi, Ferrara, Franckowiak, Fukazawa, Funk,
  Fusco, Gargano, Gasparrini, Giglietto, Giroletti, Godfrey, Gomez-Vargas,
  Grenier, Guiriec, Gustafsson, Hayashida, Hays, Hewitt, Hughes, Jogler, Kamae,
  Kn\"odlseder, Kocevski, Kuss, Larsson, Latronico, Llena~Garde, Longo,
  Loparco, Lovellette, Lubrano, Martinez, Mayer, Mazziotta, Michelson,
  Mitthumsiri, Mizuno, Moiseev, Monzani, Morselli, Moskalenko, Murgia, Nemmen,
  Nuss, Ohsugi, Orlando, Ormes, Perkins, Piron, Pivato, Porter, Rain\`o, Rando,
  Razzano, Razzaque, Reimer, Reimer, Ritz, S\'anchez-Conde, Sehgal, Sgr\`o,
  Siskind, Spinelli, Strigari, Suson, Tajima, Takahashi, Thayer, Tibaldo,
  Tinivella, Torres, Uchiyama, Usher, Vandenbroucke, Vianello, Vitale, Werner,
  Winer, Wood, Wood, Zaharijas, \& Zimmer}]{PhysRevD.89.042001}
Ackermann, M., Albert, A., Anderson, B., {et~al.} 2014, Phys. Rev. D, 89,
  042001.
\newblock \url{https://link.aps.org/doi/10.1103/PhysRevD.89.042001}

\bibitem[{{Ackermann} {et~al.}(2015){Ackermann}, {Albert}, {Anderson},
  {Atwood}, {Baldini}, {Barbiellini}, {Bastieri}, {Bechtol}, {Bellazzini},
  {Bissaldi}, {Blandford}, {Bloom}, {Bonino}, {Bottacini}, {Brandt}, {Bregeon},
  {Bruel}, {Buehler}, {Caliandro}, {Cameron}, {Caputo}, {Caragiulo}, {Caraveo},
  {Cecchi}, {Charles}, {Chekhtman}, {Chiang}, {Chiaro}, {Ciprini}, {Claus},
  {Cohen-Tanugi}, {Conrad}, {Cuoco}, {Cutini}, {D'Ammando}, {de Angelis}, {de
  Palma}, {Desiante}, {Digel}, {Di Venere}, {Drell}, {Drlica-Wagner}, {Essig},
  {Favuzzi}, {Fegan}, {Ferrara}, {Focke}, {Franckowiak}, {Fukazawa}, {Funk},
  {Fusco}, {Gargano}, {Gasparrini}, {Giglietto}, {Giordano}, {Giroletti},
  {Glanzman}, {Godfrey}, {Gomez-Vargas}, {Grenier}, {Guiriec}, {Gustafsson},
  {Hays}, {Hewitt}, {Horan}, {Jogler}, {J{\'o}hannesson}, {Kuss}, {Larsson},
  {Latronico}, {Li}, {Li}, {Llena Garde}, {Longo}, {Loparco}, {Lubrano},
  {Malyshev}, {Mayer}, {Mazziotta}, {McEnery}, {Meyer}, {Michelson}, {Mizuno},
  {Moiseev}, {Monzani}, {Morselli}, {Murgia}, {Nuss}, {Ohsugi}, {Orienti},
  {Orlando}, {Ormes}, {Paneque}, {Perkins}, {Pesce-Rollins}, {Piron}, {Pivato},
  {Porter}, {Rain{\`o}}, {Rando}, {Razzano}, {Reimer}, {Reimer}, {Ritz},
  {S{\'a}nchez-Conde}, {Schulz}, {Sehgal}, {Sgr{\`o}}, {Siskind}, {Spada},
  {Spandre}, {Spinelli}, {Strigari}, {Tajima}, {Takahashi}, {Thayer},
  {Tibaldo}, {Torres}, {Troja}, {Vianello}, {Werner}, {Winer}, {Wood}, {Wood},
  {Zaharijas}, {Zimmer}, \& {Fermi-LAT Collaboration}}]{2015PhRvL.115w1301A}
{Ackermann}, M., {Albert}, A., {Anderson}, B., {et~al.} 2015, \prl, 115, 231301

\bibitem[{{Albert} {et~al.}(2017){Albert}, {Anderson}, {Bechtol},
  {Drlica-Wagner}, {Meyer}, {S{\'a}nchez-Conde}, {Strigari}, {Wood}, {Abbott},
  {Abdalla}, {Benoit-L{\'e}vy}, {Bernstein}, {Bernstein}, {Bertin}, {Brooks},
  {Burke}, {Carnero Rosell}, {Carrasco Kind}, {Carretero}, {Crocce}, {Cunha},
  {D'Andrea}, {da Costa}, {Desai}, {Diehl}, {Dietrich}, {Doel}, {Eifler},
  {Evrard}, {Fausti Neto}, {Finley}, {Flaugher}, {Fosalba}, {Frieman},
  {Gerdes}, {Goldstein}, {Gruen}, {Gruendl}, {Honscheid}, {James}, {Kent},
  {Kuehn}, {Kuropatkin}, {Lahav}, {Li}, {Maia}, {March}, {Marshall}, {Martini},
  {Miller}, {Miquel}, {Neilsen}, {Nord}, {Ogando}, {Plazas}, {Reil}, {Romer},
  {Rykoff}, {Sanchez}, {Santiago}, {Schubnell}, {Sevilla-Noarbe}, {Smith},
  {Soares-Santos}, {Sobreira}, {Suchyta}, {Swanson}, {Tarle}, {Vikram},
  {Walker}, {Wechsler}, {Fermi-LAT Collaboration}, \& {DES
  Collaboration}}]{2017ApJ...834..110A}
{Albert}, A., {Anderson}, B., {Bechtol}, K., {et~al.} 2017, \apj, 834, 110

\bibitem[{{Bailin}(2019)}]{2019ApJS..245....5B}
{Bailin}, J. 2019, \apjs, 245, 5

\bibitem[{{Balbinot} {et~al.}(2013){Balbinot}, {Santiago}, {da Costa}, {Maia},
  {Majewski}, {Nidever}, {Rocha-Pinto}, {Thomas}, {Wechsler}, \&
  {Yanny}}]{2013ApJ...767..101B}
{Balbinot}, E., {Santiago}, B.~X., {da Costa}, L., {et~al.} 2013, \apj, 767,
  101

\bibitem[{{Battaglia} \& {Starkenburg}(2012)}]{2012A&A...539A.123B}
{Battaglia}, G., \& {Starkenburg}, E. 2012, \aap, 539, A123

\bibitem[{{Bechtol} {et~al.}(2015){Bechtol}, {Drlica-Wagner}, {Balbinot},
  {Pieres}, {Simon}, {Yanny}, {Santiago}, {Wechsler}, {Frieman}, {Walker},
  {Williams}, {Rozo}, {Rykoff}, {Queiroz}, {Luque}, {Benoit-L{\'e}vy},
  {Tucker}, {Sevilla}, {Gruendl}, {da Costa}, {Fausti Neto}, {Maia}, {Abbott},
  {Allam}, {Armstrong}, {Bauer}, {Bernstein}, {Bernstein}, {Bertin}, {Brooks},
  {Buckley-Geer}, {Burke}, {Carnero Rosell}, {Castander}, {Covarrubias},
  {D'Andrea}, {DePoy}, {Desai}, {Diehl}, {Eifler}, {Estrada}, {Evrard},
  {Fernandez}, {Finley}, {Flaugher}, {Gaztanaga}, {Gerdes}, {Girardi},
  {Gladders}, {Gruen}, {Gutierrez}, {Hao}, {Honscheid}, {Jain}, {James},
  {Kent}, {Kron}, {Kuehn}, {Kuropatkin}, {Lahav}, {Li}, {Lin}, {Makler},
  {March}, {Marshall}, {Martini}, {Merritt}, {Miller}, {Miquel}, {Mohr},
  {Neilsen}, {Nichol}, {Nord}, {Ogando}, {Peoples}, {Petravick}, {Plazas},
  {Romer}, {Roodman}, {Sako}, {Sanchez}, {Scarpine}, {Schubnell}, {Smith},
  {Soares-Santos}, {Sobreira}, {Suchyta}, {Swanson}, {Tarle}, {Thaler},
  {Thomas}, {Wester}, {Zuntz}, \& {DES Collaboration}}]{Bechtol:2015}
{Bechtol}, K., {Drlica-Wagner}, A., {Balbinot}, E., {et~al.} 2015, \apj, 807,
  50

\bibitem[{{Belokurov} {et~al.}(2014){Belokurov}, {Irwin}, {Koposov}, {Evans},
  {Gonzalez-Solares}, {Metcalfe}, \& {Shanks}}]{2014MNRAS.441.2124B}
{Belokurov}, V., {Irwin}, M.~J., {Koposov}, S.~E., {et~al.} 2014, \mnras, 441,
  2124

\bibitem[{{Belokurov} {et~al.}(2007){Belokurov}, {Zucker}, {Evans}, {Kleyna},
  {Koposov}, {Hodgkin}, {Irwin}, {Gilmore}, {Wilkinson}, {Fellhauer},
  {Bramich}, {Hewett}, {Vidrih}, {De Jong}, {Smith}, {Rix}, {Bell}, {Wyse},
  {Newberg}, {Mayeur}, {Yanny}, {Rockosi}, {Gnedin}, {Schneider}, {Beers},
  {Barentine}, {Brewington}, {Brinkmann}, {Harvanek}, {Kleinman}, {Krzesinski},
  {Long}, {Nitta}, \& {Snedden}}]{2007ApJ...654..897B}
{Belokurov}, V., {Zucker}, D.~B., {Evans}, N.~W., {et~al.} 2007, \apj, 654, 897

\bibitem[{{Bertin}(2006)}]{Bertin:2006}
{Bertin}, E. 2006, in Astronomical Society of the Pacific Conference Series,
  Vol. 351, Astronomical Data Analysis Software and Systems XV, ed.
  C.~{Gabriel}, C.~{Arviset}, D.~{Ponz}, \& S.~{Enrique}, 112

\bibitem[{{Bertin}(2011)}]{Bertin:2011}
{Bertin}, E. 2011, in Astronomical Society of the Pacific Conference Series,
  Vol. 442, Astronomical Data Analysis Software and Systems XX, ed. I.~N.
  {Evans}, A.~{Accomazzi}, D.~J. {Mink}, \& A.~H. {Rots}, San Francisco, CA,
  435

\bibitem[{{Bertin} \& {Arnouts}(1996)}]{Bertin:1996}
{Bertin}, E., \& {Arnouts}, S. 1996, \aaps, 117, 393

\bibitem[{{Binney} \& {Tremaine}(2008)}]{2008gady.book.....B}
{Binney}, J., \& {Tremaine}, S. 2008, {Galactic Dynamics: Second Edition}

\bibitem[{{Boettcher} {et~al.}(2013){Boettcher}, {Willman}, {Fadely},
  {Strader}, {Baker}, {Hopkins}, {Tasnim Ananna}, {Cunningham}, {Douglas},
  {Gilbert}, {Preston}, \& {Sturner}}]{2013AJ....146...94B}
{Boettcher}, E., {Willman}, B., {Fadely}, R., {et~al.} 2013, \aj, 146, 94

\bibitem[{{Bonnivard} {et~al.}(2015){Bonnivard}, {Combet}, {Maurin}, \&
  {Walker}}]{Bonnivard2015MNRAS.446.3002B}
{Bonnivard}, V., {Combet}, C., {Maurin}, D., \& {Walker}, M.~G. 2015, \mnras,
  446, 3002

\bibitem[{{Bovill} \& {Ricotti}(2009)}]{2009ApJ...693.1859B}
{Bovill}, M.~S., \& {Ricotti}, M. 2009, \apj, 693, 1859

\bibitem[{{Bressan} {et~al.}(2012){Bressan}, {Marigo}, {Girardi}, {Salasnich},
  {Dal Cero}, {Rubele}, \& {Nanni}}]{Bressan:2012}
{Bressan}, A., {Marigo}, P., {Girardi}, L., {et~al.} 2012, \mnras, 427, 127

\bibitem[{{Brown} {et~al.}(2014){Brown}, {Tumlinson}, {Geha}, {Simon},
  {Vargas}, {VandenBerg}, {Kirby}, {Kalirai}, {Avila}, {Gennaro}, {Ferguson},
  {Mu{\~n}oz}, {Guhathakurta}, \& {Renzini}}]{2014ApJ...796...91B}
{Brown}, T.~M., {Tumlinson}, J., {Geha}, M., {et~al.} 2014, \apj, 796, 91

\bibitem[{{Bullock} \& {Boylan-Kolchin}(2017)}]{2017ARA&A..55..343B}
{Bullock}, J.~S., \& {Boylan-Kolchin}, M. 2017, \araa, 55, 343

\bibitem[{{Bullock} \& {Johnston}(2005)}]{2005ApJ...635..931B}
{Bullock}, J.~S., \& {Johnston}, K.~V. 2005, \apj, 635, 931

\bibitem[{{Burkert}(1995)}]{1995ApJ...447L..25B}
{Burkert}, A. 1995, \apjl, 447, L25

\bibitem[{{Caldwell} {et~al.}(2017){Caldwell}, {Walker}, {Mateo}, {Olszewski},
  {Koposov}, {Belokurov}, {Torrealba}, {Geringer-Sameth}, \&
  {Johnson}}]{2017ApJ...839...20C}
{Caldwell}, N., {Walker}, M.~G., {Mateo}, M., {et~al.} 2017, \apj, 839, 20

\bibitem[{{Cantu} {et~al.}(2021){Cantu}, {Pace}, {Marshall}, {Strigari},
  {Crnojevic}, {Simon}, {Drlica-Wagner}, {Bechtol},
  {Mart{\'\i}nez-V{\'a}zquez}, {Santiago}, {Amara}, {Stringer}, {Diehl},
  {Aguena}, {Allam}, {Avila}, {Brooks}, {Carnero Rosell}, {Carrasco Kind},
  {Carretero}, {Costanzi}, {Da Costa}, {De Vicente}, {Desai}, {Doel}, {Eifler},
  {Everett}, {Frieman}, {Garc{\'\i}a-Bellido}, {Gaztanaga}, {Gruen}, {Gruendl},
  {Gschwend}, {Gutierrez}, {Hinton}, {Hollowood}, {Honscheid}, {James},
  {Kuehn}, {Maia}, {Menanteau}, {Miquel}, {Palmese}, {Paz-Chinch{\'o}n},
  {Plazas}, {Sanchez}, {Scarpine}, {Schubnell}, {Serrano}, {Sevilla-Noarbe},
  {Smith}, {Soares-Santos}, {Suchyta}, {Swanson}, {Tarle}, {Walker},
  {Wilkinson}, \& {DES Collaboration}}]{2021ApJ...916...81C}
{Cantu}, S.~A., {Pace}, A.~B., {Marshall}, J., {et~al.} 2021, \apj, 916, 81

\bibitem[{{Carlin} {et~al.}(2009){Carlin}, {Grillmair}, {Mu{\~n}oz}, {Nidever},
  \& {Majewski}}]{2009ApJ...702L...9C}
{Carlin}, J.~L., {Grillmair}, C.~J., {Mu{\~n}oz}, R.~R., {Nidever}, D.~L., \&
  {Majewski}, S.~R. 2009, \apjl, 702, L9

\bibitem[{{Carlin} {et~al.}(2017){Carlin}, {Sand}, {Mu{\~n}oz}, {Spekkens},
  {Willman}, {Crnojevi{\'c}}, {Forbes}, {Hargis}, {Kirby}, {Peter},
  {Romanowsky}, \& {Strader}}]{2017AJ....154..267C}
{Carlin}, J.~L., {Sand}, D.~J., {Mu{\~n}oz}, R.~R., {et~al.} 2017, \aj, 154,
  267

\bibitem[{{Carrera} {et~al.}(2013){Carrera}, {Pancino}, {Gallart}, \& {del
  Pino}}]{2013MNRAS.434.1681C}
{Carrera}, R., {Pancino}, E., {Gallart}, C., \& {del Pino}, A. 2013, \mnras,
  434, 1681

\bibitem[{{Cerny} {et~al.}(2021{\natexlab{a}}){Cerny}, {Pace}, {Drlica-Wagner},
  {Ferguson}, {Mau}, {Adam{\'o}w}, {Carlin}, {Choi}, {Erkal}, {Johnson}, {Li},
  {Mart{\'\i}nez-V{\'a}zquez}, {Mutlu-Pakdil}, {Nidever}, {Olsen}, {Pieres},
  {Tollerud}, {Simon}, {Vivas}, {James}, {Kuropatkin}, {Majewski},
  {Mart{\'\i}nez-Delgado}, {Massana}, {Miller}, {Neilsen}, {No{\"e}l}, {Riley},
  {Sand}, {Santana-Silva}, {Stringfellow}, {Tucker}, \& {Delve
  Collaboration}}]{2021ApJ...910...18C}
{Cerny}, W., {Pace}, A.~B., {Drlica-Wagner}, A., {et~al.} 2021{\natexlab{a}},
  \apj, 910, 18

\bibitem[{{Cerny} {et~al.}(2021{\natexlab{b}}){Cerny}, {Pace}, {Drlica-Wagner},
  {Koposov}, {Vivas}, {Mau}, {Riley}, {Bom}, {Carlin}, {Choi}, {Erkal},
  {Ferguson}, {James}, {Li}, {Mart{\'\i}nez-Delgado},
  {Mart{\'\i}nez-V{\'a}zquez}, {Munoz}, {Mutlu-Pakdil}, {Olsen}, {Pieres},
  {Sakowska}, {Sand}, {Simon}, {Smercina}, {Stringfellow}, {Tollerud},
  {Adam{\'o}w}, {Hernandez-Lang}, {Kuropatkin}, {Santana-Silva}, {Tucker},
  {Zenteno}, \& {Delve Collaboration}}]{2021ApJ...920L..44C}
---. 2021{\natexlab{b}}, \apjl, 920, L44

\bibitem[{{Chabrier}(2001)}]{Chabrier:2001}
{Chabrier}, G. 2001, \apj, 554, 1274

\bibitem[{{Chambers} {et~al.}(2016){Chambers}, {Magnier}, {Metcalfe},
  {Flewelling}, {Huber}, {Waters}, {Denneau}, {Draper}, {Farrow}, {Finkbeiner},
  {Holmberg}, {Koppenhoefer}, {Price}, {Rest}, {Saglia}, {Schlafly}, {Smartt},
  {Sweeney}, {Wainscoat}, {Burgett}, {Chastel}, {Grav}, {Heasley}, {Hodapp},
  {Jedicke}, {Kaiser}, {Kudritzki}, {Luppino}, {Lupton}, {Monet}, {Morgan},
  {Onaka}, {Shiao}, {Stubbs}, {Tonry}, {White}, {Ba{\~n}ados}, {Bell},
  {Bender}, {Bernard}, {Boegner}, {Boffi}, {Botticella}, {Calamida},
  {Casertano}, {Chen}, {Chen}, {Cole}, {Deacon}, {Frenk}, {Fitzsimmons},
  {Gezari}, {Gibbs}, {Goessl}, {Goggia}, {Gourgue}, {Goldman}, {Grant},
  {Grebel}, {Hambly}, {Hasinger}, {Heavens}, {Heckman}, {Henderson}, {Henning},
  {Holman}, {Hopp}, {Ip}, {Isani}, {Jackson}, {Keyes}, {Koekemoer}, {Kotak},
  {Le}, {Liska}, {Long}, {Lucey}, {Liu}, {Martin}, {Masci}, {McLean}, {Mindel},
  {Misra}, {Morganson}, {Murphy}, {Obaika}, {Narayan}, {Nieto-Santisteban},
  {Norberg}, {Peacock}, {Pier}, {Postman}, {Primak}, {Rae}, {Rai}, {Riess},
  {Riffeser}, {Rix}, {R{\"o}ser}, {Russel}, {Rutz}, {Schilbach}, {Schultz},
  {Scolnic}, {Strolger}, {Szalay}, {Seitz}, {Small}, {Smith}, {Soderblom},
  {Taylor}, {Thomson}, {Taylor}, {Thakar}, {Thiel}, {Thilker}, {Unger},
  {Urata}, {Valenti}, {Wagner}, {Walder}, {Walter}, {Watters}, {Werner},
  {Wood-Vasey}, \& {Wyse}}]{2016arXiv161205560C}
{Chambers}, K.~C., {Magnier}, E.~A., {Metcalfe}, N., {et~al.} 2016, arXiv
  e-prints, arXiv:1612.05560

\bibitem[{{Chiti} {et~al.}(2021){Chiti}, {Frebel}, {Simon}, {Erkal}, {Chang},
  {Necib}, {Ji}, {Jerjen}, {Kim}, \& {Norris}}]{2021NatAs...5..392C}
{Chiti}, A., {Frebel}, A., {Simon}, J.~D., {et~al.} 2021, Nature Astronomy, 5,
  392

\bibitem[{{Clementini} {et~al.}(2019){Clementini}, {Ripepi}, {Molinaro},
  {Garofalo}, {Muraveva}, {Rimoldini}, {Guy}, {Jevardat de Fombelle},
  {Nienartowicz}, {Marchal}, {Audard}, {Holl}, {Leccia}, {Marconi}, {Musella},
  {Mowlavi}, {Lecoeur-Taibi}, {Eyer}, {De Ridder}, {Regibo}, {Sarro},
  {Szabados}, {Evans}, \& {Riello}}]{2019A&A...622A..60C}
{Clementini}, G., {Ripepi}, V., {Molinaro}, R., {et~al.} 2019, \aap, 622, A60

\bibitem[{{Collins} {et~al.}(2013){Collins}, {Chapman}, {Rich}, {Ibata},
  {Martin}, {Irwin}, {Bate}, {Lewis}, {Pe{\~n}arrubia}, {Arimoto}, {Casey},
  {Ferguson}, {Koch}, {McConnachie}, \& {Tanvir}}]{2013ApJ...768..172C}
{Collins}, M. L.~M., {Chapman}, S.~C., {Rich}, R.~M., {et~al.} 2013, \apj, 768,
  172

\bibitem[{{Conn} {et~al.}(2018){Conn}, {Jerjen}, {Kim}, \&
  {Schirmer}}]{2018ApJ...852...68C}
{Conn}, B.~C., {Jerjen}, H., {Kim}, D., \& {Schirmer}, M. 2018, \apj, 852, 68

\bibitem[{{Cooper} {et~al.}(2012){Cooper}, {Newman}, {Davis}, {Finkbeiner}, \&
  {Gerke}}]{2012ascl.soft03003C}
{Cooper}, M.~C., {Newman}, J.~A., {Davis}, M., {Finkbeiner}, D.~P., \& {Gerke},
  B.~F. 2012, {spec2d: DEEP2 DEIMOS Spectral Pipeline}, , , ascl:1203.003

\bibitem[{{Correa Magnus} \& {Vasiliev}(2022)}]{2022MNRAS.511.2610C}
{Correa Magnus}, L., \& {Vasiliev}, E. 2022, \mnras, 511, 2610

\bibitem[{{Crnojevi{\'c}} {et~al.}(2016){Crnojevi{\'c}}, {Sand}, {Zaritsky},
  {Spekkens}, {Willman}, \& {Hargis}}]{2016ApJ...824L..14C}
{Crnojevi{\'c}}, D., {Sand}, D.~J., {Zaritsky}, D., {et~al.} 2016, \apjl, 824,
  L14

\bibitem[{{Dalgleish} {et~al.}(2020){Dalgleish}, {Kamann}, {Usher},
  {Baumgardt}, {Bastian}, {Veitch-Michaelis}, {Bellini}, {Martocchia}, {Da
  Costa}, {Mackey}, {Bellstedt}, {Pastorello}, \&
  {Cerulo}}]{2020MNRAS.492.3859D}
{Dalgleish}, H., {Kamann}, S., {Usher}, C., {et~al.} 2020, \mnras, 492, 3859

\bibitem[{{Desai} {et~al.}(2012){Desai}, {Armstrong}, {Mohr}, {Semler}, {Liu},
  {Bertin}, {Allam}, {Barkhouse}, {Bazin}, {Buckley-Geer}, {Cooper}, {Hansen},
  {High}, {Lin}, {Lin}, {Ngeow}, {Rest}, {Song}, {Tucker}, \&
  {Zenteno}}]{2012ApJ...757...83D}
{Desai}, S., {Armstrong}, R., {Mohr}, J.~J., {et~al.} 2012, \apj, 757, 83

\bibitem[{{Dey} {et~al.}(2019){Dey}, {Schlegel}, {Lang}, {Blum}, {Burleigh},
  {Fan}, {Findlay}, {Finkbeiner}, {Herrera}, {Juneau}, {Landriau}, {Levi},
  {McGreer}, {Meisner}, {Myers}, {Moustakas}, {Nugent}, {Patej}, {Schlafly},
  {Walker}, {Valdes}, {Weaver}, {Y{\`e}che}, {Zou}, {Zhou}, {Abareshi},
  {Abbott}, {Abolfathi}, {Aguilera}, {Alam}, {Allen}, {Alvarez}, {Annis},
  {Ansarinejad}, {Aubert}, {Beechert}, {Bell}, {BenZvi}, {Beutler}, {Bielby},
  {Bolton}, {Brice{\~n}o}, {Buckley-Geer}, {Butler}, {Calamida}, {Carlberg},
  {Carter}, {Casas}, {Castander}, {Choi}, {Comparat}, {Cukanovaite}, {Delubac},
  {DeVries}, {Dey}, {Dhungana}, {Dickinson}, {Ding}, {Donaldson}, {Duan},
  {Duckworth}, {Eftekharzadeh}, {Eisenstein}, {Etourneau}, {Fagrelius},
  {Farihi}, {Fitzpatrick}, {Font-Ribera}, {Fulmer}, {G{\"a}nsicke},
  {Gaztanaga}, {George}, {Gerdes}, {Gontcho}, {Gorgoni}, {Green}, {Guy},
  {Harmer}, {Hernandez}, {Honscheid}, {Huang}, {James}, {Jannuzi}, {Jiang},
  {Joyce}, {Karcher}, {Karkar}, {Kehoe}, {Kneib}, {Kueter-Young}, {Lan},
  {Lauer}, {Le Guillou}, {Le Van Suu}, {Lee}, {Lesser}, {Perreault Levasseur},
  {Li}, {Mann}, {Marshall}, {Mart{\'\i}nez-V{\'a}zquez}, {Martini}, {du Mas des
  Bourboux}, {McManus}, {Meier}, {M{\'e}nard}, {Metcalfe},
  {Mu{\~n}oz-Guti{\'e}rrez}, {Najita}, {Napier}, {Narayan}, {Newman}, {Nie},
  {Nord}, {Norman}, {Olsen}, {Paat}, {Palanque-Delabrouille}, {Peng},
  {Poppett}, {Poremba}, {Prakash}, {Rabinowitz}, {Raichoor}, {Rezaie},
  {Robertson}, {Roe}, {Ross}, {Ross}, {Rudnick}, {Safonova}, {Saha},
  {S{\'a}nchez}, {Savary}, {Schweiker}, {Scott}, {Seo}, {Shan}, {Silva},
  {Slepian}, {Soto}, {Sprayberry}, {Staten}, {Stillman}, {Stupak}, {Summers},
  {Sien Tie}, {Tirado}, {Vargas-Maga{\~n}a}, {Vivas}, {Wechsler}, {Williams},
  {Yang}, {Yang}, {Yapici}, {Zaritsky}, {Zenteno}, {Zhang}, {Zhang}, {Zhou}, \&
  {Zhou}}]{2019AJ....157..168D}
{Dey}, A., {Schlegel}, D.~J., {Lang}, D., {et~al.} 2019, \aj, 157, 168

\bibitem[{{Dotter}(2016)}]{2016ApJS..222....8D}
{Dotter}, A. 2016, \apjs, 222, 8

\bibitem[{{Dressler} {et~al.}(2011){Dressler}, {Bigelow}, {Hare}, {Sutin},
  {Thompson}, {Burley}, {Epps}, {Oemler}, {Bagish}, {Birk}, {Clardy},
  {Gunnels}, {Kelson}, {Shectman}, \& {Osip}}]{2011PASP..123..288D}
{Dressler}, A., {Bigelow}, B., {Hare}, T., {et~al.} 2011, \pasp, 123, 288

\bibitem[{{Drlica-Wagner} {et~al.}(2015){Drlica-Wagner}, {Bechtol}, {Rykoff},
  {Luque}, {Queiroz}, {Mao}, {Wechsler}, {Simon}, {Santiago}, {Yanny},
  {Balbinot}, {Dodelson}, {Fausti Neto}, {James}, {Li}, {Maia}, {Marshall},
  {Pieres}, {Stringer}, {Walker}, {Abbott}, {Abdalla}, {Allam},
  {Benoit-L{\'e}vy}, {Bernstein}, {Bertin}, {Brooks}, {Buckley-Geer}, {Burke},
  {Carnero Rosell}, {Carrasco Kind}, {Carretero}, {Crocce}, {da Costa},
  {Desai}, {Diehl}, {Dietrich}, {Doel}, {Eifler}, {Evrard}, {Finley},
  {Flaugher}, {Fosalba}, {Frieman}, {Gaztanaga}, {Gerdes}, {Gruen}, {Gruendl},
  {Gutierrez}, {Honscheid}, {Kuehn}, {Kuropatkin}, {Lahav}, {Martini},
  {Miquel}, {Nord}, {Ogando}, {Plazas}, {Reil}, {Roodman}, {Sako}, {Sanchez},
  {Scarpine}, {Schubnell}, {Sevilla-Noarbe}, {Smith}, {Soares-Santos},
  {Sobreira}, {Suchyta}, {Swanson}, {Tarle}, {Tucker}, {Vikram}, {Wester},
  {Zhang}, {Zuntz}, \& {DES Collaboration}}]{Drlica-Wagner:2015}
{Drlica-Wagner}, A., {Bechtol}, K., {Rykoff}, E.~S., {et~al.} 2015, \apj, 813,
  109

\bibitem[{{Drlica-Wagner} {et~al.}(2018){Drlica-Wagner}, {Sevilla-Noarbe},
  {Rykoff}, {Gruendl}, {Yanny}, {Tucker}, {Hoyle}, {Carnero Rosell},
  {Bernstein}, {Bechtol}, {Becker}, {Benoit-L{\'e}vy}, {Bertin}, {Carrasco
  Kind}, {Davis}, {de Vicente}, {Diehl}, {Gruen}, {Hartley}, {Leistedt}, {Li},
  {Marshall}, {Neilsen}, {Rau}, {Sheldon}, {Smith}, {Troxel}, {Wyatt}, {Zhang},
  {Abbott}, {Abdalla}, {Allam}, {Banerji}, {Brooks}, {Buckley-Geer}, {Burke},
  {Capozzi}, {Carretero}, {Cunha}, {D'Andrea}, {da Costa}, {DePoy}, {Desai},
  {Dietrich}, {Doel}, {Evrard}, {Fausti Neto}, {Flaugher}, {Fosalba},
  {Frieman}, {Garc{\'\i}a-Bellido}, {Gerdes}, {Giannantonio}, {Gschwend},
  {Gutierrez}, {Honscheid}, {James}, {Jeltema}, {Kuehn}, {Kuhlmann},
  {Kuropatkin}, {Lahav}, {Lima}, {Lin}, {Maia}, {Martini}, {McMahon},
  {Melchior}, {Menanteau}, {Miquel}, {Nichol}, {Ogando}, {Plazas}, {Romer},
  {Roodman}, {Sanchez}, {Scarpine}, {Schindler}, {Schubnell}, {Smith}, {Smith},
  {Soares-Santos}, {Sobreira}, {Suchyta}, {Tarle}, {Vikram}, {Walker},
  {Wechsler}, {Zuntz}, \& {DES Collaboration}}]{2018ApJS..235...33D}
{Drlica-Wagner}, A., {Sevilla-Noarbe}, I., {Rykoff}, E.~S., {et~al.} 2018,
  \apjs, 235, 33

\bibitem[{{Drlica-Wagner} {et~al.}(2020){Drlica-Wagner}, {Bechtol}, {Mau},
  {McNanna}, {Nadler}, {Pace}, {Li}, {Pieres}, {Rozo}, {Simon}, {Walker},
  {Wechsler}, {Abbott}, {Allam}, {Annis}, {Bertin}, {Brooks}, {Burke},
  {Rosell}, {Carrasco Kind}, {Carretero}, {Costanzi}, {da Costa}, {De Vicente},
  {Desai}, {Diehl}, {Doel}, {Eifler}, {Everett}, {Flaugher}, {Frieman},
  {Garc{\'\i}a-Bellido}, {Gaztanaga}, {Gruen}, {Gruendl}, {Gschwend},
  {Gutierrez}, {Honscheid}, {James}, {Krause}, {Kuehn}, {Kuropatkin}, {Lahav},
  {Maia}, {Marshall}, {Melchior}, {Menanteau}, {Miquel}, {Palmese}, {Plazas},
  {Sanchez}, {Scarpine}, {Schubnell}, {Serrano}, {Sevilla-Noarbe}, {Smith},
  {Suchyta}, {Tarle}, \& {DES Collaboration}}]{Drlica-Wagner:2020}
{Drlica-Wagner}, A., {Bechtol}, K., {Mau}, S., {et~al.} 2020, \apj, 893, 47

\bibitem[{{Drlica-Wagner} {et~al.}(2021){Drlica-Wagner}, {Carlin}, {Nidever},
  {Ferguson}, {Kuropatkin}, {Adam{\'o}w}, {Cerny}, {Choi}, {Esteves},
  {Mart{\'\i}nez-V{\'a}zquez}, {Mau}, {Miller}, {Mutlu-Pakdil}, {Neilsen},
  {Olsen}, {Pace}, {Riley}, {Sakowska}, {Sand}, {Santana-Silva}, {Tollerud},
  {Tucker}, {Vivas}, {Zaborowski}, {Zenteno}, {Abbott}, {Allam}, {Bechtol},
  {Bell}, {Bell}, {Bilaji}, {Bom}, {Carballo-Bello}, {Crnojevi{\'c}}, {Cioni},
  {Diaz-Ocampo}, {de Boer}, {Erkal}, {Gruendl}, {Hernandez-Lang}, {Hughes},
  {James}, {Johnson}, {Li}, {Mao}, {Mart{\'\i}nez-Delgado}, {Massana},
  {McNanna}, {Morgan}, {Nadler}, {No{\"e}l}, {Palmese}, {Peter}, {Rykoff},
  {S{\'a}nchez}, {Shipp}, {Simon}, {Smercina}, {Soares-Santos}, {Stringfellow},
  {Tavangar}, {van der Marel}, {Walker}, {Wechsler}, {Wu}, {Yanny},
  {Fitzpatrick}, {Huang}, {Jacques}, {Nikutta}, {Scott}, \& {Astro Data
  Lab}}]{2021ApJS..256....2D}
{Drlica-Wagner}, A., {Carlin}, J.~L., {Nidever}, D.~L., {et~al.} 2021, \apjs,
  256, 2

\bibitem[{{Eadie} \& {Harris}(2016)}]{2016ApJ...829..108E}
{Eadie}, G.~M., \& {Harris}, W.~E. 2016, \apj, 829, 108

\bibitem[{{Efron}(1982)}]{1982jbor.book.....E}
{Efron}, B. 1982, {The Jackknife, the Bootstrap and other resampling plans}

\bibitem[{{Erkal} \& {Belokurov}(2020)}]{Erkal_etal_2020}
{Erkal}, D., \& {Belokurov}, V.~A. 2020, \mnras, 495, 2554

\bibitem[{{Erkal} {et~al.}(2019){Erkal}, {Belokurov}, {Laporte}, {Koposov},
  {Li}, {Grillmair}, {Kallivayalil}, {Price-Whelan}, {Evans}, {Hawkins},
  {Hendel}, {Mateu}, {Navarro}, {del Pino}, {Slater}, {Sohn}, \& {Orphan Aspen
  Treasury Collaboration}}]{Erkal_etal_2019}
{Erkal}, D., {Belokurov}, V., {Laporte}, C.~F.~P., {et~al.} 2019, \mnras, 487,
  2685

\bibitem[{{Fadely} {et~al.}(2011){Fadely}, {Willman}, {Geha}, {Walsh},
  {Mu{\~n}oz}, {Jerjen}, {Vargas}, \& {Da Costa}}]{2011AJ....142...88F}
{Fadely}, R., {Willman}, B., {Geha}, M., {et~al.} 2011, \aj, 142, 88

\bibitem[{{Flaugher} {et~al.}(2015){Flaugher}, {Diehl}, {Honscheid}, {Abbott},
  {Alvarez}, {Angstadt}, {Annis}, {Antonik}, {Ballester}, {Beaufore},
  {Bernstein}, {Bernstein}, {Bigelow}, {Bonati}, {Boprie}, {Brooks},
  {Buckley-Geer}, {Campa}, {Cardiel-Sas}, {Castander}, {Castilla}, {Cease},
  {Cela-Ruiz}, {Chappa}, {Chi}, {Cooper}, {da Costa}, {Dede}, {Derylo},
  {DePoy}, {de Vicente}, {Doel}, {Drlica-Wagner}, {Eiting}, {Elliott}, {Emes},
  {Estrada}, {Fausti Neto}, {Finley}, {Flores}, {Frieman}, {Gerdes},
  {Gladders}, {Gregory}, {Gutierrez}, {Hao}, {Holland}, {Holm}, {Huffman},
  {Jackson}, {James}, {Jonas}, {Karcher}, {Karliner}, {Kent}, {Kessler},
  {Kozlovsky}, {Kron}, {Kubik}, {Kuehn}, {Kuhlmann}, {Kuk}, {Lahav}, {Lathrop},
  {Lee}, {Levi}, {Lewis}, {Li}, {Mandrichenko}, {Marshall}, {Martinez},
  {Merritt}, {Miquel}, {Mu{\~n}oz}, {Neilsen}, {Nichol}, {Nord}, {Ogando},
  {Olsen}, {Palaio}, {Patton}, {Peoples}, {Plazas}, {Rauch}, {Reil}, {Rheault},
  {Roe}, {Rogers}, {Roodman}, {Sanchez}, {Scarpine}, {Schindler}, {Schmidt},
  {Schmitt}, {Schubnell}, {Schultz}, {Schurter}, {Scott}, {Serrano}, {Shaw},
  {Smith}, {Soares-Santos}, {Stefanik}, {Stuermer}, {Suchyta}, {Sypniewski},
  {Tarle}, {Thaler}, {Tighe}, {Tran}, {Tucker}, {Walker}, {Wang}, {Watson},
  {Weaverdyck}, {Wester}, {Woods}, {Yanny}, \& {DES
  Collaboration}}]{2015AJ....150..150F}
{Flaugher}, B., {Diehl}, H.~T., {Honscheid}, K., {et~al.} 2015, \aj, 150, 150

\bibitem[{{Foreman-Mackey} {et~al.}(2013){Foreman-Mackey}, {Hogg}, {Lang}, \&
  {Goodman}}]{Foreman-Mackey:2013}
{Foreman-Mackey}, D., {Hogg}, D.~W., {Lang}, D., \& {Goodman}, J. 2013, \pasp,
  125, 306

\bibitem[{{Frebel}(2010)}]{2010AN....331..474F}
{Frebel}, A. 2010, Astronomische Nachrichten, 331, 474

\bibitem[{{Frebel} {et~al.}(2014){Frebel}, {Simon}, \&
  {Kirby}}]{2014ApJ...786...74F}
{Frebel}, A., {Simon}, J.~D., \& {Kirby}, E.~N. 2014, \apj, 786, 74

\bibitem[{{Fritz} {et~al.}(2018){Fritz}, {Battaglia}, {Pawlowski},
  {Kallivayalil}, {van der Marel}, {Sohn}, {Brook}, \&
  {Besla}}]{2018A&A...619A.103F}
{Fritz}, T.~K., {Battaglia}, G., {Pawlowski}, M.~S., {et~al.} 2018, \aap, 619,
  A103

\bibitem[{{Fritz} {et~al.}(2019){Fritz}, {Carrera}, {Battaglia}, \&
  {Taibi}}]{2019A&A...623A.129F}
{Fritz}, T.~K., {Carrera}, R., {Battaglia}, G., \& {Taibi}, S. 2019, \aap, 623,
  A129

\bibitem[{{Gaia Collaboration} {et~al.}(2018){Gaia Collaboration}, {Brown},
  {Vallenari}, {Prusti}, {de Bruijne}, {Babusiaux}, {Bailer-Jones}, {Biermann},
  {Evans}, {Eyer}, {Jansen}, {Jordi}, {Klioner}, {Lammers}, {Lindegren},
  {Luri}, {Mignard}, {Panem}, {Pourbaix}, {Randich}, {Sartoretti}, {Siddiqui},
  {Soubiran}, {van Leeuwen}, {Walton}, {Arenou}, {Bastian}, {Cropper},
  {Drimmel}, {Katz}, {Lattanzi}, {Bakker}, {Cacciari}, {Casta{\~n}eda},
  {Chaoul}, {Cheek}, {De Angeli}, {Fabricius}, {Guerra}, {Holl}, {Masana},
  {Messineo}, {Mowlavi}, {Nienartowicz}, {Panuzzo}, {Portell}, {Riello},
  {Seabroke}, {Tanga}, {Th{\'e}venin}, {Gracia-Abril}, {Comoretto},
  {Garcia-Reinaldos}, {Teyssier}, {Altmann}, {Andrae}, {Audard},
  {Bellas-Velidis}, {Benson}, {Berthier}, {Blomme}, {Burgess}, {Busso},
  {Carry}, {Cellino}, {Clementini}, {Clotet}, {Creevey}, {Davidson}, {De
  Ridder}, {Delchambre}, {Dell'Oro}, {Ducourant},
  {Fern{\'a}ndez-Hern{\'a}ndez}, {Fouesneau}, {Fr{\'e}mat}, {Galluccio},
  {Garc{\'\i}a-Torres}, {Gonz{\'a}lez-N{\'u}{\~n}ez}, {Gonz{\'a}lez-Vidal},
  {Gosset}, {Guy}, {Halbwachs}, {Hambly}, {Harrison}, {Hern{\'a}ndez},
  {Hestroffer}, {Hodgkin}, {Hutton}, {Jasniewicz}, {Jean-Antoine-Piccolo},
  {Jordan}, {Korn}, {Krone-Martins}, {Lanzafame}, {Lebzelter}, {L{\"o}ffler},
  {Manteiga}, {Marrese}, {Mart{\'\i}n-Fleitas}, {Moitinho}, {Mora}, {Muinonen},
  {Osinde}, {Pancino}, {Pauwels}, {Petit}, {Recio-Blanco}, {Richards},
  {Rimoldini}, {Robin}, {Sarro}, {Siopis}, {Smith}, {Sozzetti}, {S{\"u}veges},
  {Torra}, {van Reeven}, {Abbas}, {Abreu Aramburu}, {Accart}, {Aerts},
  {Altavilla}, {{\'A}lvarez}, {Alvarez}, {Alves}, {Anderson}, {Andrei},
  {Anglada Varela}, {Antiche}, {Antoja}, {Arcay}, {Astraatmadja}, {Bach},
  {Baker}, {Balaguer-N{\'u}{\~n}ez}, {Balm}, {Barache}, {Barata}, {Barbato},
  {Barblan}, {Barklem}, {Barrado}, {Barros}, {Barstow}, {Bartholom{\'e}
  Mu{\~n}oz}, {Bassilana}, {Becciani}, {Bellazzini}, {Berihuete}, {Bertone},
  {Bianchi}, {Bienaym{\'e}}, {Blanco-Cuaresma}, {Boch}, {Boeche}, {Bombrun},
  {Borrachero}, {Bossini}, {Bouquillon}, {Bourda}, {Bragaglia}, {Bramante},
  {Breddels}, {Bressan}, {Brouillet}, {Br{\"u}semeister}, {Brugaletta},
  {Bucciarelli}, {Burlacu}, {Busonero}, {Butkevich}, {Buzzi}, {Caffau},
  {Cancelliere}, {Cannizzaro}, {Cantat-Gaudin}, {Carballo}, {Carlucci},
  {Carrasco}, {Casamiquela}, {Castellani}, {Castro-Ginard}, {Charlot},
  {Chemin}, {Chiavassa}, {Cocozza}, {Costigan}, {Cowell}, {Crifo}, {Crosta},
  {Crowley}, {Cuypers}, {Dafonte}, {Damerdji}, {Dapergolas}, {David}, {David},
  {de Laverny}, {De Luise}, {De March}, {de Martino}, {de Souza}, {de Torres},
  {Debosscher}, {del Pozo}, {Delbo}, {Delgado}, {Delgado}, {Di Matteo},
  {Diakite}, {Diener}, {Distefano}, {Dolding}, {Drazinos}, {Dur{\'a}n},
  {Edvardsson}, {Enke}, {Eriksson}, {Esquej}, {Eynard Bontemps}, {Fabre},
  {Fabrizio}, {Faigler}, {Falc{\~a}o}, {Farr{\`a}s Casas}, {Federici},
  {Fedorets}, {Fernique}, {Figueras}, {Filippi}, {Findeisen}, {Fonti},
  {Fraile}, {Fraser}, {Fr{\'e}zouls}, {Gai}, {Galleti}, {Garabato},
  {Garc{\'\i}a-Sedano}, {Garofalo}, {Garralda}, {Gavel}, {Gavras}, {Gerssen},
  {Geyer}, {Giacobbe}, {Gilmore}, {Girona}, {Giuffrida}, {Glass}, {Gomes},
  {Granvik}, {Gueguen}, {Guerrier}, {Guiraud}, {Guti{\'e}rrez-S{\'a}nchez},
  {Haigron}, {Hatzidimitriou}, {Hauser}, {Haywood}, {Heiter}, {Helmi}, {Heu},
  {Hilger}, {Hobbs}, {Hofmann}, {Holland}, {Huckle}, {Hypki}, {Icardi},
  {Jan{\ss}en}, {Jevardat de Fombelle}, {Jonker}, {Juh{\'a}sz}, {Julbe},
  {Karampelas}, {Kewley}, {Klar}, {Kochoska}, {Kohley}, {Kolenberg},
  {Kontizas}, {Kontizas}, {Koposov}, {Kordopatis}, {Kostrzewa-Rutkowska},
  {Koubsky}, {Lambert}, {Lanza}, {Lasne}, {Lavigne}, {Le Fustec}, {Le
  Poncin-Lafitte}, {Lebreton}, {Leccia}, {Leclerc}, {Lecoeur-Taibi},
  {Lenhardt}, {Leroux}, {Liao}, {Licata}, {Lindstr{\o}m}, {Lister}, {Livanou},
  {Lobel}, {L{\'o}pez}, {Managau}, {Mann}, {Mantelet}, {Marchal}, {Marchant},
  {Marconi}, {Marinoni}, {Marschalk{\'o}}, {Marshall}, {Martino}, {Marton},
  {Mary}, {Massari}, {Matijevi{\v{c}}}, {Mazeh}, {McMillan}, {Messina},
  {Michalik}, {Millar}, {Molina}, {Molinaro}, {Moln{\'a}r}, {Montegriffo},
  {Mor}, {Morbidelli}, {Morel}, {Morris}, {Mulone}, {Muraveva}, {Musella},
  {Nelemans}, {Nicastro}, {Noval}, {O'Mullane}, {Ord{\'e}novic},
  {Ord{\'o}{\~n}ez-Blanco}, {Osborne}, {Pagani}, {Pagano}, {Pailler},
  {Palacin}, {Palaversa}, {Panahi}, {Pawlak}, {Piersimoni}, {Pineau}, {Plachy},
  {Plum}, {Poggio}, {Poujoulet}, {Pr{\v{s}}a}, {Pulone}, {Racero}, {Ragaini},
  {Rambaux}, {Ramos-Lerate}, {Regibo}, {Reyl{\'e}}, {Riclet}, {Ripepi}, {Riva},
  {Rivard}, {Rixon}, {Roegiers}, {Roelens}, {Romero-G{\'o}mez}, {Rowell},
  {Royer}, {Ruiz-Dern}, {Sadowski}, {Sagrist{\`a} Sell{\'e}s}, {Sahlmann},
  {Salgado}, {Salguero}, {Sanna}, {Santana-Ros}, {Sarasso}, {Savietto},
  {Schultheis}, {Sciacca}, {Segol}, {Segovia}, {S{\'e}gransan}, {Shih},
  {Siltala}, {Silva}, {Smart}, {Smith}, {Solano}, {Solitro}, {Sordo}, {Soria
  Nieto}, {Souchay}, {Spagna}, {Spoto}, {Stampa}, {Steele},
  {Steidelm{\"u}ller}, {Stephenson}, {Stoev}, {Suess}, {Surdej}, {Szabados},
  {Szegedi-Elek}, {Tapiador}, {Taris}, {Tauran}, {Taylor}, {Teixeira},
  {Terrett}, {Teyssandier}, {Thuillot}, {Titarenko}, {Torra Clotet}, {Turon},
  {Ulla}, {Utrilla}, {Uzzi}, {Vaillant}, {Valentini}, {Valette}, {van Elteren},
  {Van Hemelryck}, {van Leeuwen}, {Vaschetto}, {Vecchiato}, {Veljanoski},
  {Viala}, {Vicente}, {Vogt}, {von Essen}, {Voss}, {Votruba}, {Voutsinas},
  {Walmsley}, {Weiler}, {Wertz}, {Wevers}, {Wyrzykowski}, {Yoldas},
  {{\v{Z}}erjal}, {Ziaeepour}, {Zorec}, {Zschocke}, {Zucker}, {Zurbach}, \&
  {Zwitter}}]{2018A&A...616A...1G}
{Gaia Collaboration}, {Brown}, A.~G.~A., {Vallenari}, A., {et~al.} 2018, \aap,
  616, A1

\bibitem[{{Gaia Collaboration} {et~al.}(2021){Gaia Collaboration}, {Brown},
  {Vallenari}, {Prusti}, {de Bruijne}, {Babusiaux}, {Biermann}, {Creevey},
  {Evans}, {Eyer}, {Hutton}, {Jansen}, {Jordi}, {Klioner}, {Lammers},
  {Lindegren}, {Luri}, {Mignard}, {Panem}, {Pourbaix}, {Randich}, {Sartoretti},
  {Soubiran}, {Walton}, {Arenou}, {Bailer-Jones}, {Bastian}, {Cropper},
  {Drimmel}, {Katz}, {Lattanzi}, {van Leeuwen}, {Bakker}, {Cacciari},
  {Casta{\~n}eda}, {De Angeli}, {Ducourant}, {Fabricius}, {Fouesneau},
  {Fr{\'e}mat}, {Guerra}, {Guerrier}, {Guiraud}, {Jean-Antoine Piccolo},
  {Masana}, {Messineo}, {Mowlavi}, {Nicolas}, {Nienartowicz}, {Pailler},
  {Panuzzo}, {Riclet}, {Roux}, {Seabroke}, {Sordo}, {Tanga}, {Th{\'e}venin},
  {Gracia-Abril}, {Portell}, {Teyssier}, {Altmann}, {Andrae}, {Bellas-Velidis},
  {Benson}, {Berthier}, {Blomme}, {Brugaletta}, {Burgess}, {Busso}, {Carry},
  {Cellino}, {Cheek}, {Clementini}, {Damerdji}, {Davidson}, {Delchambre},
  {Dell'Oro}, {Fern{\'a}ndez-Hern{\'a}ndez}, {Galluccio}, {Garc{\'\i}a-Lario},
  {Garcia-Reinaldos}, {Gonz{\'a}lez-N{\'u}{\~n}ez}, {Gosset}, {Haigron},
  {Halbwachs}, {Hambly}, {Harrison}, {Hatzidimitriou}, {Heiter},
  {Hern{\'a}ndez}, {Hestroffer}, {Hodgkin}, {Holl}, {Jan{\ss}en}, {Jevardat de
  Fombelle}, {Jordan}, {Krone-Martins}, {Lanzafame}, {L{\"o}ffler}, {Lorca},
  {Manteiga}, {Marchal}, {Marrese}, {Moitinho}, {Mora}, {Muinonen}, {Osborne},
  {Pancino}, {Pauwels}, {Petit}, {Recio-Blanco}, {Richards}, {Riello},
  {Rimoldini}, {Robin}, {Roegiers}, {Rybizki}, {Sarro}, {Siopis}, {Smith},
  {Sozzetti}, {Ulla}, {Utrilla}, {van Leeuwen}, {van Reeven}, {Abbas}, {Abreu
  Aramburu}, {Accart}, {Aerts}, {Aguado}, {Ajaj}, {Altavilla}, {{\'A}lvarez},
  {{\'A}lvarez Cid-Fuentes}, {Alves}, {Anderson}, {Anglada Varela}, {Antoja},
  {Audard}, {Baines}, {Baker}, {Balaguer-N{\'u}{\~n}ez}, {Balbinot}, {Balog},
  {Barache}, {Barbato}, {Barros}, {Barstow}, {Bartolom{\'e}}, {Bassilana},
  {Bauchet}, {Baudesson-Stella}, {Becciani}, {Bellazzini}, {Bernet}, {Bertone},
  {Bianchi}, {Blanco-Cuaresma}, {Boch}, {Bombrun}, {Bossini}, {Bouquillon},
  {Bragaglia}, {Bramante}, {Breedt}, {Bressan}, {Brouillet}, {Bucciarelli},
  {Burlacu}, {Busonero}, {Butkevich}, {Buzzi}, {Caffau}, {Cancelliere},
  {C{\'a}novas}, {Cantat-Gaudin}, {Carballo}, {Carlucci}, {Carnerero},
  {Carrasco}, {Casamiquela}, {Castellani}, {Castro-Ginard}, {Castro Sampol},
  {Chaoul}, {Charlot}, {Chemin}, {Chiavassa}, {Cioni}, {Comoretto}, {Cooper},
  {Cornez}, {Cowell}, {Crifo}, {Crosta}, {Crowley}, {Dafonte}, {Dapergolas},
  {David}, {David}, {de Laverny}, {De Luise}, {De March}, {De Ridder}, {de
  Souza}, {de Teodoro}, {de Torres}, {del Peloso}, {del Pozo}, {Delbo},
  {Delgado}, {Delgado}, {Delisle}, {Di Matteo}, {Diakite}, {Diener},
  {Distefano}, {Dolding}, {Eappachen}, {Edvardsson}, {Enke}, {Esquej}, {Fabre},
  {Fabrizio}, {Faigler}, {Fedorets}, {Fernique}, {Fienga}, {Figueras},
  {Fouron}, {Fragkoudi}, {Fraile}, {Franke}, {Gai}, {Garabato},
  {Garcia-Gutierrez}, {Garc{\'\i}a-Torres}, {Garofalo}, {Gavras}, {Gerlach},
  {Geyer}, {Giacobbe}, {Gilmore}, {Girona}, {Giuffrida}, {Gomel}, {Gomez},
  {Gonzalez-Santamaria}, {Gonz{\'a}lez-Vidal}, {Granvik},
  {Guti{\'e}rrez-S{\'a}nchez}, {Guy}, {Hauser}, {Haywood}, {Helmi}, {Hidalgo},
  {Hilger}, {H{\l}adczuk}, {Hobbs}, {Holland}, {Huckle}, {Jasniewicz},
  {Jonker}, {Juaristi Campillo}, {Julbe}, {Karbevska}, {Kervella}, {Khanna},
  {Kochoska}, {Kontizas}, {Kordopatis}, {Korn}, {Kostrzewa-Rutkowska},
  {Kruszy{\'n}ska}, {Lambert}, {Lanza}, {Lasne}, {Le Campion}, {Le Fustec},
  {Lebreton}, {Lebzelter}, {Leccia}, {Leclerc}, {Lecoeur-Taibi}, {Liao},
  {Licata}, {Lindstr{\o}m}, {Lister}, {Livanou}, {Lobel}, {Madrero Pardo},
  {Managau}, {Mann}, {Marchant}, {Marconi}, {Marcos Santos}, {Marinoni},
  {Marocco}, {Marshall}, {Martin Polo}, {Mart{\'\i}n-Fleitas}, {Masip},
  {Massari}, {Mastrobuono-Battisti}, {Mazeh}, {McMillan}, {Messina},
  {Michalik}, {Millar}, {Mints}, {Molina}, {Molinaro}, {Moln{\'a}r},
  {Montegriffo}, {Mor}, {Morbidelli}, {Morel}, {Morris}, {Mulone}, {Munoz},
  {Muraveva}, {Murphy}, {Musella}, {Noval}, {Ord{\'e}novic}, {Orr{\`u}},
  {Osinde}, {Pagani}, {Pagano}, {Palaversa}, {Palicio}, {Panahi}, {Pawlak},
  {Pe{\~n}alosa Esteller}, {Penttil{\"a}}, {Piersimoni}, {Pineau}, {Plachy},
  {Plum}, {Poggio}, {Poretti}, {Poujoulet}, {Pr{\v{s}}a}, {Pulone}, {Racero},
  {Ragaini}, {Rainer}, {Raiteri}, {Rambaux}, {Ramos}, {Ramos-Lerate}, {Re
  Fiorentin}, {Regibo}, {Reyl{\'e}}, {Ripepi}, {Riva}, {Rixon}, {Robichon},
  {Robin}, {Roelens}, {Rohrbasser}, {Romero-G{\'o}mez}, {Rowell}, {Royer},
  {Rybicki}, {Sadowski}, {Sagrist{\`a} Sell{\'e}s}, {Sahlmann}, {Salgado},
  {Salguero}, {Samaras}, {Sanchez Gimenez}, {Sanna}, {Santove{\~n}a},
  {Sarasso}, {Schultheis}, {Sciacca}, {Segol}, {Segovia}, {S{\'e}gransan},
  {Semeux}, {Shahaf}, {Siddiqui}, {Siebert}, {Siltala}, {Slezak}, {Smart},
  {Solano}, {Solitro}, {Souami}, {Souchay}, {Spagna}, {Spoto}, {Steele},
  {Steidelm{\"u}ller}, {Stephenson}, {S{\"u}veges}, {Szabados}, {Szegedi-Elek},
  {Taris}, {Tauran}, {Taylor}, {Teixeira}, {Thuillot}, {Tonello}, {Torra},
  {Torra}, {Turon}, {Unger}, {Vaillant}, {van Dillen}, {Vanel}, {Vecchiato},
  {Viala}, {Vicente}, {Voutsinas}, {Weiler}, {Wevers}, {Wyrzykowski}, {Yoldas},
  {Yvard}, {Zhao}, {Zorec}, {Zucker}, {Zurbach}, \&
  {Zwitter}}]{Gaia_Brown_2021A&A...649A...1G}
---. 2021, \aap, 649, A1

\bibitem[{{Gatto} {et~al.}(2021){Gatto}, {Ripepi}, {Bellazzini}, {Tosi},
  {Tortora}, {Cignoni}, {Spavone}, {Dall'ora}, {Clementini}, {Cusano}, {Longo},
  {Musella}, {Marconi}, \& {Schipani}}]{2021RNAAS...5..159G}
{Gatto}, M., {Ripepi}, V., {Bellazzini}, M., {et~al.} 2021, Research Notes of
  the American Astronomical Society, 5, 159

\bibitem[{{Geringer-Sameth} {et~al.}(2015){Geringer-Sameth}, {Koushiappas}, \&
  {Walker}}]{GeringerSameth2015ApJ...801...74G}
{Geringer-Sameth}, A., {Koushiappas}, S.~M., \& {Walker}, M. 2015, \apj, 801,
  74

\bibitem[{{G{\'o}mez} {et~al.}(2015){G{\'o}mez}, {Besla}, {Carpintero},
  {Villalobos}, {O'Shea}, \& {Bell}}]{Gomez_etal_2015}
{G{\'o}mez}, F.~A., {Besla}, G., {Carpintero}, D.~D., {et~al.} 2015, \apj, 802,
  128

\bibitem[{{G{\'o}rski} {et~al.}(2005){G{\'o}rski}, {Hivon}, {Banday},
  {Wandelt}, {Hansen}, {Reinecke}, \& {Bartelmann}}]{Gorski:2005}
{G{\'o}rski}, K.~M., {Hivon}, E., {Banday}, A.~J., {et~al.} 2005, \apj, 622,
  759

\bibitem[{{Greco} {et~al.}(2008){Greco}, {Dall'Ora}, {Clementini}, {Ripepi},
  {Di Fabrizio}, {Kinemuchi}, {Marconi}, {Musella}, {Smith}, {Rodgers},
  {Kuehn}, {Beers}, {Catelan}, \& {Pritzl}}]{2008ApJ...675L..73G}
{Greco}, C., {Dall'Ora}, M., {Clementini}, G., {et~al.} 2008, \apjl, 675, L73

\bibitem[{{Hargis} {et~al.}(2014){Hargis}, {Willman}, \&
  {Peter}}]{2014ApJ...795L..13H}
{Hargis}, J.~R., {Willman}, B., \& {Peter}, A. H.~G. 2014, \apjl, 795, L13

\bibitem[{{Harris}(1996)}]{Harris:1996}
{Harris}, W.~E. 1996, \aj, 112, 1487

\bibitem[{{Hendricks} {et~al.}(2014){Hendricks}, {Koch}, {Walker}, {Johnson},
  {Pe{\~n}arrubia}, \& {Gilmore}}]{2014A&A...572A..82H}
{Hendricks}, B., {Koch}, A., {Walker}, M., {et~al.} 2014, \aap, 572, A82

\bibitem[{{Hernquist}(1990)}]{Hernquist_1990}
{Hernquist}, L. 1990, \apj, 356, 359

\bibitem[{{Holl} {et~al.}(2018){Holl}, {Audard}, {Nienartowicz}, {Jevardat de
  Fombelle}, {Marchal}, {Mowlavi}, {Clementini}, {De Ridder}, {Evans}, {Guy},
  {Lanzafame}, {Lebzelter}, {Rimoldini}, {Roelens}, {Zucker}, {Distefano},
  {Garofalo}, {Lecoeur-Ta{\"\i}bi}, {Lopez}, {Molinaro}, {Muraveva}, {Panahi},
  {Regibo}, {Ripepi}, {Sarro}, {Aerts}, {Anderson}, {Charnas}, {Barblan},
  {Blanco-Cuaresma}, {Busso}, {Cuypers}, {De Angeli}, {Glass}, {Grenon},
  {Juh{\'a}sz}, {Kochoska}, {Koubsky}, {Lanza}, {Leccia}, {Lorenz}, {Marconi},
  {Marschalk{\'o}}, {Mazeh}, {Messina}, {Mignard}, {Moitinho}, {Moln{\'a}r},
  {Morgenthaler}, {Musella}, {Ordenovic}, {Ord{\'o}{\~n}ez}, {Pagano},
  {Palaversa}, {Pawlak}, {Plachy}, {Pr{\v{s}}a}, {Riello}, {S{\"u}veges},
  {Szabados}, {Szegedi-Elek}, {Votruba}, \& {Eyer}}]{2018A&A...618A..30H}
{Holl}, B., {Audard}, M., {Nienartowicz}, K., {et~al.} 2018, \aap, 618, A30

\bibitem[{{Homma} {et~al.}(2016){Homma}, {Chiba}, {Okamoto}, {Komiyama},
  {Tanaka}, {Tanaka}, {Ishigaki}, {Akiyama}, {Arimoto}, {Garmilla}, {Lupton},
  {Strauss}, {Furusawa}, {Miyazaki}, {Murayama}, {Nishizawa}, {Takada},
  {Usuda}, \& {Wang}}]{Homma:2016}
{Homma}, D., {Chiba}, M., {Okamoto}, S., {et~al.} 2016, \apj, 832, 21

\bibitem[{{Homma} {et~al.}(2018){Homma}, {Chiba}, {Okamoto}, {Komiyama},
  {Tanaka}, {Tanaka}, {Ishigaki}, {Hayashi}, {Arimoto}, {Garmilla}, {Lupton},
  {Strauss}, {Miyazaki}, {Wang}, \& {Murayama}}]{Homma:2018}
---. 2018, \pasj, 70, S18

\bibitem[{{Homma} {et~al.}(2019){Homma}, {Chiba}, {Komiyama}, {Tanaka},
  {Okamoto}, {Tanaka}, {Ishigaki}, {Hayashi}, {Arimoto}, {Carlsten}, {Lupton},
  {Strauss}, {Miyazaki}, {Torrealba}, {Wang}, \& {Murayama}}]{Homma:2019}
{Homma}, D., {Chiba}, M., {Komiyama}, Y., {et~al.} 2019, \pasj, 71, 94

\bibitem[{{Ivezi{\'c}} {et~al.}(2019){Ivezi{\'c}}, {Kahn}, {Tyson}, {Abel},
  {Acosta}, {Allsman}, {Alonso}, {AlSayyad}, {Anderson}, {Andrew}, {Angel},
  {Angeli}, {Ansari}, {Antilogus}, {Araujo}, {Armstrong}, {Arndt}, {Astier},
  {Aubourg}, {Auza}, {Axelrod}, {Bard}, {Barr}, {Barrau}, {Bartlett}, {Bauer},
  {Bauman}, {Baumont}, {Bechtol}, {Bechtol}, {Becker}, {Becla}, {Beldica},
  {Bellavia}, {Bianco}, {Biswas}, {Blanc}, {Blazek}, {Blandford}, {Bloom},
  {Bogart}, {Bond}, {Booth}, {Borgland}, {Borne}, {Bosch}, {Boutigny},
  {Brackett}, {Bradshaw}, {Brandt}, {Brown}, {Bullock}, {Burchat}, {Burke},
  {Cagnoli}, {Calabrese}, {Callahan}, {Callen}, {Carlin}, {Carlson},
  {Chandrasekharan}, {Charles-Emerson}, {Chesley}, {Cheu}, {Chiang}, {Chiang},
  {Chirino}, {Chow}, {Ciardi}, {Claver}, {Cohen-Tanugi}, {Cockrum}, {Coles},
  {Connolly}, {Cook}, {Cooray}, {Covey}, {Cribbs}, {Cui}, {Cutri}, {Daly},
  {Daniel}, {Daruich}, {Daubard}, {Daues}, {Dawson}, {Delgado}, {Dellapenna},
  {de Peyster}, {de Val-Borro}, {Digel}, {Doherty}, {Dubois},
  {Dubois-Felsmann}, {Durech}, {Economou}, {Eifler}, {Eracleous}, {Emmons},
  {Fausti Neto}, {Ferguson}, {Figueroa}, {Fisher-Levine}, {Focke}, {Foss},
  {Frank}, {Freemon}, {Gangler}, {Gawiser}, {Geary}, {Gee}, {Geha}, {Gessner},
  {Gibson}, {Gilmore}, {Glanzman}, {Glick}, {Goldina}, {Goldstein}, {Goodenow},
  {Graham}, {Gressler}, {Gris}, {Guy}, {Guyonnet}, {Haller}, {Harris},
  {Hascall}, {Haupt}, {Hernandez}, {Herrmann}, {Hileman}, {Hoblitt}, {Hodgson},
  {Hogan}, {Howard}, {Huang}, {Huffer}, {Ingraham}, {Innes}, {Jacoby}, {Jain},
  {Jammes}, {Jee}, {Jenness}, {Jernigan}, {Jevremovi{\'c}}, {Johns}, {Johnson},
  {Johnson}, {Jones}, {Juramy-Gilles}, {Juri{\'c}}, {Kalirai}, {Kallivayalil},
  {Kalmbach}, {Kantor}, {Karst}, {Kasliwal}, {Kelly}, {Kessler}, {Kinnison},
  {Kirkby}, {Knox}, {Kotov}, {Krabbendam}, {Krughoff}, {Kub{\'a}nek},
  {Kuczewski}, {Kulkarni}, {Ku}, {Kurita}, {Lage}, {Lambert}, {Lange},
  {Langton}, {Le Guillou}, {Levine}, {Liang}, {Lim}, {Lintott}, {Long},
  {Lopez}, {Lotz}, {Lupton}, {Lust}, {MacArthur}, {Mahabal}, {Mandelbaum},
  {Markiewicz}, {Marsh}, {Marshall}, {Marshall}, {May}, {McKercher}, {McQueen},
  {Meyers}, {Migliore}, {Miller}, {Mills}, {Miraval}, {Moeyens}, {Moolekamp},
  {Monet}, {Moniez}, {Monkewitz}, {Montgomery}, {Morrison}, {Mueller},
  {Muller}, {Mu{\~n}oz Arancibia}, {Neill}, {Newbry}, {Nief}, {Nomerotski},
  {Nordby}, {O'Connor}, {Oliver}, {Olivier}, {Olsen}, {O'Mullane}, {Ortiz},
  {Osier}, {Owen}, {Pain}, {Palecek}, {Parejko}, {Parsons}, {Pease},
  {Peterson}, {Peterson}, {Petravick}, {Libby Petrick}, {Petry},
  {Pierfederici}, {Pietrowicz}, {Pike}, {Pinto}, {Plante}, {Plate}, {Plutchak},
  {Price}, {Prouza}, {Radeka}, {Rajagopal}, {Rasmussen}, {Regnault}, {Reil},
  {Reiss}, {Reuter}, {Ridgway}, {Riot}, {Ritz}, {Robinson}, {Roby}, {Roodman},
  {Rosing}, {Roucelle}, {Rumore}, {Russo}, {Saha}, {Sassolas}, {Schalk},
  {Schellart}, {Schindler}, {Schmidt}, {Schneider}, {Schneider}, {Schoening},
  {Schumacher}, {Schwamb}, {Sebag}, {Selvy}, {Sembroski}, {Seppala}, {Serio},
  {Serrano}, {Shaw}, {Shipsey}, {Sick}, {Silvestri}, {Slater}, {Smith},
  {Smith}, {Sobhani}, {Soldahl}, {Storrie-Lombardi}, {Stover}, {Strauss},
  {Street}, {Stubbs}, {Sullivan}, {Sweeney}, {Swinbank}, {Szalay}, {Takacs},
  {Tether}, {Thaler}, {Thayer}, {Thomas}, {Thornton}, {Thukral}, {Tice},
  {Trilling}, {Turri}, {Van Berg}, {Vanden Berk}, {Vetter}, {Virieux},
  {Vucina}, {Wahl}, {Walkowicz}, {Walsh}, {Walter}, {Wang}, {Wang}, {Warner},
  {Wiecha}, {Willman}, {Winters}, {Wittman}, {Wolff}, {Wood-Vasey}, {Wu},
  {Xin}, {Yoachim}, \& {Zhan}}]{2019ApJ...873..111I}
{Ivezi{\'c}}, {\v{Z}}., {Kahn}, S.~M., {Tyson}, J.~A., {et~al.} 2019, \apj,
  873, 111

\bibitem[{{Jenkins} {et~al.}(2021){Jenkins}, {Li}, {Pace}, {Ji}, {Koposov}, \&
  {Mutlu-Pakdil}}]{2021ApJ...920...92J}
{Jenkins}, S.~A., {Li}, T.~S., {Pace}, A.~B., {et~al.} 2021, \apj, 920, 92

\bibitem[{{Jethwa} {et~al.}(2016){Jethwa}, {Erkal}, \&
  {Belokurov}}]{Jethwa_etal_2016}
{Jethwa}, P., {Erkal}, D., \& {Belokurov}, V. 2016, \mnras, 461, 2212

\bibitem[{{Ji} {et~al.}(2019){Ji}, {Simon}, {Frebel}, {Venn}, \&
  {Hansen}}]{2019ApJ...870...83J}
{Ji}, A.~P., {Simon}, J.~D., {Frebel}, A., {Venn}, K.~A., \& {Hansen}, T.~T.
  2019, \apj, 870, 83

\bibitem[{{Joo} {et~al.}(2018){Joo}, {Kyeong}, {Yang}, {Han}, {Sung}, {Kim},
  {Jeong}, {Ree}, {Rey}, {Jerjen}, {Kim}, {Cha}, \&
  {Lee}}]{2018ApJ...861...23J}
{Joo}, S.-J., {Kyeong}, J., {Yang}, S.-C., {et~al.} 2018, \apj, 861, 23

\bibitem[{{Joo} {et~al.}(2019){Joo}, {Kyeong}, {Yang}, {Han}, {Sung}, {Rey},
  {Jerjen}, {Kim}, {Kim}, {Jeong}, {Ree}, {Cha}, \&
  {Lee}}]{2019ApJ...875..120J}
---. 2019, \apj, 875, 120

\bibitem[{{Kallivayalil} {et~al.}(2013){Kallivayalil}, {van der Marel},
  {Besla}, {Anderson}, \& {Alcock}}]{Kallivayalil_etal_2013}
{Kallivayalil}, N., {van der Marel}, R.~P., {Besla}, G., {Anderson}, J., \&
  {Alcock}, C. 2013, \apj, 764, 161

\bibitem[{{Kim} \& {Jerjen}(2015)}]{2015ApJ...808L..39K}
{Kim}, D., \& {Jerjen}, H. 2015, \apjl, 808, L39

\bibitem[{{Kim} {et~al.}(2015{\natexlab{a}}){Kim}, {Jerjen}, {Mackey}, {Da
  Costa}, \& {Milone}}]{2015ApJ...804L..44K}
{Kim}, D., {Jerjen}, H., {Mackey}, D., {Da Costa}, G.~S., \& {Milone}, A.~P.
  2015{\natexlab{a}}, \apjl, 804, L44

\bibitem[{{Kim} {et~al.}(2016{\natexlab{a}}){Kim}, {Jerjen}, {Mackey}, {Da
  Costa}, \& {Milone}}]{2016ApJ...820..119K}
---. 2016{\natexlab{a}}, \apj, 820, 119

\bibitem[{{Kim} {et~al.}(2015{\natexlab{b}}){Kim}, {Jerjen}, {Milone},
  {Mackey}, \& {Da Costa}}]{2015ApJ...803...63K}
{Kim}, D., {Jerjen}, H., {Milone}, A.~P., {Mackey}, D., \& {Da Costa}, G.~S.
  2015{\natexlab{b}}, \apj, 803, 63

\bibitem[{{Kim} {et~al.}(2016{\natexlab{b}}){Kim}, {Jerjen}, {Geha}, {Chiti},
  {Milone}, {Da Costa}, {Mackey}, {Frebel}, \& {Conn}}]{2016ApJ...833...16K}
{Kim}, D., {Jerjen}, H., {Geha}, M., {et~al.} 2016{\natexlab{b}}, \apj, 833, 16

\bibitem[{{Kirby} {et~al.}(2015){Kirby}, {Simon}, \&
  {Cohen}}]{2015ApJ...810...56K}
{Kirby}, E.~N., {Simon}, J.~D., \& {Cohen}, J.~G. 2015, \apj, 810, 56

\bibitem[{{Kirby} {et~al.}(2008){Kirby}, {Simon}, {Geha}, {Guhathakurta}, \&
  {Frebel}}]{2008ApJ...685L..43K}
{Kirby}, E.~N., {Simon}, J.~D., {Geha}, M., {Guhathakurta}, P., \& {Frebel}, A.
  2008, \apjl, 685, L43

\bibitem[{{Kleyna} {et~al.}(2005){Kleyna}, {Wilkinson}, {Evans}, \&
  {Gilmore}}]{2005ApJ...630L.141K}
{Kleyna}, J.~T., {Wilkinson}, M.~I., {Evans}, N.~W., \& {Gilmore}, G. 2005,
  \apjl, 630, L141

\bibitem[{{Koposov} {et~al.}(2015{\natexlab{a}}){Koposov}, {Belokurov},
  {Torrealba}, \& {Evans}}]{Koposov_2015}
{Koposov}, S.~E., {Belokurov}, V., {Torrealba}, G., \& {Evans}, N.~W.
  2015{\natexlab{a}}, \apj, 805, 130

\bibitem[{{Koposov} {et~al.}(2015{\natexlab{b}}){Koposov}, {Casey},
  {Belokurov}, {Lewis}, {Gilmore}, {Worley}, {Hourihane}, {Randich}, {Bensby},
  {Bragaglia}, {Bergemann}, {Carraro}, {Costado}, {Flaccomio}, {Francois},
  {Heiter}, {Hill}, {Jofre}, {Lando}, {Lanzafame}, {de Laverny}, {Monaco},
  {Morbidelli}, {Sbordone}, {Mikolaitis}, \& {Ryde}}]{2015ApJ...811...62K}
{Koposov}, S.~E., {Casey}, A.~R., {Belokurov}, V., {et~al.} 2015{\natexlab{b}},
  \apj, 811, 62

\bibitem[{{Koposov} {et~al.}(2018){Koposov}, {Walker}, {Belokurov}, {Casey},
  {Geringer-Sameth}, {Mackey}, {Da Costa}, {Erkal}, {Jethwa}, {Mateo},
  {Olszewski}, \& {Bailey}}]{2018MNRAS.479.5343K}
{Koposov}, S.~E., {Walker}, M.~G., {Belokurov}, V., {et~al.} 2018, \mnras, 479,
  5343

\bibitem[{{Laevens} {et~al.}(2015){Laevens}, {Martin}, {Ibata}, {Rix},
  {Bernard}, {Bell}, {Sesar}, {Ferguson}, {Schlafly}, {Slater}, {Burgett},
  {Chambers}, {Flewelling}, {Hodapp}, {Kaiser}, {Kudritzki}, {Lupton},
  {Magnier}, {Metcalfe}, {Morgan}, {Price}, {Tonry}, {Wainscoat}, \&
  {Waters}}]{2015ApJ...802L..18L}
{Laevens}, B. P.~M., {Martin}, N.~F., {Ibata}, R.~A., {et~al.} 2015, \apjl,
  802, L18

\bibitem[{{Law} \& {Majewski}(2010)}]{2010ApJ...714..229L}
{Law}, D.~R., \& {Majewski}, S.~R. 2010, \apj, 714, 229

\bibitem[{{Li} {et~al.}(2021{\natexlab{a}}){Li}, {Hammer}, {Babusiaux},
  {Pawlowski}, {Yang}, {Arenou}, {Du}, \& {Wang}}]{HLi:2021}
{Li}, H., {Hammer}, F., {Babusiaux}, C., {et~al.} 2021{\natexlab{a}}, \apj,
  916, 8

\bibitem[{{Li} {et~al.}(2017){Li}, {Simon}, {Drlica-Wagner}, {Bechtol}, {Wang},
  {Garc{\'\i}a-Bellido}, {Frieman}, {Marshall}, {James}, {Strigari}, {Pace},
  {Balbinot}, {Zhang}, {Abbott}, {Allam}, {Benoit-L{\'e}vy}, {Bernstein},
  {Bertin}, {Brooks}, {Burke}, {Carnero Rosell}, {Carrasco Kind}, {Carretero},
  {Cunha}, {D'Andrea}, {da Costa}, {DePoy}, {Desai}, {Diehl}, {Eifler},
  {Flaugher}, {Goldstein}, {Gruen}, {Gruendl}, {Gschwend}, {Gutierrez},
  {Krause}, {Kuehn}, {Lin}, {Maia}, {March}, {Menanteau}, {Miquel}, {Plazas},
  {Romer}, {Sanchez}, {Santiago}, {Schubnell}, {Sevilla-Noarbe}, {Smith},
  {Sobreira}, {Suchyta}, {Tarle}, {Thomas}, {Tucker}, {Walker}, {Wechsler},
  {Wester}, {Yanny}, \& {DES Collaboration}}]{2017ApJ...838....8L}
{Li}, T.~S., {Simon}, J.~D., {Drlica-Wagner}, A., {et~al.} 2017, \apj, 838, 8

\bibitem[{{Li} {et~al.}(2018){Li}, {Simon}, {Pace}, {Torrealba}, {Kuehn},
  {Drlica-Wagner}, {Bechtol}, {Vivas}, {van der Marel}, {Wood}, {Yanny},
  {Belokurov}, {Jethwa}, {Zucker}, {Lewis}, {Kron}, {Nidever},
  {S{\'a}nchez-Conde}, {Ji}, {Conn}, {James}, {Martin}, {Martinez-Delgado},
  {No{\"e}l}, \& {MagLiteS Collaboration}}]{2018ApJ...857..145L}
{Li}, T.~S., {Simon}, J.~D., {Pace}, A.~B., {et~al.} 2018, \apj, 857, 145

\bibitem[{{Li} {et~al.}(2021{\natexlab{b}}){Li}, {Ji}, {Pace}, {Erkal},
  {Koposov}, {Shipp}, {Da Costa}, {Cullinane}, {Kuehn}, {Lewis}, {Mackey},
  {Simpson}, {Zucker}, {Ferguson}, {Martell}, {Bland-Hawthorn}, {Balbinot},
  {Tavangar}, {Drlica-Wagner}, {De Silva1}, {Simon}, \& {S5
  Collaboration}}]{2021arXiv211006950L}
{Li}, T.~S., {Ji}, A.~P., {Pace}, A.~B., {et~al.} 2021{\natexlab{b}}, arXiv
  e-prints, arXiv:2110.06950

\bibitem[{{Longeard} {et~al.}(2019){Longeard}, {Martin}, {Ibata}, {Collins},
  {Laevens}, {Bell}, \& {Mackey}}]{2019MNRAS.490.1498L}
{Longeard}, N., {Martin}, N., {Ibata}, R.~A., {et~al.} 2019, \mnras, 490, 1498

\bibitem[{{Longeard} {et~al.}(2018){Longeard}, {Martin}, {Starkenburg},
  {Ibata}, {Collins}, {Geha}, {Laevens}, {Rich}, {Aguado}, {Arentsen},
  {Carlberg}, {C{\^o}t{\'e}}, {Hill}, {Jablonka}, {Gonz{\'a}lez Hern{\'a}ndez},
  {Navarro}, {S{\'a}nchez-Janssen}, {Tolstoy}, {Venn}, \&
  {Youakim}}]{2018MNRAS.480.2609L}
{Longeard}, N., {Martin}, N., {Starkenburg}, E., {et~al.} 2018, \mnras, 480,
  2609

\bibitem[{{Longeard} {et~al.}(2021){Longeard}, {Jablonka}, {Arentsen},
  {Thomas}, {Aguado}, {Carlberg}, {Lucchesi}, {Malhan}, {Martin},
  {McConnachie}, {Navarro}, {S{\'a}nchez-Janssen}, {Sestito}, {Starkenburg}, \&
  {Yuan}}]{2021arXiv210710849L}
{Longeard}, N., {Jablonka}, P., {Arentsen}, A., {et~al.} 2021, arXiv e-prints,
  arXiv:2107.10849

\bibitem[{{Lovell} {et~al.}(2012){Lovell}, {Eke}, {Frenk}, {Gao}, {Jenkins},
  {Theuns}, {Wang}, {White}, {Boyarsky}, \& {Ruchayskiy}}]{2012MNRAS.420.2318L}
{Lovell}, M.~R., {Eke}, V., {Frenk}, C.~S., {et~al.} 2012, \mnras, 420, 2318

\bibitem[{{Luque} {et~al.}(2017){Luque}, {Pieres}, {Santiago}, {Yanny},
  {Vivas}, {Queiroz}, {Drlica-Wagner}, {Morganson}, {Balbinot}, {Marshall},
  {Li}, {Neto}, {da Costa}, {Maia}, {Bechtol}, {Kim}, {Bernstein}, {Dodelson},
  {Whiteway}, {Diehl}, {Finley}, {Abbott}, {Abdalla}, {Allam}, {Annis},
  {Benoit-L{\'e}vy}, {Bertin}, {Brooks}, {Burke}, {Rosell}, {Kind},
  {Carretero}, {Cunha}, {D'Andrea}, {Desai}, {Doel}, {Evrard}, {Flaugher},
  {Fosalba}, {Gerdes}, {Goldstein}, {Gruen}, {Gruendl}, {Gutierrez}, {James},
  {Kuehn}, {Kuropatkin}, {Lahav}, {Martini}, {Miquel}, {Nord}, {Ogando},
  {Plazas}, {Romer}, {Sanchez}, {Scarpine}, {Schubnell}, {Sevilla-Noarbe},
  {Smith}, {Soares-Santos}, {Sobreira}, {Suchyta}, {Swanson}, {Tarle},
  {Thomas}, \& {Walker}}]{2017MNRAS.468...97L}
{Luque}, E., {Pieres}, A., {Santiago}, B., {et~al.} 2017, \mnras, 468, 97

\bibitem[{{Luque} {et~al.}(2018){Luque}, {Santiago}, {Pieres}, {Marshall},
  {Pace}, {Kron}, {Drlica-Wagner}, {Queiroz}, {Balbinot}, {dal Ponte}, {Fausti
  Neto}, {da Costa}, {Maia}, {Walker}, {Abdalla}, {Allam}, {Annis}, {Bechtol},
  {Benoit-L{\'e}vy}, {Bertin}, {Brooks}, {Carnero Rosell}, {Carrasco Kind},
  {Carretero}, {Crocce}, {Davis}, {Doel}, {Eifler}, {Flaugher},
  {Garc{\'\i}a-Bellido}, {Gerdes}, {Gruen}, {Gruendl}, {Gutierrez},
  {Honscheid}, {James}, {Kuehn}, {Kuropatkin}, {Miquel}, {Nichol}, {Plazas},
  {Sanchez}, {Scarpine}, {Schindler}, {Sevilla-Noarbe}, {Smith},
  {Soares-Santos}, {Sobreira}, {Suchyta}, {Tarle}, \&
  {Thomas}}]{2018MNRAS.478.2006L}
{Luque}, E., {Santiago}, B., {Pieres}, A., {et~al.} 2018, \mnras, 478, 2006

\bibitem[{{Manwadkar} \& {Kravtsov}(2021)}]{2021arXiv211204511M}
{Manwadkar}, V., \& {Kravtsov}, A. 2021, arXiv e-prints, arXiv:2112.04511

\bibitem[{{Martin} {et~al.}(2008){Martin}, {de Jong}, \& {Rix}}]{Martin:2008}
{Martin}, N.~F., {de Jong}, J.~T.~A., \& {Rix}, H.-W. 2008, \apj, 684, 1075

\bibitem[{{Martin} {et~al.}(2015){Martin}, {Nidever}, {Besla}, {Olsen},
  {Walker}, {Vivas}, {Gruendl}, {Kaleida}, {Mu{\~n}oz}, {Blum}, {Saha}, {Conn},
  {Bell}, {Chu}, {Cioni}, {de Boer}, {Gallart}, {Jin}, {Kunder}, {Majewski},
  {Martinez-Delgado}, {Monachesi}, {Monelli}, {Monteagudo}, {No{\"e}l},
  {Olszewski}, {Stringfellow}, {van der Marel}, \&
  {Zaritsky}}]{2015ApJ...804L...5M}
{Martin}, N.~F., {Nidever}, D.~L., {Besla}, G., {et~al.} 2015, \apjl, 804, L5

\bibitem[{{Martin} {et~al.}(2016){Martin}, {Jungbluth}, {Nidever}, {Bell},
  {Besla}, {Blum}, {Cioni}, {Conn}, {Kaleida}, {Gallart}, {Jin}, {Majewski},
  {Martinez-Delgado}, {Monachesi}, {Mu{\~n}oz}, {No{\"e}l}, {Olsen},
  {Stringfellow}, {van der Marel}, {Vivas}, {Walker}, \&
  {Zaritsky}}]{2016ApJ...830L..10M}
{Martin}, N.~F., {Jungbluth}, V., {Nidever}, D.~L., {et~al.} 2016, \apjl, 830,
  L10

\bibitem[{{Mart{\'\i}nez-V{\'a}zquez}
  {et~al.}(2019){Mart{\'\i}nez-V{\'a}zquez}, {Vivas}, {Gurevich}, {Walker},
  {McCarthy}, {Pace}, {Stringer}, {Santiago}, {Hounsell}, {Macri}, {Li},
  {Bechtol}, {Riley}, {Kim}, {Simon}, {Drlica-Wagner}, {Nadler}, {Marshall},
  {Annis}, {Avila}, {Bertin}, {Brooks}, {Buckley-Geer}, {Burke}, {Carnero
  Rosell}, {Carrasco Kind}, {da Costa}, {De Vicente}, {Desai}, {Diehl}, {Doel},
  {Everett}, {Frieman}, {Garc{\'\i}a-Bellido}, {Gaztanaga}, {Gruen}, {Gruendl},
  {Gschwend}, {Gutierrez}, {Hollowood}, {Honscheid}, {James}, {Kuehn},
  {Kuropatkin}, {Maia}, {Menanteau}, {Miller}, {Miquel}, {Paz-Chinch{\'o}n},
  {Plazas}, {Sanchez}, {Scarpine}, {Serrano}, {Sevilla-Noarbe}, {Smith},
  {Soares-Santos}, {Sobreira}, {Swanson}, {Tarle}, {Vikram}, \& {DES
  Collaboration}}]{2019MNRAS.490.2183M}
{Mart{\'\i}nez-V{\'a}zquez}, C.~E., {Vivas}, A.~K., {Gurevich}, M., {et~al.}
  2019, \mnras, 490, 2183

\bibitem[{{Mart{\'\i}nez-V{\'a}zquez}
  {et~al.}(2021){Mart{\'\i}nez-V{\'a}zquez}, {Cerny}, {Vivas}, {Drlica-Wagner},
  {Pace}, {Simon}, {Munoz}, {Walker}, {Allam}, {Tucker}, {Adam{\'o}w},
  {Carlin}, {Choi}, {Ferguson}, {Ji}, {Kuropatkin}, {Li},
  {Mart{\'\i}nez-Delgado}, {Mau}, {Mutlu-Pakdil}, {Nidever}, {Riley},
  {Sakowska}, {Sand}, {Stringfellow}, \& {Stringfellow}}]{2021AJ....162..253M}
{Mart{\'\i}nez-V{\'a}zquez}, C.~E., {Cerny}, W., {Vivas}, A.~K., {et~al.} 2021,
  \aj, 162, 253

\bibitem[{{Massari} \& {Helmi}(2018)}]{2018A&A...620A.155M}
{Massari}, D., \& {Helmi}, A. 2018, \aap, 620, A155

\bibitem[{{Mateu} {et~al.}(2020){Mateu}, {Holl}, {De Ridder}, \&
  {Rimoldini}}]{2020MNRAS.496.3291M}
{Mateu}, C., {Holl}, B., {De Ridder}, J., \& {Rimoldini}, L. 2020, \mnras, 496,
  3291

\bibitem[{{Mau} {et~al.}(2019){Mau}, {Drlica-Wagner}, {Bechtol}, {Pace}, {Li},
  {Soares-Santos}, {Kuropatkin}, {Allam}, {Tucker}, {Santana-Silva}, {Yanny},
  {Jethwa}, {Palmese}, {Vivas}, {Burgad}, {Chen}, \& {BLISS
  Collaboration}}]{2019ApJ...875..154M}
{Mau}, S., {Drlica-Wagner}, A., {Bechtol}, K., {et~al.} 2019, \apj, 875, 154

\bibitem[{{Mau} {et~al.}(2020){Mau}, {Cerny}, {Pace}, {Choi}, {Drlica-Wagner},
  {Santana-Silva}, {Riley}, {Erkal}, {Stringfellow}, {Adam{\'o}w}, {Carlin},
  {Gruendl}, {Hernandez-Lang}, {Kuropatkin}, {Li}, {Mart{\'\i}nez-V{\'a}zquez},
  {Morganson}, {Mutlu-Pakdil}, {Neilsen}, {Nidever}, {Olsen}, {Sand},
  {Tollerud}, {Tucker}, {Yanny}, {Zenteno}, {Allam}, {Barkhouse}, {Bechtol},
  {Bell}, {Balaji}, {Crnojevi{\'c}}, {Esteves}, {Ferguson}, {Gallart},
  {Hughes}, {James}, {Jethwa}, {Johnson}, {Kuehn}, {Majewski}, {Mao},
  {Massana}, {McNanna}, {Monachesi}, {Nadler}, {No{\"e}l}, {Palmese},
  {Paz-Chinchon}, {Pieres}, {Sanchez}, {Shipp}, {Simon}, {Soares-Santos},
  {Tavangar}, {van der Marel}, {Vivas}, {Walker}, \&
  {Wechsler}}]{2020ApJ...890..136M}
{Mau}, S., {Cerny}, W., {Pace}, A.~B., {et~al.} 2020, \apj, 890, 136

\bibitem[{{McConnachie}(2012)}]{2012AJ....144....4M}
{McConnachie}, A.~W. 2012, \aj, 144, 4

\bibitem[{{McMillan}(2017)}]{McMillan_2017}
{McMillan}, P.~J. 2017, \mnras, 465, 76

\bibitem[{{Medina} {et~al.}(2017){Medina}, {Mu{\~n}oz}, {Vivas}, {F{\"o}rster},
  {Carlin}, {Martinez}, {Galbany}, {Gonz{\'a}lez-Gait{\'a}n}, {Hamuy}, {de
  Jaeger}, {Maureira}, \& {San Mart{\'\i}n}}]{2017ApJ...845L..10M}
{Medina}, G.~E., {Mu{\~n}oz}, R.~R., {Vivas}, A.~K., {et~al.} 2017, \apjl, 845,
  L10

\bibitem[{{Medina} {et~al.}(2018){Medina}, {Mu{\~n}oz}, {Vivas}, {Carlin},
  {F{\"o}rster}, {Mart{\'\i}nez}, {Galbany}, {Gonz{\'a}lez-Gait{\'a}n},
  {Hamuy}, {de Jaeger}, {Maureira}, \& {San Mart{\'\i}n}}]{2018ApJ...855...43M}
---. 2018, \apj, 855, 43

\bibitem[{{Miyamoto} \& {Nagai}(1975)}]{1975PASJ...27..533M}
{Miyamoto}, M., \& {Nagai}, R. 1975, \pasj, 27, 533

\bibitem[{{Morganson} {et~al.}(2018){Morganson}, {Gruendl}, {Menanteau},
  {Carrasco Kind}, {Chen}, {Daues}, {Drlica-Wagner}, {Friedel}, {Gower},
  {Johnson}, {Johnson}, {Kessler}, {Paz-Chinch{\'o}n}, {Petravick}, {Pond},
  {Yanny}, {Allam}, {Armstrong}, {Barkhouse}, {Bechtol}, {Benoit-L{\'e}vy},
  {Bernstein}, {Bertin}, {Buckley-Geer}, {Covarrubias}, {Desai}, {Diehl},
  {Goldstein}, {Gruen}, {Li}, {Lin}, {Marriner}, {Mohr}, {Neilsen}, {Ngeow},
  {Paech}, {Rykoff}, {Sako}, {Sevilla-Noarbe}, {Sheldon}, {Sobreira}, {Tucker},
  {Wester}, \& {DES Collaboration}}]{2018PASP..130g4501M}
{Morganson}, E., {Gruendl}, R.~A., {Menanteau}, F., {et~al.} 2018, \pasp, 130,
  074501

\bibitem[{{Moskowitz} \& {Walker}(2020)}]{2020ApJ...892...27M}
{Moskowitz}, A.~G., \& {Walker}, M.~G. 2020, \apj, 892, 27

\bibitem[{{Mu{\~n}oz} {et~al.}(2006){Mu{\~n}oz}, {Carlin}, {Frinchaboy},
  {Nidever}, {Majewski}, \& {Patterson}}]{2006ApJ...650L..51M}
{Mu{\~n}oz}, R.~R., {Carlin}, J.~L., {Frinchaboy}, P.~M., {et~al.} 2006, \apjl,
  650, L51

\bibitem[{{Mu{\~n}oz} {et~al.}(2018){Mu{\~n}oz}, {C{\^o}t{\'e}}, {Santana},
  {Geha}, {Simon}, {Oyarz{\'u}n}, {Stetson}, \&
  {Djorgovski}}]{2018ApJ...860...66M}
{Mu{\~n}oz}, R.~R., {C{\^o}t{\'e}}, P., {Santana}, F.~A., {et~al.} 2018, \apj,
  860, 66

\bibitem[{{Muraveva} {et~al.}(2018){Muraveva}, {Delgado}, {Clementini},
  {Sarro}, \& {Garofalo}}]{2018MNRAS.481.1195M}
{Muraveva}, T., {Delgado}, H.~E., {Clementini}, G., {Sarro}, L.~M., \&
  {Garofalo}, A. 2018, \mnras, 481, 1195

\bibitem[{{Mutlu-Pakdil} {et~al.}(2018){Mutlu-Pakdil}, {Sand}, {Carlin},
  {Spekkens}, {Caldwell}, {Crnojevi{\'c}}, {Hughes}, {Willman}, \&
  {Zaritsky}}]{2018ApJ...863...25M}
{Mutlu-Pakdil}, B., {Sand}, D.~J., {Carlin}, J.~L., {et~al.} 2018, \apj, 863,
  25

\bibitem[{{Mutlu-Pakdil} {et~al.}(2021){Mutlu-Pakdil}, {Sand}, {Crnojevi{\'c}},
  {Drlica-Wagner}, {Caldwell}, {Guhathakurta}, {Seth}, {Simon}, {Strader}, \&
  {Toloba}}]{2021ApJ...918...88M}
{Mutlu-Pakdil}, B., {Sand}, D.~J., {Crnojevi{\'c}}, D., {et~al.} 2021, \apj,
  918, 88

\bibitem[{{Nadler} {et~al.}(2020){Nadler}, {Wechsler}, {Bechtol}, {Mao},
  {Green}, {Drlica-Wagner}, {McNanna}, {Mau}, {Pace}, {Simon}, {Kravtsov},
  {Dodelson}, {Li}, {Riley}, {Wang}, {Abbott}, {Aguena}, {Allam}, {Annis},
  {Avila}, {Bernstein}, {Bertin}, {Brooks}, {Burke}, {Rosell}, {Kind},
  {Carretero}, {Costanzi}, {da Costa}, {De Vicente}, {Desai}, {Evrard},
  {Flaugher}, {Fosalba}, {Frieman}, {Garc{\'\i}a-Bellido}, {Gaztanaga},
  {Gerdes}, {Gruen}, {Gschwend}, {Gutierrez}, {Hartley}, {Hinton}, {Honscheid},
  {Krause}, {Kuehn}, {Kuropatkin}, {Lahav}, {Maia}, {Marshall}, {Menanteau},
  {Miquel}, {Palmese}, {Paz-Chinch{\'o}n}, {Plazas}, {Romer}, {Sanchez},
  {Santiago}, {Scarpine}, {Serrano}, {Smith}, {Soares-Santos}, {Suchyta},
  {Tarle}, {Thomas}, {Varga}, {Walker}, \& {DES
  Collaboration}}]{2020ApJ...893...48N}
{Nadler}, E.~O., {Wechsler}, R.~H., {Bechtol}, K., {et~al.} 2020, \apj, 893, 48

\bibitem[{{Nadler} {et~al.}(2021){Nadler}, {Drlica-Wagner}, {Bechtol}, {Mau},
  {Wechsler}, {Gluscevic}, {Boddy}, {Pace}, {Li}, {McNanna}, {Riley},
  {Garc{\'\i}a-Bellido}, {Mao}, {Green}, {Burke}, {Peter}, {Jain}, {Abbott},
  {Aguena}, {Allam}, {Annis}, {Avila}, {Brooks}, {Carrasco Kind}, {Carretero},
  {Costanzi}, {da Costa}, {De Vicente}, {Desai}, {Diehl}, {Doel}, {Everett},
  {Evrard}, {Flaugher}, {Frieman}, {Gerdes}, {Gruen}, {Gruendl}, {Gschwend},
  {Gutierrez}, {Hinton}, {Honscheid}, {Huterer}, {James}, {Krause}, {Kuehn},
  {Kuropatkin}, {Lahav}, {Maia}, {Marshall}, {Menanteau}, {Miquel}, {Palmese},
  {Paz-Chinch{\'o}n}, {Plazas}, {Romer}, {Sanchez}, {Scarpine}, {Serrano},
  {Sevilla-Noarbe}, {Smith}, {Soares-Santos}, {Suchyta}, {Swanson}, {Tarle},
  {Tucker}, {Walker}, {Wester}, \& {DES Collaboration}}]{2021PhRvL.126i1101N}
{Nadler}, E.~O., {Drlica-Wagner}, A., {Bechtol}, K., {et~al.} 2021, \prl, 126,
  091101

\bibitem[{{Navarro} {et~al.}(1996){Navarro}, {Frenk}, \&
  {White}}]{Navarro1996ApJ...462..563N}
{Navarro}, J.~F., {Frenk}, C.~S., \& {White}, S.~D.~M. 1996, \apj, 462, 563

\bibitem[{Neilsen {et~al.}(2015)Neilsen, Bernstein, Gruendl, \&
  Kent}]{Neilsen:2015}
Neilsen, E., Bernstein, G., Gruendl, R., \& Kent, S. 2015, ``Limiting
  magnitude, $\tau$, $T_{eff}$, and image quality in DES Year 1'', Tech. Rep.
  FERMILAB-TM-2610-AE-CD, Fermi National Accelerator Laboratory

\bibitem[{{Newman} {et~al.}(2013){Newman}, {Cooper}, {Davis}, {Faber}, {Coil},
  {Guhathakurta}, {Koo}, {Phillips}, {Conroy}, {Dutton}, {Finkbeiner}, {Gerke},
  {Rosario}, {Weiner}, {Willmer}, {Yan}, {Harker}, {Kassin}, {Konidaris},
  {Lai}, {Madgwick}, {Noeske}, {Wirth}, {Connolly}, {Kaiser}, {Kirby},
  {Lemaux}, {Lin}, {Lotz}, {Luppino}, {Marinoni}, {Matthews}, {Metevier}, \&
  {Schiavon}}]{2013ApJS..208....5N}
{Newman}, J.~A., {Cooper}, M.~C., {Davis}, M., {et~al.} 2013, \apjs, 208, 5

\bibitem[{{Newton} {et~al.}(2018){Newton}, {Cautun}, {Jenkins}, {Frenk}, \&
  {Helly}}]{2018MNRAS.479.2853N}
{Newton}, O., {Cautun}, M., {Jenkins}, A., {Frenk}, C.~S., \& {Helly}, J.~C.
  2018, \mnras, 479, 2853

\bibitem[{{Oemler} {et~al.}(2017){Oemler}, {Clardy}, {Kelson}, {Walth}, \&
  {Villanueva}}]{2017ascl.soft05001O}
{Oemler}, A., {Clardy}, K., {Kelson}, D., {Walth}, G., \& {Villanueva}, E.
  2017, {COSMOS: Carnegie Observatories System for MultiObject Spectroscopy}, ,
  , ascl:1705.001

\bibitem[{{Pace} \& {Li}(2019)}]{Pace2019ApJ...875...77P}
{Pace}, A.~B., \& {Li}, T.~S. 2019, \apj, 875, 77

\bibitem[{{Pace} \& {Strigari}(2019)}]{Pace2019MNRAS.482.3480P}
{Pace}, A.~B., \& {Strigari}, L.~E. 2019, \mnras, 482, 3480

\bibitem[{{Pace} {et~al.}(2020){Pace}, {Kaplinghat}, {Kirby}, {Simon},
  {Tollerud}, {Mu{\~n}oz}, {C{\^o}t{\'e}}, {Djorgovski}, \&
  {Geha}}]{2020MNRAS.495.3022P}
{Pace}, A.~B., {Kaplinghat}, M., {Kirby}, E., {et~al.} 2020, \mnras, 495, 3022

\bibitem[{{Patel} {et~al.}(2020){Patel}, {Kallivayalil}, {Garavito-Camargo},
  {Besla}, {Weisz}, {van der Marel}, {Boylan-Kolchin}, {Pawlowski}, \&
  {G{\'o}mez}}]{Patel_etal_2020}
{Patel}, E., {Kallivayalil}, N., {Garavito-Camargo}, N., {et~al.} 2020, \apj,
  893, 121

\bibitem[{{Pawlowski} {et~al.}(2015){Pawlowski}, {McGaugh}, \&
  {Jerjen}}]{2015MNRAS.453.1047P}
{Pawlowski}, M.~S., {McGaugh}, S.~S., \& {Jerjen}, H. 2015, \mnras, 453, 1047

\bibitem[{{Pawlowski} {et~al.}(2012){Pawlowski}, {Pflamm-Altenburg}, \&
  {Kroupa}}]{2012MNRAS.423.1109P}
{Pawlowski}, M.~S., {Pflamm-Altenburg}, J., \& {Kroupa}, P. 2012, \mnras, 423,
  1109

\bibitem[{{Pietrzy{\'n}ski} {et~al.}(2019){Pietrzy{\'n}ski}, {Graczyk},
  {Gallenne}, {Gieren}, {Thompson}, {Pilecki}, {Karczmarek}, {G{\'o}rski},
  {Suchomska}, {Taormina}, {Zgirski}, {Wielg{\'o}rski}, {Ko{\l}aczkowski},
  {Konorski}, {Villanova}, {Nardetto}, {Kervella}, {Bresolin}, {Kudritzki},
  {Storm}, {Smolec}, \& {Narloch}}]{Pietrzyski_etal_2019}
{Pietrzy{\'n}ski}, G., {Graczyk}, D., {Gallenne}, A., {et~al.} 2019, \nat, 567,
  200

\bibitem[{{Plummer}(1911)}]{1911MNRAS..71..460P}
{Plummer}, H.~C. 1911, \mnras, 71, 460

\bibitem[{{Price-Whelan}(2017)}]{2017JOSS....2..388P}
{Price-Whelan}, A.~M. 2017, The Journal of Open Source Software, 2, 388

\bibitem[{{Robin} {et~al.}(2003){Robin}, {Reyl{\'e}}, {Derri{\`e}re}, \&
  {Picaud}}]{2003A&A...409..523R}
{Robin}, A.~C., {Reyl{\'e}}, C., {Derri{\`e}re}, S., \& {Picaud}, S. 2003,
  \aap, 409, 523

\bibitem[{{Schlafly} \& {Finkbeiner}(2011)}]{Schlafly:2011}
{Schlafly}, E.~F., \& {Finkbeiner}, D.~P. 2011, \apj, 737, 103

\bibitem[{{Schlegel} {et~al.}(1998){Schlegel}, {Finkbeiner}, \&
  {Davis}}]{Schlegel:1998}
{Schlegel}, D.~J., {Finkbeiner}, D.~P., \& {Davis}, M. 1998, \apj, 500, 525

\bibitem[{{Sch{\"o}rck} {et~al.}(2009){Sch{\"o}rck}, {Christlieb}, {Cohen},
  {Beers}, {Shectman}, {Thompson}, {McWilliam}, {Bessell}, {Norris},
  {Mel{\'e}ndez}, {Ram{\'\i}rez}, {Haynes}, {Cass}, {Hartley}, {Russell},
  {Watson}, {Zickgraf}, {Behnke}, {Fechner}, {Fuhrmeister}, {Barklem},
  {Edvardsson}, {Frebel}, {Wisotzki}, \& {Reimers}}]{2009A&A...507..817S}
{Sch{\"o}rck}, T., {Christlieb}, N., {Cohen}, J.~G., {et~al.} 2009, \aap, 507,
  817

\bibitem[{{Sesar} {et~al.}(2017){Sesar}, {Hernitschek}, {Mitrovi{\'c}},
  {Ivezi{\'c}}, {Rix}, {Cohen}, {Bernard}, {Grebel}, {Martin}, {Schlafly},
  {Burgett}, {Draper}, {Flewelling}, {Kaiser}, {Kudritzki}, {Magnier},
  {Metcalfe}, {Tonry}, \& {Waters}}]{2017AJ....153..204S}
{Sesar}, B., {Hernitschek}, N., {Mitrovi{\'c}}, S., {et~al.} 2017, \aj, 153,
  204

\bibitem[{Simon(2018)}]{Simon_2018}
Simon, J.~D. 2018, The Astrophysical Journal, 863, 89.
\newblock \url{https://doi.org/10.3847/1538-4357/aacdfb}

\bibitem[{{Simon}(2019)}]{2019ARA&A..57..375S}
{Simon}, J.~D. 2019, \araa, 57, 375

\bibitem[{{Simon} \& {Geha}(2007)}]{2007ApJ...670..313S}
{Simon}, J.~D., \& {Geha}, M. 2007, \apj, 670, 313

\bibitem[{{Simon} {et~al.}(2015){Simon}, {Drlica-Wagner}, {Li}, {Nord}, {Geha},
  {Bechtol}, {Balbinot}, {Buckley-Geer}, {Lin}, {Marshall}, {Santiago},
  {Strigari}, {Wang}, {Wechsler}, {Yanny}, {Abbott}, {Bauer}, {Bernstein},
  {Bertin}, {Brooks}, {Burke}, {Capozzi}, {Carnero Rosell}, {Carrasco Kind},
  {D'Andrea}, {da Costa}, {DePoy}, {Desai}, {Diehl}, {Dodelson}, {Cunha},
  {Estrada}, {Evrard}, {Fausti Neto}, {Fernandez}, {Finley}, {Flaugher},
  {Frieman}, {Gaztanaga}, {Gerdes}, {Gruen}, {Gruendl}, {Honscheid}, {James},
  {Kent}, {Kuehn}, {Kuropatkin}, {Lahav}, {Maia}, {March}, {Martini}, {Miller},
  {Miquel}, {Ogando}, {Romer}, {Roodman}, {Rykoff}, {Sako}, {Sanchez},
  {Schubnell}, {Sevilla}, {Smith}, {Soares-Santos}, {Sobreira}, {Suchyta},
  {Swanson}, {Tarle}, {Thaler}, {Tucker}, {Vikram}, {Walker}, {Wester}, \& {DES
  Collaboration}}]{2015ApJ...808...95S}
{Simon}, J.~D., {Drlica-Wagner}, A., {Li}, T.~S., {et~al.} 2015, \apj, 808, 95

\bibitem[{{Simon} {et~al.}(2017){Simon}, {Li}, {Drlica-Wagner}, {Bechtol},
  {Marshall}, {James}, {Wang}, {Strigari}, {Balbinot}, {Kuehn}, {Walker},
  {Abbott}, {Allam}, {Annis}, {Benoit-L{\'e}vy}, {Brooks}, {Buckley-Geer},
  {Burke}, {Carnero Rosell}, {Carrasco Kind}, {Carretero}, {Cunha}, {D'Andrea},
  {da Costa}, {DePoy}, {Desai}, {Doel}, {Fernandez}, {Flaugher}, {Frieman},
  {Garc{\'\i}a-Bellido}, {Gaztanaga}, {Goldstein}, {Gruen}, {Gutierrez},
  {Kuropatkin}, {Maia}, {Martini}, {Menanteau}, {Miller}, {Miquel}, {Neilsen},
  {Nord}, {Ogando}, {Plazas}, {Romer}, {Rykoff}, {Sanchez}, {Santiago},
  {Scarpine}, {Schubnell}, {Sevilla-Noarbe}, {Smith}, {Sobreira}, {Suchyta},
  {Swanson}, {Tarle}, {Whiteway}, {Yanny}, \& {DES Collaboration}}]{JDS17}
{Simon}, J.~D., {Li}, T.~S., {Drlica-Wagner}, A., {et~al.} 2017, \apj, 838, 11

\bibitem[{{Simon} {et~al.}(2020){Simon}, {Li}, {Erkal}, {Pace},
  {Drlica-Wagner}, {James}, {Marshall}, {Bechtol}, {Hansen}, {Kuehn}, {Lidman},
  {Allam}, {Annis}, {Avila}, {Bertin}, {Brooks}, {Burke}, {Rosell}, {Carrasco
  Kind}, {Carretero}, {da Costa}, {De Vicente}, {Desai}, {Doel}, {Eifler},
  {Everett}, {Fosalba}, {Frieman}, {Garc{\'\i}a-Bellido}, {Gaztanaga},
  {Gerdes}, {Gruen}, {Gruendl}, {Gschwend}, {Gutierrez}, {Hollowood},
  {Honscheid}, {Krause}, {Kuropatkin}, {MacCrann}, {Maia}, {March}, {Miquel},
  {Palmese}, {Paz-Chinch{\'o}n}, {Plazas}, {Reil}, {Roodman}, {Sanchez},
  {Santiago}, {Scarpine}, {Schubnell}, {Serrano}, {Smith}, {Suchyta}, {Tarle},
  {Walker}, \& {DES Collaboration}}]{2020ApJ...892..137S}
{Simon}, J.~D., {Li}, T.~S., {Erkal}, D., {et~al.} 2020, \apj, 892, 137

\bibitem[{{Sohn} {et~al.}(2007){Sohn}, {Majewski}, {Mu{\~n}oz}, {Kunkel},
  {Johnston}, {Ostheimer}, {Guhathakurta}, {Patterson}, {Siegel}, \&
  {Cooper}}]{2007ApJ...663..960S}
{Sohn}, S.~T., {Majewski}, S.~R., {Mu{\~n}oz}, R.~R., {et~al.} 2007, \apj, 663,
  960

\bibitem[{{Strigari}(2018)}]{2018RPPh...81e6901S}
{Strigari}, L.~E. 2018, Reports on Progress in Physics, 81, 056901

\bibitem[{{Stringer} {et~al.}(2021){Stringer}, {Drlica-Wagner}, {Macri},
  {Mart{\'\i}nez-V{\'a}zquez}, {Vivas}, {Ferguson}, {Pace}, {Walker},
  {Neilsen}, {Tavangar}, {Wester}, {Abbott}, {Aguena}, {Allam}, {Bacon},
  {Bechtol}, {Bertin}, {Brooks}, {Burke}, {Carnero Rosell}, {Carrasco Kind},
  {Carretero}, {Costanzi}, {Crocce}, {da Costa}, {Pereira}, {De Vicente},
  {Desai}, {Diehl}, {Doel}, {Ferrero}, {Garc{\'\i}a-Bellido}, {Gaztanaga},
  {Gerdes}, {Gruen}, {Gruendl}, {Gschwend}, {Gutierrez}, {Hinton}, {Hollowood},
  {Honscheid}, {Hoyle}, {James}, {Kuehn}, {Kuropatkin}, {Li}, {Maia},
  {Marshall}, {Menanteau}, {Miquel}, {Morgan}, {Ogando}, {Palmese},
  {Paz-Chinch{\'o}n}, {Plazas}, {Roodman}, {Sanchez}, {Schubnell}, {Serrano},
  {Sevilla-Noarbe}, {Smith}, {Soares-Santos}, {Suchyta}, {Tarle}, {Thomas},
  {To}, {Varga}, {Wilkinson}, {Zhang}, \& {DES
  Collaboration}}]{2021ApJ...911..109S}
{Stringer}, K.~M., {Drlica-Wagner}, A., {Macri}, L., {et~al.} 2021, \apj, 911,
  109

\bibitem[{{Tonry} {et~al.}(2018){Tonry}, {Denneau}, {Flewelling}, {Heinze},
  {Onken}, {Smartt}, {Stalder}, {Weiland}, \& {Wolf}}]{Tonry:2018}
{Tonry}, J.~L., {Denneau}, L., {Flewelling}, H., {et~al.} 2018, \apj, 867, 105

\bibitem[{{Torrealba} {et~al.}(2019{\natexlab{a}}){Torrealba}, {Belokurov}, \&
  {Koposov}}]{2019MNRAS.484.2181T}
{Torrealba}, G., {Belokurov}, V., \& {Koposov}, S.~E. 2019{\natexlab{a}},
  \mnras, 484, 2181

\bibitem[{{Torrealba} {et~al.}(2016){Torrealba}, {Koposov}, {Belokurov},
  {Irwin}, {Collins}, {Spencer}, {Ibata}, {Mateo}, {Bonaca}, \&
  {Jethwa}}]{2016MNRAS.463..712T}
{Torrealba}, G., {Koposov}, S.~E., {Belokurov}, V., {et~al.} 2016, \mnras, 463,
  712

\bibitem[{{Torrealba} {et~al.}(2018){Torrealba}, {Belokurov}, {Koposov},
  {Bechtol}, {Drlica-Wagner}, {Olsen}, {Vivas}, {Yanny}, {Jethwa}, {Walker},
  {Li}, {Allam}, {Conn}, {Gallart}, {Gruendl}, {James}, {Johnson}, {Kuehn},
  {Kuropatkin}, {Martin}, {Martinez-Delgado}, {Nidever}, {No{\"e}l}, {Simon},
  {Stringfellow}, \& {Tucker}}]{2018MNRAS.475.5085T}
{Torrealba}, G., {Belokurov}, V., {Koposov}, S.~E., {et~al.} 2018, \mnras, 475,
  5085

\bibitem[{{Torrealba} {et~al.}(2019{\natexlab{b}}){Torrealba}, {Belokurov},
  {Koposov}, {Li}, {Walker}, {Sanders}, {Geringer-Sameth}, {Zucker}, {Kuehn},
  {Evans}, \& {Dehnen}}]{Torrealba:2019}
---. 2019{\natexlab{b}}, \mnras, 488, 2743

\bibitem[{{Trujillo} {et~al.}(2021){Trujillo}, {D'Onofrio}, {Zaritsky},
  {Madrigal-Aguado}, {Chamba}, {Golini}, {Akhlaghi}, {Sharbaf},
  {Infante-Sainz}, {Rom{\'a}n}, {Morales-Socorro}, {Sand}, \&
  {Martin}}]{2021A&A...654A..40T}
{Trujillo}, I., {D'Onofrio}, M., {Zaritsky}, D., {et~al.} 2021, \aap, 654, A40

\bibitem[{{van der Marel} {et~al.}(2002){van der Marel}, {Alves}, {Hardy}, \&
  {Suntzeff}}]{van_der_Marel_etal_2002}
{van der Marel}, R.~P., {Alves}, D.~R., {Hardy}, E., \& {Suntzeff}, N.~B. 2002,
  \aj, 124, 2639

\bibitem[{{Vasiliev} {et~al.}(2021){Vasiliev}, {Belokurov}, \&
  {Erkal}}]{2021MNRAS.501.2279V}
{Vasiliev}, E., {Belokurov}, V., \& {Erkal}, D. 2021, \mnras, 501, 2279

\bibitem[{{Vivas} {et~al.}(2020){Vivas}, {Mart{\'\i}nez-V{\'a}zquez}, \&
  {Walker}}]{2020ApJS..247...35V}
{Vivas}, A.~K., {Mart{\'\i}nez-V{\'a}zquez}, C., \& {Walker}, A.~R. 2020,
  \apjs, 247, 35

\bibitem[{{Walker} {et~al.}(2006){Walker}, {Mateo}, {Olszewski}, {Bernstein},
  {Wang}, \& {Woodroofe}}]{2006AJ....131.2114W}
{Walker}, M.~G., {Mateo}, M., {Olszewski}, E.~W., {et~al.} 2006, \aj, 131, 2114

\bibitem[{{Walsh} {et~al.}(2007){Walsh}, {Jerjen}, \&
  {Willman}}]{2007ApJ...662L..83W}
{Walsh}, S.~M., {Jerjen}, H., \& {Willman}, B. 2007, \apjl, 662, L83

\bibitem[{{Wang} {et~al.}(2019){Wang}, {de Boer}, {Pieres}, {Li},
  {Drlica-Wagner}, {Koposov}, {Vivas}, {Pace}, {Santiago}, {Walker}, {Tucker},
  {Strigari}, {Marshall}, {Yanny}, {DePoy}, {Bechtol}, {Roodman}, {Abbott},
  {Abdalla}, {Allam}, {Annis}, {Avila}, {Bertin}, {Brooks}, {Burke}, {Carnero
  Rosell}, {Carrasco Kind}, {Cunha}, {D'Andrea}, {da Costa}, {De Vicente},
  {Desai}, {Eifler}, {Estrada}, {Flaugher}, {Frieman}, {Garc{\'\i}a-Bellido},
  {Gerdes}, {Gruen}, {Gruendl}, {Gutierrez}, {Hollowood}, {Honscheid}, {James},
  {Kuehn}, {Kuropatkin}, {Lahav}, {Maia}, {Miquel}, {Sanchez}, {Scarpine},
  {Sevilla-Noarbe}, {Smith}, {Smith}, {Sobreira}, {Suchyta}, {Swanson},
  {Tarle}, \& {DES Collaboration}}]{2019ApJ...881..118W}
{Wang}, M.~Y., {de Boer}, T., {Pieres}, A., {et~al.} 2019, \apj, 881, 118

\bibitem[{Wang \& Chen(2019)}]{Wang_2019}
Wang, S., \& Chen, X. 2019, The Astrophysical Journal, 877, 116.
\newblock \url{https://doi.org/10.3847/1538-4357/ab1c61}

\bibitem[{{Weisz} {et~al.}(2016){Weisz}, {Koposov}, {Dolphin}, {Belokurov},
  {Gieles}, {Mateo}, {Olszewski}, {Sills}, \& {Walker}}]{2016ApJ...822...32W}
{Weisz}, D.~R., {Koposov}, S.~E., {Dolphin}, A.~E., {et~al.} 2016, \apj, 822,
  32

\bibitem[{{Wenger} {et~al.}(2000){Wenger}, {Ochsenbein}, {Egret}, {Dubois},
  {Bonnarel}, {Borde}, {Genova}, {Jasniewicz}, {Lalo{\"e}}, {Lesteven}, \&
  {Monier}}]{2000A&AS..143....9W}
{Wenger}, M., {Ochsenbein}, F., {Egret}, D., {et~al.} 2000, \aaps, 143, 9

\bibitem[{{Willman} {et~al.}(2011){Willman}, {Geha}, {Strader}, {Strigari},
  {Simon}, {Kirby}, {Ho}, \& {Warres}}]{2011AJ....142..128W}
{Willman}, B., {Geha}, M., {Strader}, J., {et~al.} 2011, \aj, 142, 128

\bibitem[{Willman \& Strader(2012)}]{Willman:2012}
Willman, B., \& Strader, J. 2012, The Astronomical Journal, 144, 76.
\newblock \url{https://doi.org/10.1088%2F0004-6256%2F144%2F3%2F76}

\bibitem[{{Willman} {et~al.}(2005){Willman}, {Blanton}, {West}, {Dalcanton},
  {Hogg}, {Schneider}, {Wherry}, {Yanny}, \& {Brinkmann}}]{2005AJ....129.2692W}
{Willman}, B., {Blanton}, M.~R., {West}, A.~A., {et~al.} 2005, \aj, 129, 2692

\bibitem[{{Wolf} {et~al.}(2010){Wolf}, {Martinez}, {Bullock}, {Kaplinghat},
  {Geha}, {Mu{\~n}oz}, {Simon}, \& {Avedo}}]{2010MNRAS.406.1220W}
{Wolf}, J., {Martinez}, G.~D., {Bullock}, J.~S., {et~al.} 2010, \mnras, 406,
  1220

\bibitem[{{York} {et~al.}(2000){York}, {Adelman}, {Anderson}, {Anderson},
  {Annis}, {Bahcall}, {Bakken}, {Barkhouser}, {Bastian}, {Berman}, {Boroski},
  {Bracker}, {Briegel}, {Briggs}, {Brinkmann}, {Brunner}, {Burles}, {Carey},
  {Carr}, {Castander}, {Chen}, {Colestock}, {Connolly}, {Crocker}, {Csabai},
  {Czarapata}, {Davis}, {Doi}, {Dombeck}, {Eisenstein}, {Ellman}, {Elms},
  {Evans}, {Fan}, {Federwitz}, {Fiscelli}, {Friedman}, {Frieman}, {Fukugita},
  {Gillespie}, {Gunn}, {Gurbani}, {de Haas}, {Haldeman}, {Harris}, {Hayes},
  {Heckman}, {Hennessy}, {Hindsley}, {Holm}, {Holmgren}, {Huang}, {Hull},
  {Husby}, {Ichikawa}, {Ichikawa}, {Ivezi{\'c}}, {Kent}, {Kim}, {Kinney},
  {Klaene}, {Kleinman}, {Kleinman}, {Knapp}, {Korienek}, {Kron}, {Kunszt},
  {Lamb}, {Lee}, {Leger}, {Limmongkol}, {Lindenmeyer}, {Long}, {Loomis},
  {Loveday}, {Lucinio}, {Lupton}, {MacKinnon}, {Mannery}, {Mantsch}, {Margon},
  {McGehee}, {McKay}, {Meiksin}, {Merelli}, {Monet}, {Munn}, {Narayanan},
  {Nash}, {Neilsen}, {Neswold}, {Newberg}, {Nichol}, {Nicinski}, {Nonino},
  {Okada}, {Okamura}, {Ostriker}, {Owen}, {Pauls}, {Peoples}, {Peterson},
  {Petravick}, {Pier}, {Pope}, {Pordes}, {Prosapio}, {Rechenmacher}, {Quinn},
  {Richards}, {Richmond}, {Rivetta}, {Rockosi}, {Ruthmansdorfer}, {Sandford},
  {Schlegel}, {Schneider}, {Sekiguchi}, {Sergey}, {Shimasaku}, {Siegmund},
  {Smee}, {Smith}, {Snedden}, {Stone}, {Stoughton}, {Strauss}, {Stubbs},
  {SubbaRao}, {Szalay}, {Szapudi}, {Szokoly}, {Thakar}, {Tremonti}, {Tucker},
  {Uomoto}, {Vanden Berk}, {Vogeley}, {Waddell}, {Wang}, {Watanabe},
  {Weinberg}, {Yanny}, {Yasuda}, \& {SDSS Collaboration}}]{2000AJ....120.1579Y}
{York}, D.~G., {Adelman}, J., {Anderson}, John~E., J., {et~al.} 2000, \aj, 120,
  1579

\bibitem[{{Youakim} {et~al.}(2020){Youakim}, {Starkenburg}, {Martin},
  {Matijevi{\v{c}}}, {Aguado}, {Allende Prieto}, {Arentsen}, {Bonifacio},
  {Carlberg}, {Gonz{\'a}lez Hern{\'a}ndez}, {Hill}, {Kordopatis}, {Lardo},
  {Navarro}, {Jablonka}, {S{\'a}nchez Janssen}, {Sestito}, {Thomas}, \&
  {Venn}}]{2020MNRAS.492.4986Y}
{Youakim}, K., {Starkenburg}, E., {Martin}, N.~F., {et~al.} 2020, \mnras, 492,
  4986

\bibitem[{{Zoutendijk} {et~al.}(2021{\natexlab{a}}){Zoutendijk}, {Brinchmann},
  {Bouch{\'e}}, {den Brok}, {Krajnovi{\'c}}, {Kuijken}, {Maseda}, \&
  {Schaye}}]{2021A&A...651A..80Z}
{Zoutendijk}, S.~L., {Brinchmann}, J., {Bouch{\'e}}, N.~F., {et~al.}
  2021{\natexlab{a}}, \aap, 651, A80

\bibitem[{{Zoutendijk} {et~al.}(2021{\natexlab{b}}){Zoutendijk}, {J{\'u}lio},
  {Brinchmann}, {Read}, {Vaz}, {Boogaard}, {Bouch{\'e}}, {Krajnovi{\'c}},
  {Kuijken}, {Schaye}, \& {Steinmetz}}]{2021arXiv211209374Z}
{Zoutendijk}, S.~L., {J{\'u}lio}, M.~P., {Brinchmann}, J., {et~al.}
  2021{\natexlab{b}}, arXiv e-prints, arXiv:2112.09374

\bibitem[{{Zucker} {et~al.}(2006){Zucker}, {Belokurov}, {Evans}, {Kleyna},
  {Irwin}, {Wilkinson}, {Fellhauer}, {Bramich}, {Gilmore}, {Newberg}, {Yanny},
  {Smith}, {Hewett}, {Bell}, {Rix}, {Gnedin}, {Vidrih}, {Wyse}, {Willman},
  {Grebel}, {Schneider}, {Beers}, {Kniazev}, {Barentine}, {Brewington},
  {Brinkmann}, {Harvanek}, {Kleinman}, {Krzesinski}, {Long}, {Nitta}, \&
  {Snedden}}]{2006ApJ...650L..41Z}
{Zucker}, D.~B., {Belokurov}, V., {Evans}, N.~W., {et~al.} 2006, \apjl, 650,
  L41

\end{thebibliography}
\appendix

\section{C\lowercase{a}T Fits for Member Stars with Measured Metallicities}
\label{sec:spectra}
On the following page, we show our fits to the Calcium Triplet lines of the 5 stars for which we reported metallicities (top 5 rows of \tabref{specmem}). In each panel, we specifically plot the normalized spectrum of each star in blue, the best-fit model in black, and include the residuals for these fits in orange. For some stars, rectangular features in the spectra (associated with chip gaps) and/or residual emission-like or absorption-like features (associated with imperfect sky line subtraction) are visible. We note that wavelength ranges with chip gaps were masked during the fitting process, and therefore exerted no influence on the resulting fits.

\begin{figure}
    \centering
    \includegraphics[width = .925\textwidth]{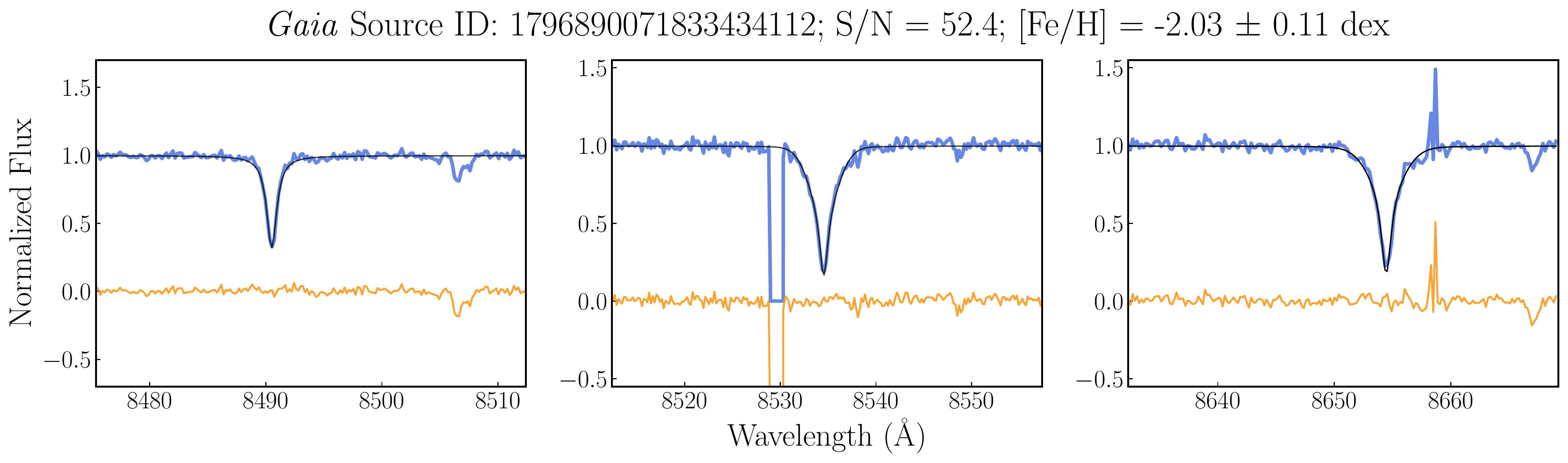}
    \includegraphics[width = .925\textwidth]{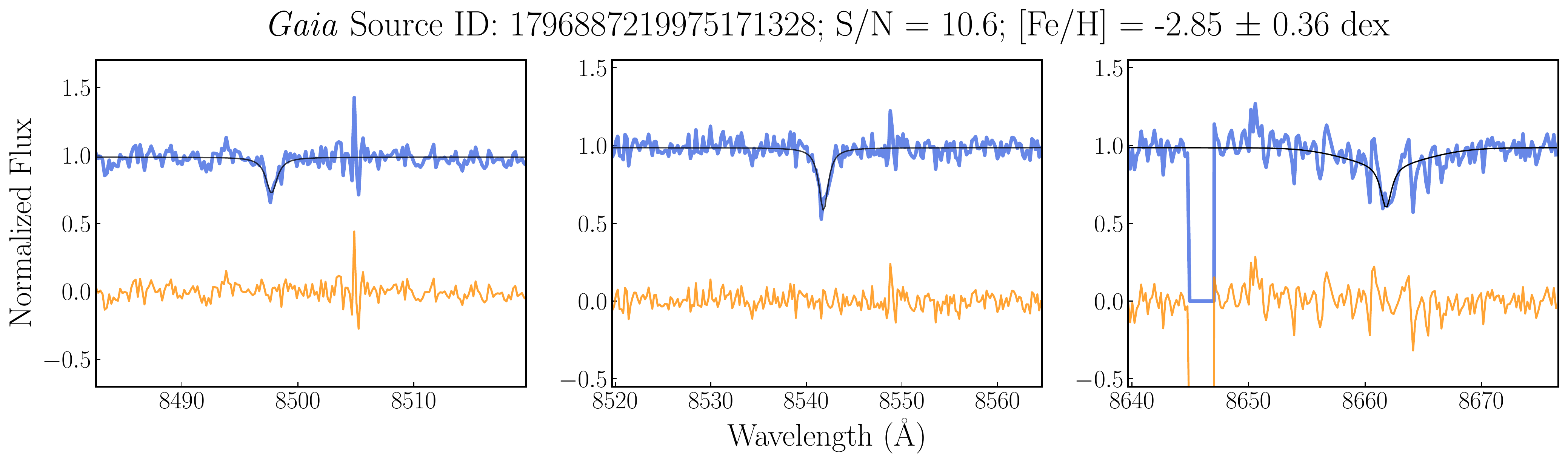}
    \includegraphics[width = .925\textwidth]{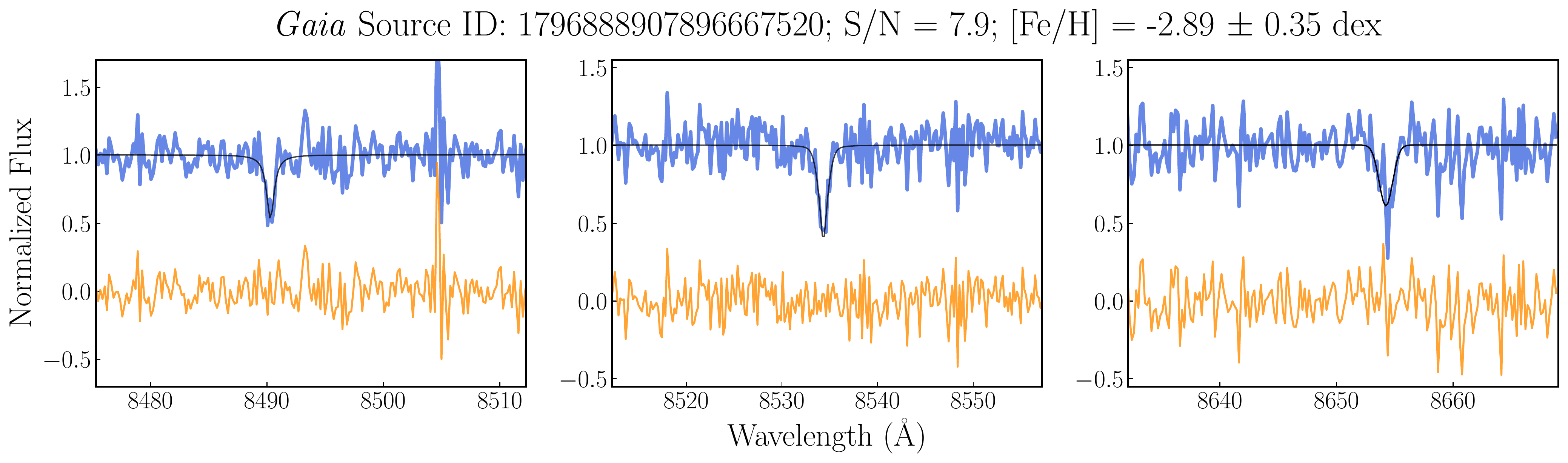}
    \includegraphics[width = .925\textwidth]{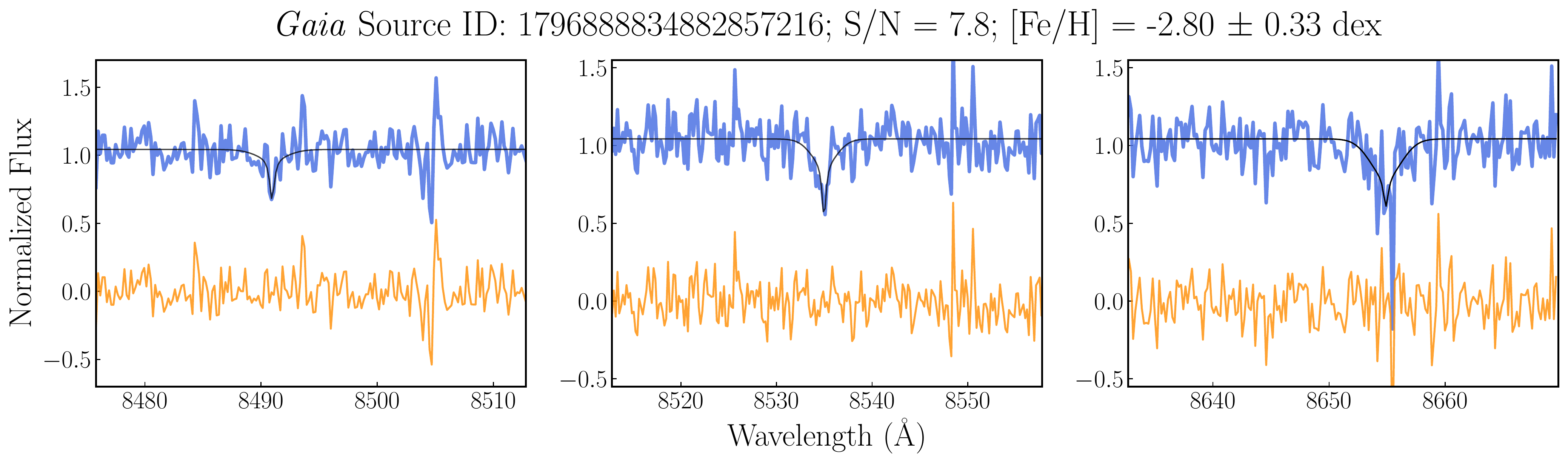}
    \includegraphics[width = .925\textwidth]{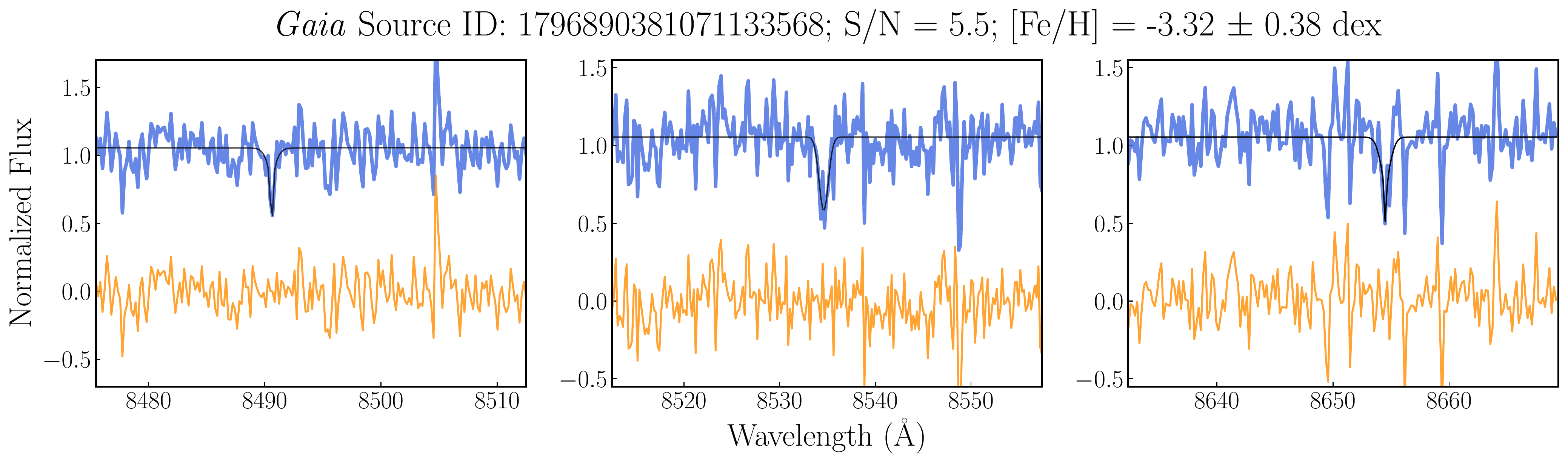}
    \caption{Spectra for the five stars with $S/N > 5$ for which we measured metallicities.}
\end{figure}

\section{References for Literature Data Presented in Figure 5}
\label{sec:refs}
The left panel of \figref{pop_comparison} shows the populations of ``classical" Milky Way globular clusters, recently discovered halo star clusters, and dwarf galaxies in the  $M_V$-$r_{1/2}$ plane. The globular cluster measurements are taken from \citet[2010 edition]{Harris:1996}. The faint star cluster measurements are taken from \citet{2011AJ....142...88F, 2013ApJ...767..101B,2015ApJ...803...63K,2016ApJ...820..119K,2016ApJ...822...32W,2016ApJ...830L..10M,2017MNRAS.468...97L,2018ApJ...860...66M,2018MNRAS.478.2006L,2018ApJ...852...68C,2019MNRAS.490.1498L,2019MNRAS.484.2181T,2019ApJ...875..154M,Homma:2019,2020ApJ...890..136M,2021RNAAS...5..159G}. We also include the DELVE 2 stellar system \citep{2021ApJ...910...18C} in this category, although this system's true classification remains unknown.
\par The dwarf galaxy measurements for the same panel are taken from \citet{2012AJ....144....4M,Koposov_2015,2015ApJ...804L...5M,2015ApJ...808L..39K,2016ApJ...833...16K,2016ApJ...824L..14C,2016MNRAS.463..712T,2017AJ....154..267C,2018ApJ...860...66M,2018MNRAS.475.5085T,Homma:2018,2018ApJ...863...25M,2018MNRAS.480.2609L, Torrealba:2019,Homma:2019,2019ApJ...881..118W,2020ApJ...892..137S,2020ApJ...892...27M,2020ApJ...890..136M,2021ApJ...916...81C,2021ApJ...920L..44C}.

The right panel of the same figure shows the $\rm [Fe/H]-r_{1/2}$ plane, including only dynamically-confirmed Milky Way dwarf galaxies (solid blue triangles in the left panel). The metallicity measurements for these systems are taken from \citet{2009ApJ...702L...9C,2015ApJ...808...95S, 2011AJ....142..128W, 2015ApJ...811...62K,2015ApJ...810...56K, 2016MNRAS.463..712T,2016ApJ...833...16K,2017ApJ...838....8L,2017ApJ...839...20C, 2018ApJ...857..145L,2018MNRAS.479.5343K, Torrealba:2019, 2019ARA&A..57..375S,2020ApJ...892..137S,2020MNRAS.495.3022P, 2021NatAs...5..392C,2021ApJ...920...92J,2021arXiv210710849L}.

\end{document}